\documentclass[oldversion]{aa}
\usepackage{txfonts}
\usepackage{natbib}
\usepackage{graphicx}
\bibpunct{(}{)}{;}{a}{}{,} 
\newcommand{\epsa}{\ensuremath{\varepsilon} Indi A}
\newcommand{\eps}{\ensuremath{\varepsilon} Indi}

\newcommand{\epsba}{\ensuremath{\varepsilon} Indi Ba}
\newcommand{\epsbb}{\ensuremath{\varepsilon} Indi Bb}
\newcommand{\micron}{\,\ensuremath{\mu}\rm{m}}
\newcommand{\mjup}{\,\ensuremath{\rm{M_{Jup}}}}
\newcommand{\asppix}{\mbox{\arcsec\,pixel$^{-1}$}}           

\begin{document} 

\title{$\varepsilon$ Indi Ba, Bb: a detailed study of the nearest known brown dwarfs \thanks{Based on observations collected with the
    ESO VLT, Paranal, Chile under program 072.C-0689}}
\author{Robert R. King\thanks{Email: rob@astro.ex.ac.uk} \inst{1}
\and Mark J. McCaughrean \inst{1,2}
\and Derek Homeier \inst{3}
\and \\France Allard \inst{4}
\and Ralf-Dieter Scholz \inst{5}
\and Nicolas Lodieu \inst{6}}

\institute{School of Physics, University of Exeter, Stocker Road,
  Exeter EX4 4QL, UK
 \and Research \& Scientific Support Department, ESA ESTEC, Keplerlaan 1, 2200 AG Noordwijk,
 The Netherlands
  \and Institut f\"{u}r Astrophysik, Georg-August-Universit\"{a}t,
  Friedrich-Hund-Platz 1, 37077 G\"{o}ttingen, Germany
  \and Centre de Recherche Astrophysique de Lyon, UMR 5574: CNRS,
  Universit\'{e} de Lyon, \'{E}cole Normale Sup\'{e}rieure de Lyon,
  \\46 all\'{e}e d'Italie, 69364 Lyon C\'{e}dex 07, France
  \and Astrophysikalisches Institut Potsdam, An der Sternwarte 16,
  14482 Potsdam, Germany
  \and Instituto de Astrof\'{i}sica de Canarias, V\'{i}a L\'{a}ctea
  s/n, E-38200 La Laguna, Tenerife, Spain}

\date{Received 24 July 2009 / Accepted 06 Nov 2009}

\abstract {The discovery of \epsba, Bb, a binary brown dwarf system
  very close to the Sun, makes possible a concerted campaign to
  characterise the physical parameters of two T dwarfs. Recent
  observations suggest substellar atmospheric and evolutionary models
  may be inconsistent with observations, but there have been few
  conclusive tests to date. We therefore aim to characterise these
  benchmark brown dwarfs to place constraints on such models. We have
  obtained high angular resolution optical, near-infrared, and
  thermal-infrared imaging and medium-resolution (up to R$\sim$5\,000)
  spectroscopy of \epsba, Bb with the ESO VLT and present
  $VRIzJHKL'M'$ broad-band photometry and 0.63--5.1\micron\
  spectroscopy of the individual components. The photometry and
  spectroscopy of the two partially blended sources were extracted
  with a custom algorithm. Furthermore, we use deep AO-imaging to
  place upper limits on the (model-dependent) mass of any further
  system members. We derive luminosities of log
  L/L$_{\sun}$=$-4.699\pm0.017$ and $-5.232\pm0.020$ for \epsba, Bb,
  respectively, and using the dynamical system mass and COND03
  evolutionary models predict a system age of 3.7--4.3\,Gyr, in excess
  of previous estimates and recent predictions from observations of
  these brown dwarfs.  Moreover, the effective temperatures
  of 1352--1385\,K and 976--1011\,K predicted from the COND03
  evolutionary models, for \epsba\ and Bb respectively,
  are in disagreement with those derived from the comparison of our
  data with the BT-Settl atmospheric models where we find effective
  temperatures of 1300--1340\,K and 880--940\,K, for \epsba\ and Bb
  respectively, with surface gravities of log g=5.25 and 5.50.
  Finally, we show that spectroscopically determined
  effective temperatures and surface gravities for ultra-cool dwarfs
  can lead to underestimated masses even where precise luminosity
  constraints are available.\thanks{The full resolution spectra of both 
  brown dwarfs are available in electronic form at the CDS via anonymous 
  ftp to cdsarc.u-strasbg.fr}}



\titlerunning{\epsba, Bb: a detailed study} 
\maketitle

\section{Introduction}
\label{sec:intro}

The characterisation of low-mass stars and brown dwarfs is important
for studies of substellar and planetary atmospheres, the reliable
application of low-mass evolutionary models, and the derivation of the
full initial mass function. With over five hundred L and over one
hundred T dwarfs now known\footnote{http://dwarfarchives.org - the M,
  L, and T dwarf compendium maintained by Chris Gelino, Davy Kirkpatrick, and Adam
  Burgasser.}, statistical studies of global properties and detailed
studies of the closest objects are now possible.

Binary systems have an important role to play. They allow the
determination of dynamical masses, provide a laboratory in which
objects with the same age and chemical composition may be compared,
and, where they have main-sequence companions, provide external
constraints of metallicity and age which isolated objects lack,
breaking the substellar mass-luminosity-age degeneracy.


To fully constrain the evolutionary models of substellar objects 
\citep[e.g.][]{Burrows:1997, Baraffe:2003, Saumon:2008}, it would be
most useful to determine the bolometric luminosity, radius, mass, and
age of a range of such objects. Bolometric luminosities can be
determined from photometric and spectroscopic observations across a
large wavelength range. Masses can be determined in systems where an
orbit may be monitored, and finally, the age and metallicity can be
inferred from better characterised stars in the same system. To
constrain the atmospheric models of brown dwarfs, it is necessary to
acquire high signal-to-noise spectra over as wide a wavelength range as
possible, allowing robust estimates of the effective temperature and
surface gravity to be made.

The discovery of a distant companion (projected separation
$\sim$1500\,AU) to the high proper-motion ($\sim$4.7\,arcsec/yr) K4.5V
star, $\varepsilon$ Indi, was reported by \citet{Scholz:2003}. One of
our nearest neighbours, \epsa\ has a well-constrained parallax from
HIPPARCOS \citep{Perryman:1997} as refined by \citet{vanleeuwen:2007},
putting the system at a distance of 3.6224$\pm$0.0037\,pc. This was
followed by the discovery of the companion's binary nature
\citep{McCaughrean:2004}. The proximity of \epsba, Bb to the Earth
means Ba is more than a magnitude brighter than any other known T
dwarf, and allows unprecedented, detailed spectroscopic studies of
these important template objects.


\epsba, Bb are uniquely suited to provide key insights into the
physics, chemistry, and evolution of substellar sources. Although there
are a number of other T dwarfs in binary systems, such as the
M4/T8.5 binary Wolf\,940 \citep{Burningham:2009} and the T5/T5.5 binary
2MASS\,1534$-$2952 \citep{Liu:2008}, \epsba, Bb has a very
well-determined distance, a main-sequence primary star with which to
constrain age and metallicity, and a short enough orbit
\citep[nominally $\sim$15\,years, ][]{McCaughrean:2004} such that the
system and individual dynamical masses can soon be determined
\citep[][in prep.]{McCaughrean:2009, Cardoso:2010}. They are also
relatively bright, close enough, and sufficiently separated to allow
detailed photometric and spectroscopic studies of both components. 
Importantly, these two objects roughly straddle the L to T transition
\citep[cf.][]{Burgasser:2009} where the atmospheres of substellar
objects alter dramatically. The study of these two coeval objects on
either side of the transition will help in understanding the processes
effecting the change from cloudy to cloud-free atmospheres.
Characterisation of this system allows the mass-luminosity-age
relation at low masses and intermediate age to be tested, investigation
of the atmospheric chemistry, including vertical up-mixing, and
detailed investigation of the species in the atmosphere. 




To date, spectroscopic observations of T dwarfs have predominantly been
either at low-resolution \citep[e.g., ][]{Burgasser:2002,Chiu:2006},
which allows spectral classification and overall spectral energy
distribution modelling to determine luminosities, or high-resolution
studies of relatively small wavelength regions to investigate gravity
and effective temperature-sensitive features. For example,
\citet{McLean:2003} presented near-IR spectra at a spectral
resolution of R$\sim$2000 of objects spanning spectral types M6 to T8,
and discussed broad changes in spectral morphology and dominant
absorbers through the spectral sequence. This was complemented by
R$\sim$20\,000 $J$-band spectra presented in \citet{McLean:2007} where
many H$_2$O and FeH features were identified and the progression of the
$J$-band potassium doublet from M to T dwarfs charted.


Previous studies of ultra-cool dwarfs attempting to constrain low-mass
evolutionary models have been hampered by ambiguous ages, possible
unresolved binarity, and the difficulty associated with acquiring
observations of close, faint companions. For example, observations of
AB\,Dor\,C have roused some controversy over the applicability of
current low-mass evolutionary models, with the assumed age of the
system being a major source of disagreement. \citet{Close:2005}
determined a dynamical mass for AB\,Dor\,C and, using an assumed age of
30--100\,Myr, argued that evolutionary models predicted a higher
luminosity than was observed. However, \citet{Luhman:2005} countered
with an analysis based on an age of 75--150\,Myr, finding no
significant discrepancy between the observations and models.
\citet{Nielsen:2005} further argued that using their slightly revised
age of 70$\pm$30\,Myr, the models still under-estimated the mass of
this object.  Again, this was disputed by \citet{Luhman:2006} after a
re-reduction of the same data used by \citet{Close:2005}, and then
\citet{Close:2007} concluded that, based on newly acquired spectra,
there was no discrepancy between the observations and models. Despite
this apparent rapprochement, the situation may nevertheless be further
complicated by the suggestion of \citet{Marois:2005} that AB Dor C may
itself be an unresolved binary.  More recently, \citet{Dupuy:2009}
presented a dynamical mass for the binary L dwarf system HD\,130948BC
which, along with an age estimated from the rotation of the
main-sequence parent star, suggests that the evolutionary models
predict luminosities 2--3 times higher than those observed.

\citet{Leggett:2008} used 0.8--4.0\micron\ spectra at R$\sim$100--460
and near- to mid-IR photometry of HN Peg B, a T2.5 dwarf companion to
a nearby G0V star, to investigate physical properties including dust
grain properties and vertical mixing. In the near-IR, the resolution
was too low to study spectral lines in detail. However, by fitting the
overall spectral morphology and making use of the longer-wavelength
data, they were able to place important constraints on vertical mixing
and sedimentation. \citet{Leggett:2009} also reported the physical
properties of four T8--9 dwarfs from fitting observed near- to mid-IR
spectral energy distributions with the atmospheric and evolutionary
models of \citet{Saumon:2008}. They discussed the effects of vertical
transport of CO and N$_2$ and demonstrated the complementary effects
of increasing metallicity and surface gravity.


\citet{Reiners:2007} analysed high-resolution (R$\sim$33\,000) optical
spectra of three L dwarfs and the combined \epsba, Bb system,
concluding that although some individual features are not well-matched
by the model atmospheres and that significant differences remain for
some molecular species and alkali metal features, general features are
reproduced. \citet{Smith:2003} also acquired high resolution
(R$\sim$50\,000) near-IR spectra of (only) \epsba\ in the wavelength
ranges 1.553--1.559\micron\ and 2.308--2.317\micron. These were fit
with the unified cloud models of \citet{Tsuji:2002} and effective
temperatures of 1400\,K and 1600\,K were derived for the two spectral
regions. Mid-IR spectroscopy of the unresolved \epsba, Bb system was
also acquired by \citet{Roellig:2004}, who use evolutionary models
along with the luminosities of \citet{McCaughrean:2004} and an assumed
age of 0.8--2.0\,Gyr to derive effective temperatures and surface
gravities and then compared composite spectral models to the observed
spectrum. Their predictions were revised by \citet{Mainzer:2007} who
derived effective temperatures of 1210--1250\,K and 840\,K, for \epsba\
and Bb respectively, under the assumption of a system age of
$\sim$1\,Gyr.

Finally, \citet{Kasper:2009} presented R$\sim$400 near-IR NACO/VLT
spectroscopy of \epsba, Bb which were compared to the evolutionary
models of \citet{Burrows:1997} and the atmospheric models of
\citet{Burrows:2006}. They derived effective temperatures of
1250--1300\,K and 875--925\,K, and surface gravities of log g
(cm\,s$^{-1}$) 5.2--5.3 and 4.9--5.1, for \epsba\ and Bb respectively,
by comparing their observed spectra with their spectral models scaled
using the distance and a radius predicted by their evolutionary
models. We will discuss these results further in
Sect.\,\ref{sec:model_diffs} in contrast to our new data.

In this paper we present high signal-to-noise photometry from the $V$-
to $M'$-band (0.5--4.9\micron) and medium resolution spectroscopy from
0.6--5.1\micron\ of the individual components of the \epsba, Bb
system. In Sect.\,\ref{sec:obs}, we describe the observations and data
reduction, including the routines employed to extract the
partially-blended photometry and spectroscopy. We re-derive the
spectral types of both objects according to the updated classification
scheme of \citet{Burgasser:2006} in Sect.\,\ref{sec:spec_class} and
discuss constraints imposed by the parent main-sequence star in
Sect.\,\ref{sec:EpsA_constrain}. We derive the luminosities of both
sources in Sect.\,\ref{sec:lum} and discuss the preliminary dynamical
mass measurement of \citet[][in prep.]{McCaughrean:2009} in
Sect.\,\ref{sec:mass}. Our observations are compared to evolutionary
models in Sect.\,\ref{sec:evo_model_comp} and to atmospheric models in
Sect.\,\ref{sec:atm_model_comp}. We then put limits on the masses of
lower-mass companions in Sect.\,\ref{sec:Bc_mass_limits}, and finally
the predictions of evolutionary and atmospheric models and previous
determinations are compared in Sect.\,\ref{sec:model_diffs}.



\section{Observations and Reduction} 
\label{sec:obs} 

\epsba, Bb were observed with the ESO VLT using FORS2/UT1
\citep{Appenzeller:1998} for optical photometry and spectroscopy,
ISAAC/UT1 \citep{Moorwood:1998} for near- to thermal-IR
photometry and spectroscopy, and NACO/UT4
\citep{Lenzen:2003,Rousset:2003} for deep near-IR AO imaging. In all
observations except those using NACO, the point-spread functions (PSFs)
of the two sources were partially blended, even under excellent
observing conditions with seeing always less than 0.7$\arcsec$.


\subsection{optical photometry} 
\label{sec:opt_phot} 

\begin{figure}
  \centering{}
 
\resizebox{\hsize}{!}{\fbox{\includegraphics{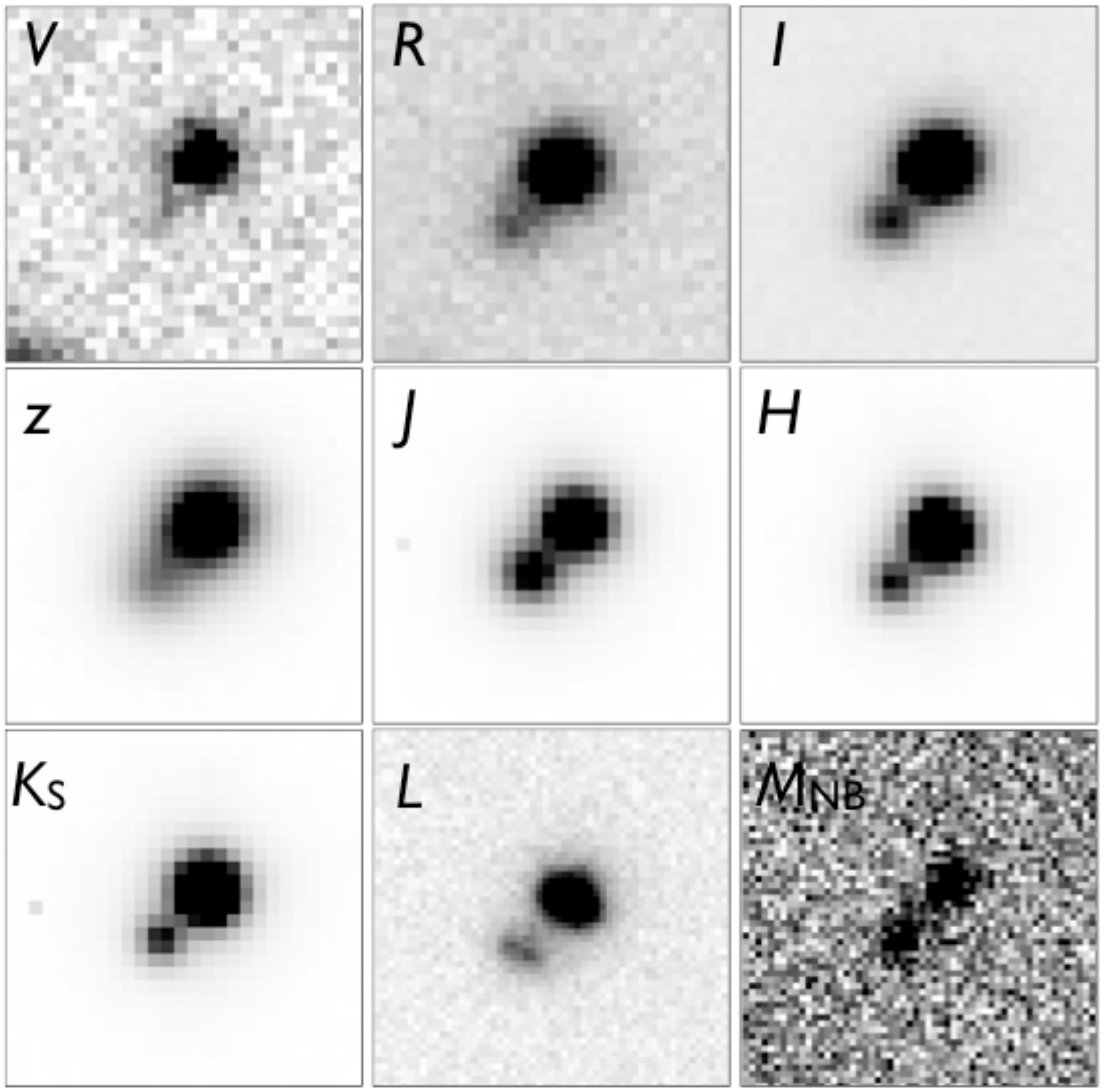}}}
  \caption{From left to right and top to bottom, FORS2 $VRIz$ and
    ISAAC $JHK_{\rm{S}}LM_{\rm{NB}}$ images of \epsba, Bb. Each image
    is a 4\arcsec\,$\times$\,4\arcsec\ sub-section of the full image.
    North is up, East left. The object to the south-east of the
    $V$-band image is a faint galaxy not seen in the other
    pass-bands. The $L$-band profiles are seen to be slightly
    elliptical. The $V$ and $M_{\rm{NB}}$-band images are the stacked
    images of all observations in those pass-bands.}
    \label{fig:images}
\end{figure}

Broadband $VRIz$ photometry was obtained on June 19 and July 20 2004
(UT) using FORS2 (2 CCDs each 2048\,$\times$\,4096 pixels) in high
resolution mode with 2\,$\times$\,2 binning resulting in a plate-scale
of 0.125\asppix\ and a field-of-view of
4.25\arcmin\,$\times$\,4.25\arcmin. \epsba\ and Bb were separated by
$\sim$0.84$\arcsec$ under photometric conditions with median seeing of
0.55$\arcsec$ FWHM\@. Five images dithered by 1\arcmin\ from a central
position were obtained in each filter except the $R$-band where twelve
dithered images were taken. Individual exposure times were 500\,s,
60\,s, 20\,s, and 10\,s in the $VRIz$ bands, respectively, giving
total integration times of 42 min, 12 min, 100\,s, and 50\,s. Sky
subtraction and flat-fielding were carried out with standard IRAF
programs. As seen in Fig.\,\ref{fig:images}, both components of the
binary are well-detected in the $RIz$ bands, but \epsbb\ is only
marginally detected in the $V$-band. We used the FORS2 Bessell $V$,
Special $R$, Bessell $I$, and Gunn $z$ broadband filters. Observations
of the standard star fields PG~2213-006 and Mark-A
\citep{Landolt:1992} were taken for photometric calibration which is
discussed in detail in Sect.\,\ref{sec:phot_calib}.

The large field-of-view meant there were sufficient bright field stars
with which to model the PSF and so DAOPHOT/IRAF PSF-fitting was
employed to extract the individual fluxes of the two brown dwarfs. For
the $RIz$ bands, each of the images were fit separately to allow a
determination of the accuracy of the profile fitting which is included
in the uncertainties of the derived magnitudes.  In the $V$-band,
\epsbb\ was only marginally detected, so we were unable to fit PSFs to
the two components of the binary.  To extract the photometry of both
sources, we used a small aperture to measure the flux of the brighter
source ensuring there was no appreciable contaminant flux from the
fainter source. We then extracted the photometry of the combined source
with a larger, circular aperture and, using the curve of growth
of brighter stars in the field, derived the excess flux due to \epsbb\
and thus an upper limit on the $V$-band flux.

The measured flux ratio and the central wavelength and width for each
of the observed filters is listed in Table~\ref{tab:ratio_errors},
while Table~\ref{tab:mags} lists the derived photometry.

\begin{table}
  \centering
  \caption{The flux ratio (Ba/Bb) of \epsba, Bb in the observed FORS2
and ISAAC filters and the effective wavelength and half-power width of those
filters. See Sect.\,\ref{sec:phot_calib} for a description of
photometric calibration and transformation to standard systems. All
except the $V$-band ratio were measured using PSF-fitting (see
Sect.\,\ref{sec:opt_phot}).}
\label{tab:ratio_errors}
\begin{tabular}{clll}
  \hline
  \hline
  Observed Passband &  Ratio (Ba/Bb)  & $\lambda_{\rm{eff}}\,(\mu$m)  &  $\Delta\lambda\,(\mu$m)  \\
  \hline
  $V$ & 9.82 $\pm$ 0.66             & 0.554 & 0.112 \\ 
  $R$ & 4.81 $\pm$ 0.07             & 0.655 & 0.165 \\
  $I$ & 5.04 $\pm$ 0.05             & 0.768 & 0.138 \\
  $z$ & 3.85 $\pm$ 0.03             & 0.910 & 0.131 \\
  $J$ & 2.28 $\pm$ 0.02             & 1.25  & 0.29  \\
  $H$ & 5.27 $\pm$ 0.06             & 1.65  & 0.30  \\
  $K_{\rm{S}}$ & 7.12 $\pm$ 0.02     & 2.16  & 0.27  \\
  $L$ & 4.50 $\pm$ 0.16             & 3.78  & 0.58  \\
  $M_{\rm{NB}}$ & 1.35 $\pm$ 0.15    & 4.66 & 0.10 \\
  \hline
\end{tabular}
\end{table}

\begin{table}
  \centering

  \caption{The derived apparent magnitudes for \epsba, Bb. The optical
    magnitudes are in the FORS2 system as described in
    Sect.\,\ref{sec:phot_calib}, and the near-IR magnitudes have been
    transformed into the MKO $JHKL'M'$ filters via synthetic
    photometry. The uncertainties for both objects are dominated by
    the uncertainties on the measured flux ratio and the standard star
    magnitudes. Absolute magnitudes may be derived using the distance
    modulus of $-2.205$.}

  \label{tab:mags}
  \begin{tabular}{lll}
	\hline
	\hline
Passband &  Ba         & Bb                 \\
	\hline	
$V$     & 24.12 $\pm$ 0.03    & $\geq$ 26.60$\pm$ 0.05 \\
$R$     & 20.65 $\pm$ 0.01    & 22.35 $\pm$ 0.02 \\
$I$     & 17.15 $\pm$ 0.02    & 18.91 $\pm$ 0.02 \\
$z$     & 15.07 $\pm$ 0.02    & 16.53 $\pm$ 0.02 \\
$J$     & 12.20 $\pm$ 0.03    & 12.96 $\pm$ 0.03 \\
$H$     & 11.60 $\pm$ 0.02    & 13.40 $\pm$ 0.03 \\
$K$     & 11.42 $\pm$ 0.02    & 13.64 $\pm$ 0.02 \\
$L'$    & \ \ 9.71 $\pm$ 0.06 & 11.33 $\pm$ 0.06 \\
$M'$    & 10.67 $\pm$ 0.23    & 11.04 $\pm$ 0.23 \\

	\hline		

\end{tabular}
\end{table}

\subsection{near- and thermal-IR photometry} 
\label{sec:nir_phot} 

The ISAAC imager was used on November 5 and 11 2003 (UT) to obtain
photometry of \epsba, Bb in the ISAAC $JHK_{\rm{S}}LM_{\rm{NB}}$
filters. ISAAC was used with the ALADDIN array (1024\,$\times$\,1024
pixels) in long-wavelength imaging modes LWI3 and LWI4 for the
$JHK_{\rm{S}}$ and $LM_{\rm{NB}}$ imaging, respectively, resulting in
plate-scales of 0.148\asppix\ and 0.071\asppix\ and fields-of-view of
151\,$\times$\,151\arcsec\,and 73\,$\times$\,73\arcsec.

\epsba\ and Bb were separated by $\sim$0.77$\arcsec$ at this epoch and
were observed under photometric conditions with typical seeing of
$\sim$0.45$\arcsec$ FWHM\@. The $JHK_{\rm{S}}$ images were taken at
three positions dithered by $\sim$14\arcsec\ with total integration
times of 3.5\,min in each of the three filters. The three offset
images were combined to remove the sky flux and then flat-fielded with
standard IRAF programs. The $L$ and $M_{\mathrm{NB}}$ images were
taken in chop-nod mode with a throw of $\sim$20\arcsec\ and total
exposure times of 2\,min in the $L$-band, and 4\,min in the
$M_{\mathrm{NB}}$-band. The images were flat-fielded and the
half-cycle frames subtracted in the standard manner producing
three sky-subtracted images at different positions on the array.

The ISAAC field-of-view was found to be mostly empty, and most
importantly with no stars bright enough to fit a model PSF\@.
Therefore, a custom profile-fitting routine was implemented to
determine the flux ratio of the partially-blended objects (see
Sect.\,\ref{sec:image_fitting}). The total flux of the two objects was
then measured by aperture photometry. The centre and size of a
circular aperture were chosen so as to minimise the sky noise
contribution while ensuring the total flux was not biased toward either
object by including substantially more of the profile wings of one
object than the other.  The photometric calibration is discussed in
detail in Sect.\,\ref{sec:phot_calib} and the derived near-IR
photometry is listed in Table~\ref{tab:mags}.

\subsection{photometric variability}

\citet{Koen:2005a} detected $I_C$-band variability in optical
photometry of the combined $\varepsilon$ Indi Ba, Bb system on
time-scales of hours, reporting a linear rise in the $I_C$-band of
$\sim$0.16$^{\rm{m}}$ over the course of 3.6 hours. Table
\ref{tab:Iband_mags} shows the $I$-band magnitudes for the combined
system from different studies. For comparison, our observations have
been combined to give the magnitude of the unresolved system. However,
a direct comparison is hindered by the different response functions
employed in these measurements. We find that a shift in filter
response of 100\,\AA\ is sufficient to explain the spread in the
$I$-band magnitudes due to the steep rise in T dwarf optical spectra,
and so may shield any intrinsic variability.

Evidence for variability of \epsba, Bb of $\sim$0.05$^{\rm{m}}$ was
also found by \citet{Koen:2005} in the near-IR. Table
\ref{tab:mags_2mass} shows the 2MASS $JHK_{\rm{S}}$ magnitudes
extracted from our flux calibrated spectra (see
Sect.\,\ref{sec:phot_calib}) and the 2MASS magnitudes of
\citet{McCaughrean:2004} and \citet{Kasper:2009}. These mostly agree
within the stated uncertainties with the exception of the $H$-band
magnitude of \epsbb. However, the photometry of
\citeauthor{McCaughrean:2004} and \citeauthor{Kasper:2009} was
acquired in the VLT/NACO system with an $H$-band filter which
extends into the region of high telluric water absorption, possibly
accounting for the differing results. The uncertainties on our near-IR
photometry are of similar magnitude to the proposed variability.

We plan to use the near-IR photometry ($JHK_{\rm{S}}$) obtained as
part of the astrometric monitoring of this binary to further
investigate any variability over the monitoring epochs ($\sim$monthly
from August 2003 to present) to probe longer-term variability than
that detected by \citet{Koen:2005}.

\begin{table}
\centering
\caption{The $I$-band magnitudes adopted for the combined \epsba, Bb system by different
  studies. We have used the FORS2 Bessell $I$-band filter which is similar to the DENIS $I$-band, while \citet{Koen:2005a} used the Cousins $I$-band. }
\label{tab:Iband_mags}
\begin{tabular}{ccc}
  \hline
  \hline
  Date & $\varepsilon$ Indi Ba, Bb  &  Reference\\
  \hline
  \hline	
  1997.771  &   16.59$\pm$0.10  & SSS-UK \\
  1999.666  &   16.77$\pm$0.10  & SSS-UK \\
  2000.781  &   16.90$\pm$0.12  & DENIS  \\
  2004.462  &   16.70$\pm$0.04  & \citet{Koen:2005a} \\
  2004.546  &   16.95$\pm$0.03  & This work \\
  \hline
\end{tabular}
\end{table}

\begin{table}
\centering
\caption{The 2MASS magnitudes adopted for \epsba, Bb in different studies. McC04 refers to  \citet{McCaughrean:2004} for which we have assumed the uncertainties on the photometry to be equal to the uncertainty on the 2MASS photometry of the unresolved system. K09 is the study of \citet{Kasper:2009}.}
\label{tab:mags_2mass}
\begin{tabular}{llll}
  \hline
  \hline
  Passband &  \multicolumn{3}{l}{Ba} \\
  \hline
           &  McC04 & K09 & This work   \\
  \hline

  2MASS $J$         & 12.29 $\pm$ 0.02 & 12.33 $\pm$ 0.02 & 12.29 $\pm$ 0.03  \\
  2MASS $H$         & 11.51 $\pm$ 0.02 & 11.54 $\pm$ 0.02 & 11.50 $\pm$ 0.03  \\
  2MASS $K_{\rm{S}}$ & 11.35 $\pm$ 0.02 & 11.37 $\pm$ 0.02 & 11.38 $\pm$ 0.03  \\
  \hline
  & && \\
  \hline
  \hline
  Passband &  \multicolumn{3}{l}{Bb} \\
  \hline
           &  McC04 & K09 & This work   \\
  \hline

  2MASS $J$         & 13.23 $\pm$ 0.02 & 13.19 $\pm$ 0.03 & 13.23 $\pm$ 0.03 \\
  2MASS $H$         & 13.27 $\pm$ 0.02 & 13.36 $\pm$ 0.03 & 13.19 $\pm$ 0.03 \\
  2MASS $K_{\rm{S}}$ & 13.53 $\pm$ 0.02 & 13.51 $\pm$ 0.02 & 13.49 $\pm$ 0.03 \\
  \hline
\end{tabular}
\end{table}

\subsection{AO deep companion search} 
\label{deep_comp_search}

NAOS/CONICA (NACO) was used on November 7 2003 (UT) to obtain
deep adaptive-optics (AO) $H$-band imaging of the field around \epsba,
Bb. The N90C10 dichroic was used to send 10\% of the source flux to the
science camera and 90\% to the IR wavefront sensor (IR WFS). The median
natural seeing conditions were 0.58$\arcsec$ FWHM at the time of our
observations and with the AO correction we obtained a FWHM of
0.12$\arcsec$ in the final combined image.

The S27 camera was used with a plate-scale of 0.027\asppix\
resulting in a field-of-view of 27.7$\arcsec\,\times$\,27.7$\arcsec$.
We obtained a total of 72 individual images with exposure times of
135\,s, giving a total integration time of 162\,min. The 9-point
dither pattern used resulted in a continuous radial coverage of
10.94$\arcsec$ measured from the position of \epsba, corresponding to
39.6\,AU at 3.622\,pc. Sky subtraction, flat-fielding, image alignment
and stacking were standard.


As the two sources were well-separated, we were able to employ
DAOPHOT/IRAF to fit a PSF to \epsba\ and use this as a model to
subtract Bb, iteratively fitting the PSF and subtracting until Bb was
well-removed. This image of Ba alone was then used to add scaled and
offset objects into the original image to investigate our detection
limits. We find no sources from 7\,AU (71\,pixels, 1.93\arcsec) out to
the image edge, corresponding to 39.7\,AU, with flux greater than
0.1\% of \epsba, corresponding to a peak pixel flux 5$\sigma$ above
the background, or a source $H$-band magnitude of 19.1$^{\rm{m}}$
which would have been detectable in our deep image.

Closer to \epsba, Bb, the noise increases due to the Poisson noise
from the source flux and so the brightness limits on any companions
are higher. The pixelation of the profile also acts to suppress the
visibility of close companions. We find no sources with a flux greater
than 2\% of the \epsba\ flux ($H$=15.8$^{\rm{m}}$) down to 0.83\,AU
(8.5\,pixels, 0.23\arcsec) from Ba or Bb, and down to 0.4\,AU
(4.3\,pixels, 0.12\arcsec) from each object, we can discount any
sources with more than 10\% of the \epsba\ flux ($H$=14.1$^{\rm{m}}$).
In Sect.\,\ref{sec:Bc_mass_limits} we use these flux limits to derive
the mass limits on any possible companion.

\subsection{optical spectroscopy}
\label{sec:opt_spec}

FORS2 (2 CCDs each with 2048\,$\times$\,4096 pixels) was used on June
16 2004 (UT) to obtain optical (0.63--1.07\micron) spectroscopy of
$\varepsilon$ Indi Ba, Bb in long-slit mode with a 30\arcsec\ dither
along the slit. We used a 0.5$\arcsec$ wide slit, the HR collimator,
and 2\,$\times$\,2 binning mode, resulting in a plate-scale of
0.125\asppix. The 600\,RI and 600\,Z grisms were used to obtain the
full 0.63--1.07\micron\ spectrum, yielding resolutions of R$\sim$1000
at 6780\,\AA\ and R$\sim$2000 at 9700\,\AA, while the median seeing of
0.47$\arcsec$ FWHM allowed the spectra of the two objects to be
resolved at this epoch when the separation was
$\sim$0.84$\arcsec$. Total integration times of 80 min with the
600\,RI grism and 38 min with the 600\,Z grism were obtained by
co-adding 6$\times$800\,s and 5$\times$460\,s individual exposures,
respectively. 

The partially-blended spectra were extracted using a profile-fitting
routine as described in Sect.\,\ref{sec:spec_fit}. Wavelength
calibration used arc spectra for the 600\,RI grism and skylines for
the 600\,Z grism, with excellent agreement in the crossover
region. Our spectra have vacuum wavelengths.  There was no visible
difference between the two resolutions in the crossover region
(0.75--0.85\micron), so they were combined without
smoothing. Flat-fielding was carried out with lamp flats. Observations
were also made of the white dwarf LTT9491 for relative flux
calibration. However, these data were taken with a wider slit to the
$\varepsilon$ Indi Ba, Bb spectra and so could not be used for removal
of telluric absorption. A spline was fit to the observed white dwarf
spectrum which was used along with the known spectrum of LTT9491
\citep{Oke:1990, Hamuy:1994} to model the wavelength dependence of the
flat-field and detector response and so remove this from our observed
target spectra.


We scaled the NSO Kitt Peak atmospheric transmission spectrum of
\citet{Hinkle:2003} to construct a grid of atmospheric spectra with
varying absorption strengths (assuming all absorbing species vary
similarly with airmass) to model the atmosphere above Paranal at the
time of our observations. These were Gaussian smoothed to the
resolution of our observed spectra and the atmospheric model scaled to
maximise removal of known features. The telluric features appear to be
fully removed in the sections of the spectrum where we expect few
features intrinsic to T dwarfs, which leads us to believe that we have
acceptable telluric removal even across the 9300--9800\,\AA\ region
where there are blended telluric and intrinsic features.

\begin{figure*}
   
\resizebox{\hsize}{!}{\includegraphics{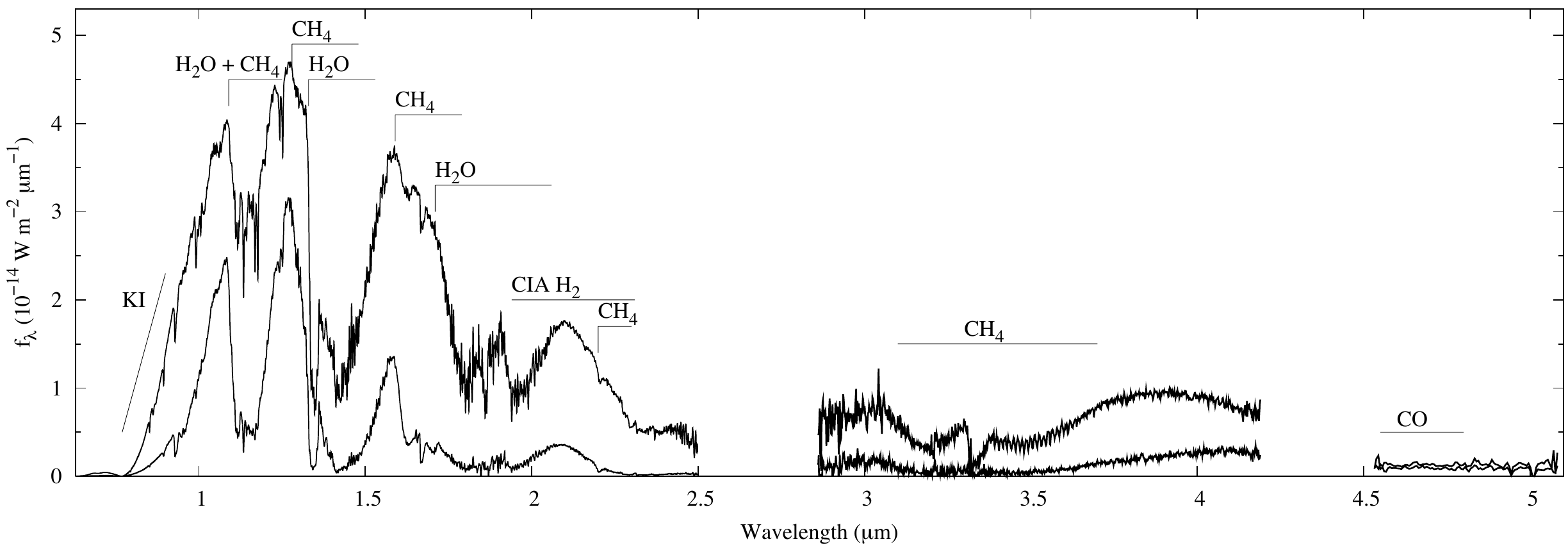}}   
    \caption{Full 0.6--5.1\micron\ spectrum of \epsba\ and Bb (upper
      and lower lines respectively). Here the full resolution optical
      and near-IR spectra have been smoothed to 17\,\AA~FWHM to allow
      inspection of the broad features without being dominated by the
      many fine features seen in the full resolution spectra (see
      Figs.\,\ref{fig:Bab_0.6-1.3}--\ref{fig:Bab_1.9-2.5}).}
    \label{fig:full_obs_spec}
\end{figure*}

\begin{figure*}
    \resizebox{\hsize}{!}{\includegraphics{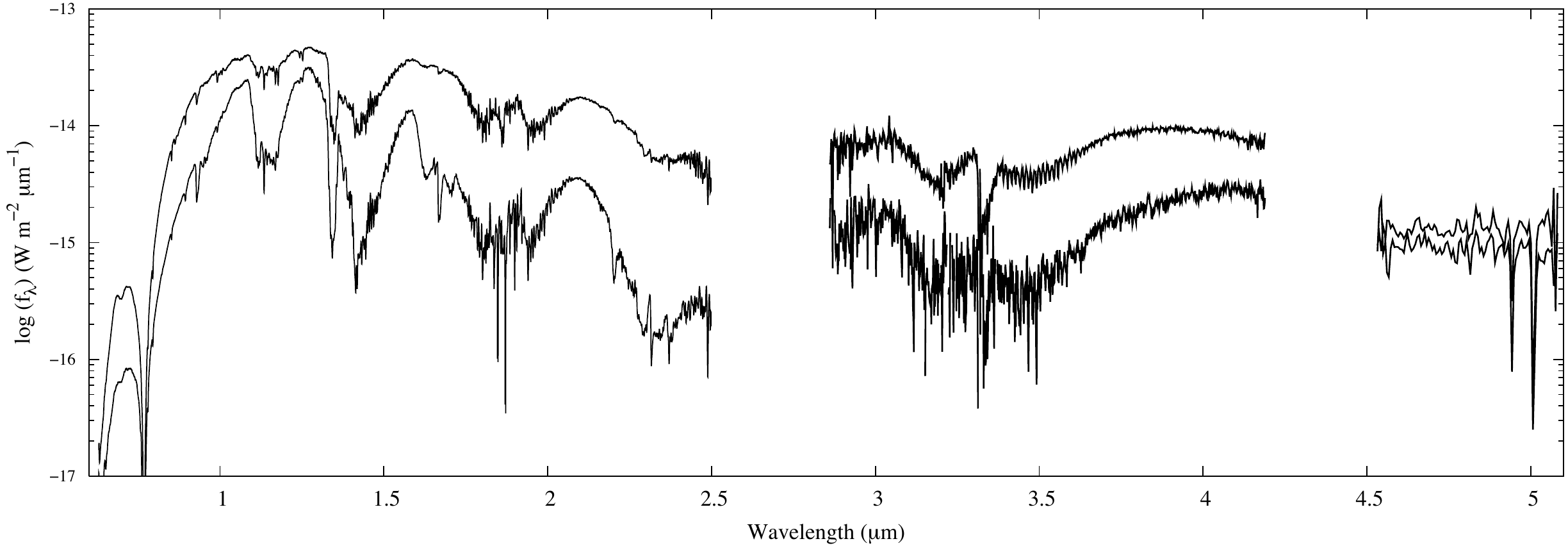}}   
    \caption{The same as Fig.~\ref{fig:full_obs_spec} but with the
      flux on a logarithmic scale. }
    \label{fig:full_obs_spec_log}
\end{figure*}

The final optical spectra have a peak signal-to-noise per pixel of
$\sim$300 and $\sim$200 for \epsba\ and Bb, respectively, falling to
$\sim$35 and $\sim$20 at 7000\AA. The 0.6--5.1\micron\ spectra of
\epsba, and Bb are shown in Figs.\,\ref{fig:full_obs_spec} and
\ref{fig:full_obs_spec_log} in both linear and logarithmic units, and
the full resolution spectra are shown in Figs.~\ref{fig:Bab_0.6-1.3},
\ref{fig:Bab_1.3-1.9}, and \ref{fig:Bab_1.9-2.5}.

\subsection{near-IR spectroscopy}
\label{sec:nir_spec-JHK}

ISAAC and its HAWAII array (1024\,$\times$\,1024 pixels) with a
plate-scale of 0.146\asppix\ was used on November 8 2003 (UT) in
short-wavelength medium-resolution (SWS1-MR) mode to obtain near-IR
spectroscopy from 0.9--2.5\micron\ of \epsba, Bb with a resolution of
R$\sim$5000. By turning the instrument rotator, both sources (at a
separation of $\sim$0.77$\arcsec$ at this epoch) were simultaneously
placed on the 0.6$\arcsec$ wide slit with median seeing of
0.50$\arcsec$ FWHM\@.  Three 60\,s exposures dithered by 20\arcsec\
along the 120\arcsec\ long slit were taken in each of twenty-one
wavelength regions to cover the entire 0.9--2.5\micron\ range.  We
observed three spectral regions in the 0.98--1.10\micron\ domain 
each 0.046\micron\ wide, and six spectral regions in each of the
1.10--1.40\micron, 1.40--1.82\micron, and 1.82--2.50\micron\ spectral
domains each covering a range of 0.059\micron, 0.079\micron, and
0.122\micron, respectively, with a minimum cross-over between regions
of 0.004\micron\ allowing full 0.9--2.5\micron\ spectra to be compiled
with no gaps in coverage. 

\begin{figure} 
\centering
\resizebox{\hsize}{!}{\includegraphics{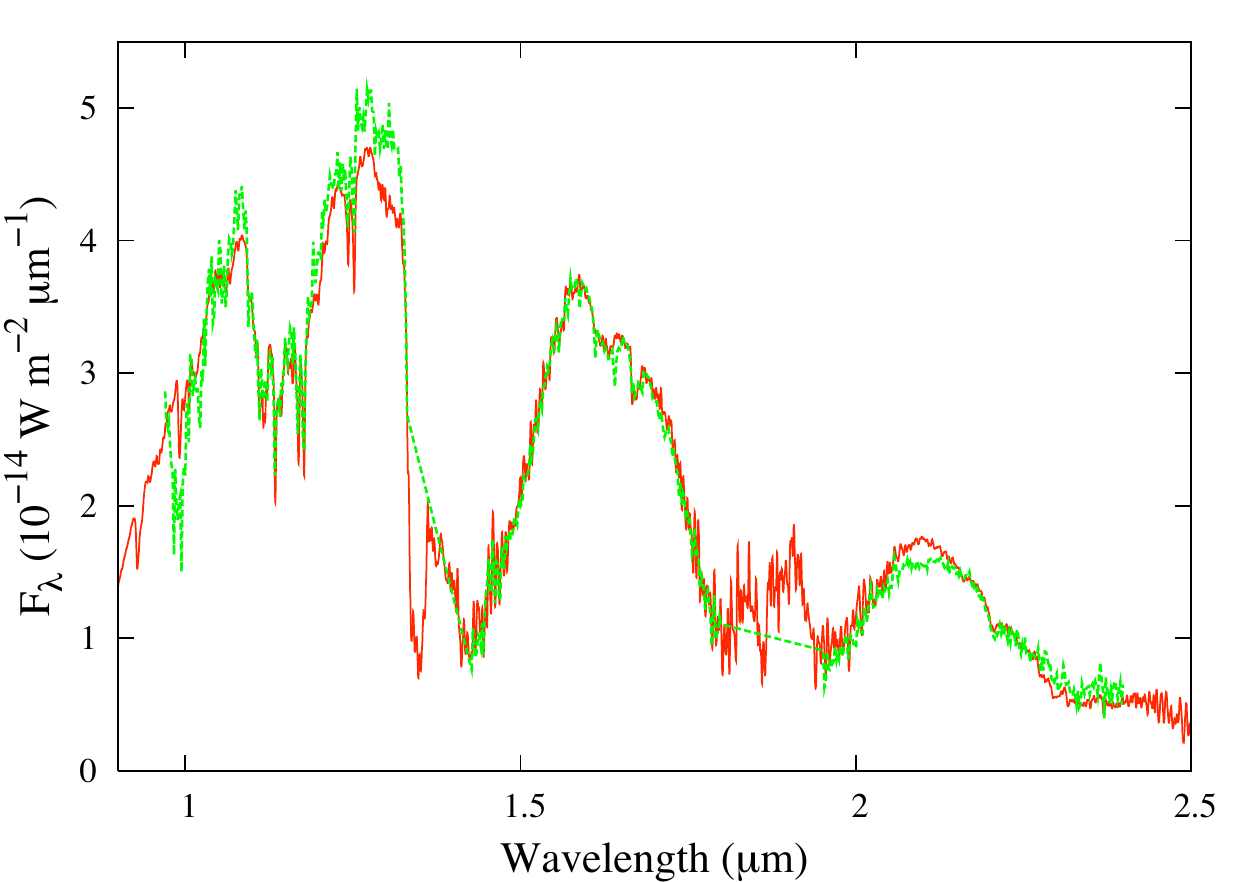}}
\resizebox{\hsize}{!}{\includegraphics{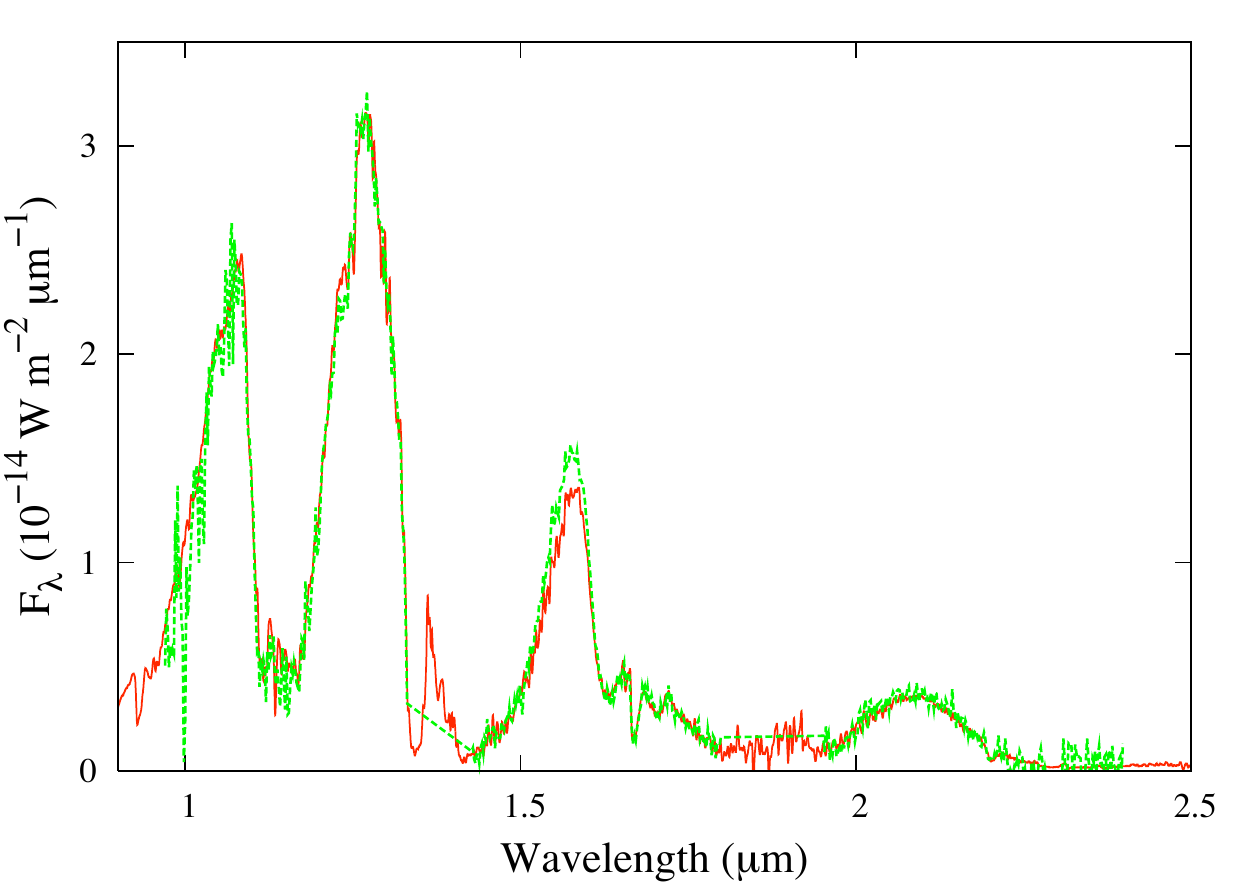}}

\caption{Our spectra of \epsba\ and Bb (red lines), smoothed to a
resolution of 17\AA\ FWHM and median filtered in the regions of high
telluric absorption, compared to the lower resolution spectroscopy of
\citet{Kasper:2009} (green lines). It is seen that the overall match of
the absolute scale is reasonable. Here we have absolutely
flux-calibrated using our ISAAC photometric observations, while
\citet{Kasper:2009} use the 2MASS point source catalogue magnitudes.}

\label{fig:spec_compare_Bab} 
\end{figure}

Observations were also made of the nearby G2 dwarf HD209552 in each of
the wavelength regions interspersed with the target observations in
order to flux calibrate and correct for telluric absorption as
first described by \citet{Maiolino:1996}. The airmass difference
between the observations of \epsba, Bb and the telluric standard were
in the range 0.07--0.17. Tungsten-illuminated spectral flats were
taken in the same configuration at the end of the night and wavelength
calibration was achieved through use of OH skylines for all but four of
the wavelength regions which had too few (known) skylines for a
reasonable fit. In these cases, XeAr arc spectra were used and
cross-over regions ensured a good match throughout.

For each of the wavelength regions, the three dithered spectral images
were combined to subtract the sky background. This process leaves
residuals of the order of a few percent of the observed counts in the
spectrum caused by temporal variations in the strength of the sky
emission, negligible for all but the strongest sky lines.  The residual
sky was further subtracted by taking apertures above and below the
spectra and subtracting the sigma-clipped mean sky level. The images
were then divided by the spectral dome flats.  The partially-blended
spectra were extracted in the same manner as the optical (see
Sect.\,\ref{sec:spec_fit}). The extracted spectra of \epsba\ and Bb
were divided by the standard star spectrum and corrected using the
solar spectrum of \citet{Wallace:1996} smoothed to the resolution of
our observations to remove the wavelength dependence of the detector
and the flat-field.  In the regions of very high telluric absorption
between the $J$ and $H$ and $H$ and $K$ bands where there were gaps in
the observed solar spectrum (13517--14032\,\AA\ and 18024--19317\,\AA),
we substituted the Kurucz IRRADIANCE
model\footnote{http://kurucz.harvard.edu/sun/irradiance/solarirr.tab}
\citep{Kurucz:2005} to avoid any breaks in the coverage of our
spectrum. These regions are generally omitted from published spectra,
but since we have high-enough signal-to-noise and spectral resolution,
we retain them to allow a comparison of the continuum flux in these
regions to model predictions. 

The known spectral type of the standard star, response of
the filter systems used, and magnitudes of the standard stars allowed us to
flux calibrate a template spectrum for each standard star in each of the
$JHKL'M'$ passbands. We then employed the ratio of the observed fluxes of
the targets and standard stars in the ISAAC system, and the known filter
responses to flux calibrate our observed \epsba\ and Bb spectra.  This was
done separately for each passband as each region is bounded by high
telluric absorption, but the optical spectra were flux calibrated by
scaling to match the $J$-band spectrum in the 9782--9999\,\AA\ cross-over
region. Figure\,\ref{fig:spec_compare_Bab} shows a comparison of our final
near-IR spectra for both brown dwarfs with those from \citet{Kasper:2009}.
No scaling has been applied to these spectra which, nonetheless, have
similar absolute flux calibrations.  While there are some small
discrepancies between the spectra, the overall match suggests that no
large systematic error has been introduced by the combination of many
short wavelength regions to produce our full near-IR spectrum. The final
spectra have a signal-to-noise of $\sim$80/60 (Ba/Bb) per pixel at the
$J$-band peak, 100/70 at the $H$-band peak, and $\sim$100/50 at the peak
of the $K$-band spectrum.  The full resolution observed spectra of both
objects are available at CDS and on the author's
web-pages.\footnote{http://www.astro.ex.ac.uk/people/rob/Research}

\subsection{thermal-IR spectroscopy}
\label{sec:nir_spec-LM}

ISAAC and its ALADDIN array (1024\,$\times$\,1024 pixels) with a
plate-scale of 0.146\asppix\ was used on November 6--7 2003 (UT)
in long-wavelength, low-resolution (LWS3-LR) mode to obtain R$\sim$600
and R$\sim$500 spectroscopy of \epsba, Bb in the spectral ranges
2.86--4.19\micron\ ($L$-band) and 4.53--5.08\micron\ ($M$-band),
respectively, using slit widths of 0.6$\arcsec$ and 1.0$\arcsec$.  The
$M$-band spectra had to be binned before they could be extracted giving
a final resolution of R$\sim$220 at 4.75\micron.  Both sources were
placed on the 120\arcsec\ long slit and chop-nod mode was used
resulting in a total on-source time of 30 min and 35 min for the $L$-
and $M$-bands, respectively, with median seeing of 0.54$\arcsec$
FWHM\@.  The half-cycle frames were subtracted in the standard manner
producing three sky-subtracted spectra at different positions on the
array.  Observations were also made of the solar-type star HD210272 in
the same manner to allow flux calibration and correction for telluric
absorption as for the 0.9--2.5\micron\ spectra.  The
\citet{Wallace:1996} solar spectrum was incomplete with a gap in the
range 4.17--4.55\micron\  which affected the last $\sim$0.02\micron\
 of the $L$-band spectrum and the first $\sim$0.02\micron\  of
the $M$-band spectrum. Again, the gap was filled by the Kurucz
IRRADIANCE model. Tungsten flats were taken with the same slits at the
end of the night and wavelength calibration was achieved through use of
XeAr arc spectra. Section\,\ref{sec:spec_fit} explains in detail how
the partially-blended spectra were extracted.  The final spectra
have a peak signal-to-noise of $\sim$40/15 (Ba/Bb) per pixel in the
$L$-band and $\sim$10 per pixel for both objects in the $M$-band.

\begin{figure*}
    \resizebox{\hsize}{23cm}{\includegraphics{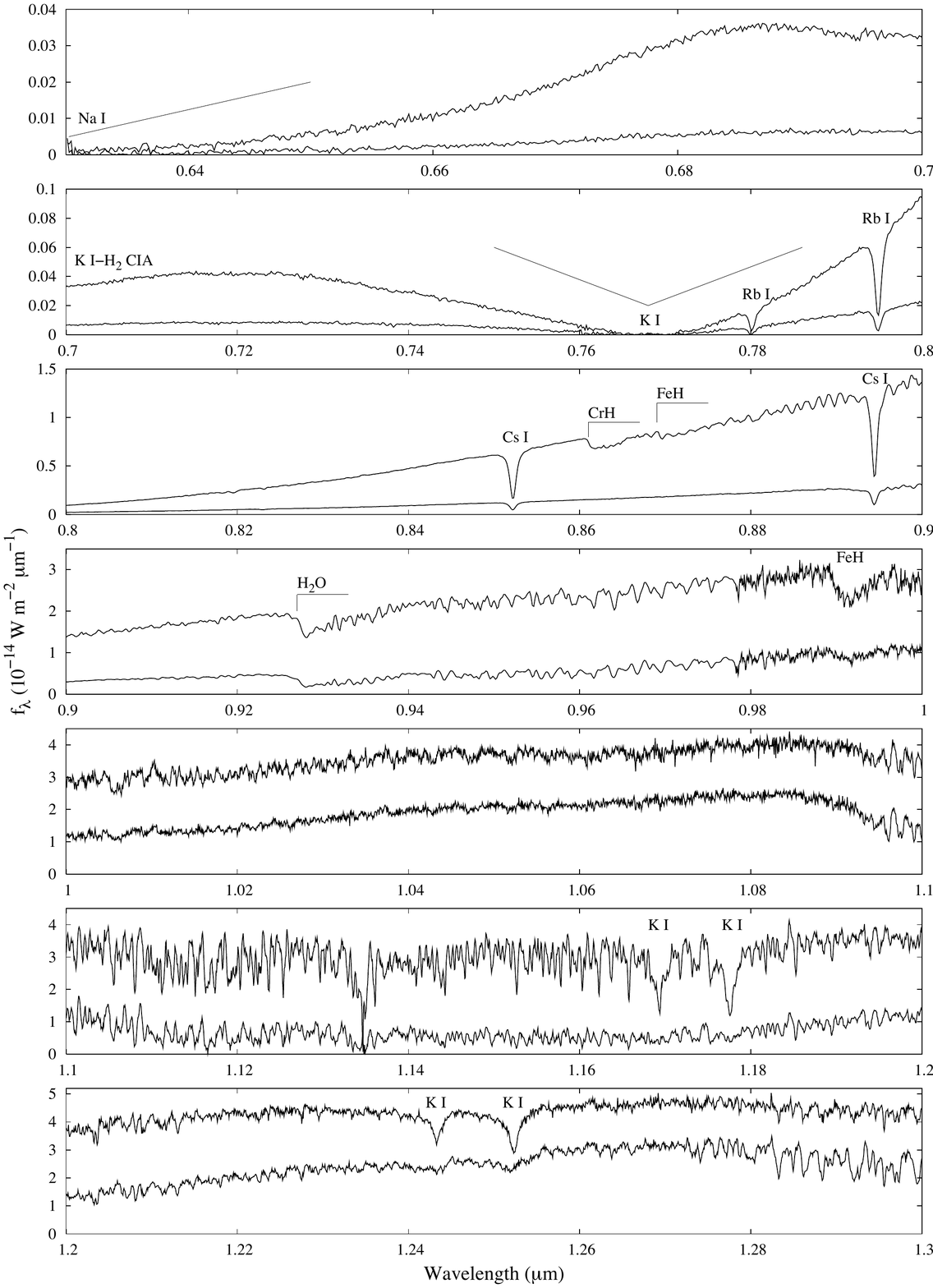}}   
    \caption{The full resolution spectra of \epsba\ (top line) and Bb
      for 0.63--1.3\micron. The FORS2 optical data is joined to the
      higher-resolution ISAAC near-IR data at 0.978\micron\ where we
      start to see even finer spectral features due to the higher
      spectral resolution and high signal-to-noise. Comparison of the
      two spectra reveals much of the apparent ``noise'' to be real
      spectral features present in both objects.  The resolution is
      6.8\,\AA\ FWHM up to $\lambda=0.79\micron$, 6.5\,\AA~FWHM for
      $\lambda=$0.79--0.978\micron, 1.8\,\AA\ FWHM for
      $\lambda=$0.978--1.1\micron, and 2.4\,\AA\ FWHM for
      $\lambda=$1.1--1.3\micron. Key spectral features are
      labelled. See \citet{McLean:2003} for a detailed discussion of
      features identified in brown dwarf spectra.}
    \label{fig:Bab_0.6-1.3}
\end{figure*}

\begin{figure*}
    \resizebox{\hsize}{23cm}{\includegraphics{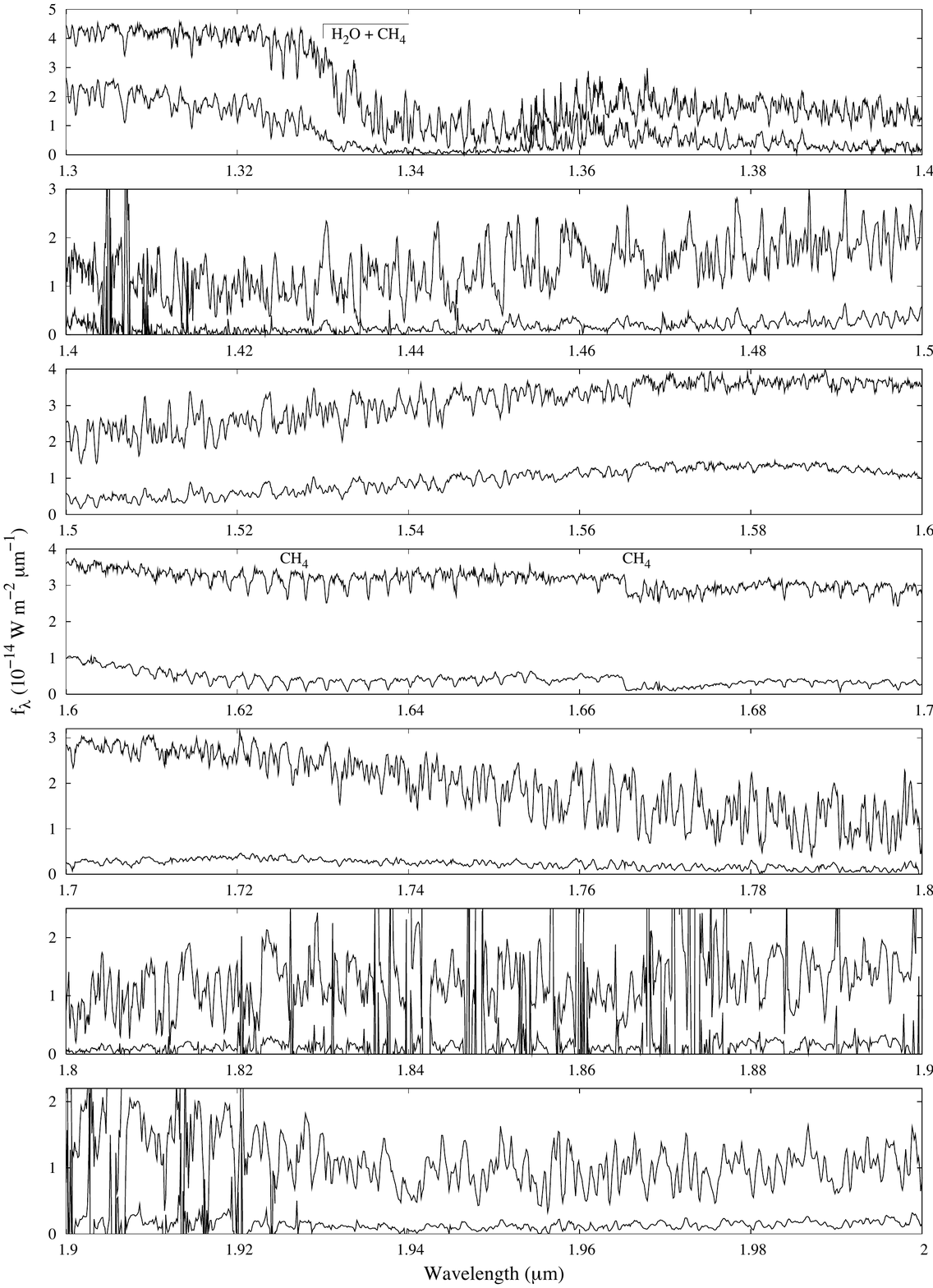}}
    \caption{Same as Fig. \ref{fig:Bab_0.6-1.3} but resolution is
      2.4\,\AA\ FWHM for $\lambda=$1.3--1.4\micron, 3.2\,\AA\ FWHM for
      $\lambda=$1.4--1.82\micron, and 4.9\,\AA\ FWHM for
      $\lambda=$1.82--2.0\micron. The regions 1.40--1.41\micron\ and
      1.82--1.92\micron\ have many artefacts due to high levels of
      telluric contamination. However, the high signal-to-noise and
      spectral resolution of these data allow us to extract the object
      spectra between the water lines, and so even in these regions we
      have a measure of the continuum flux.}
    \label{fig:Bab_1.3-1.9}
\end{figure*}

\begin{figure*}
    \resizebox{\hsize}{23cm}{\includegraphics{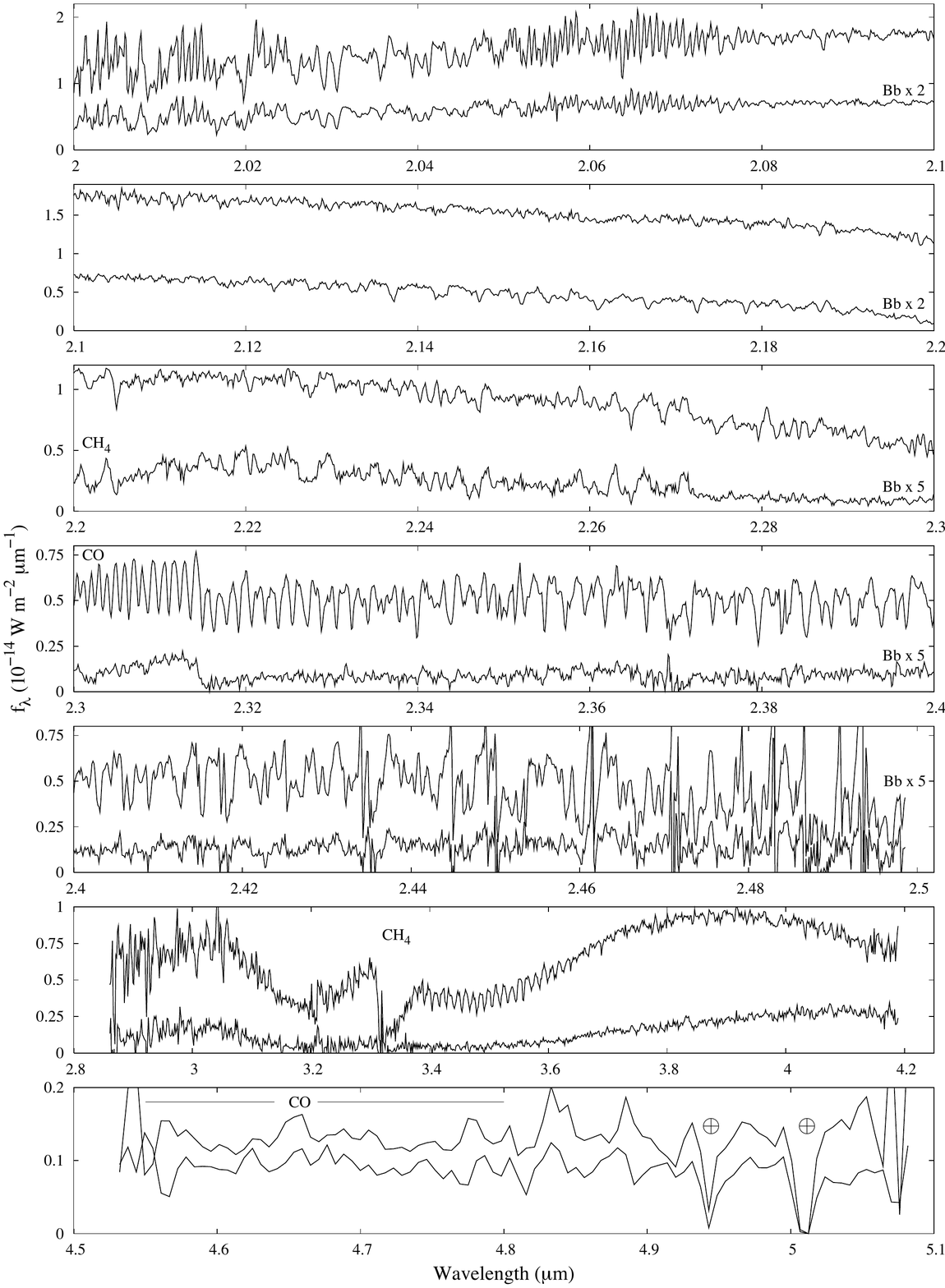}}
    \caption{Same as Fig. \ref{fig:Bab_0.6-1.3} but the observed
      spectra have gaps 2.50--2.86\micron\ and 4.19--4.53\micron\
      between the near-IR and the $L$- and $M$-bands.  The resolution
      is 4.9\,\AA\ FWHM for $\lambda=$2.0--2.5\micron, 60\,\AA\ FWHM
      for $\lambda=$2.86--4.19\micron, and 216\,\AA\ FWHM for
      $\lambda=$4.5--5.1\micron. For clarity, the flux of \epsbb\ has
      been increased by a factor of 2 in the range 2.0--2.2\micron\
      and by a factor 5 in the range 2.2--2.5\micron. The telluric
      features in the $M'$-band (marked with $\oplus$) are due to very
      bright sky emission which dominates the source signal. The
      artefacts seen in the region 2.43--2.50\micron\ are again due to
      high levels of telluric contamination.}
    \label{fig:Bab_1.9-2.5}
\end{figure*}

\section{Spectral Classification}
\label{sec:spec_class}


\begin{figure}
    \resizebox{\hsize}{!}{\includegraphics{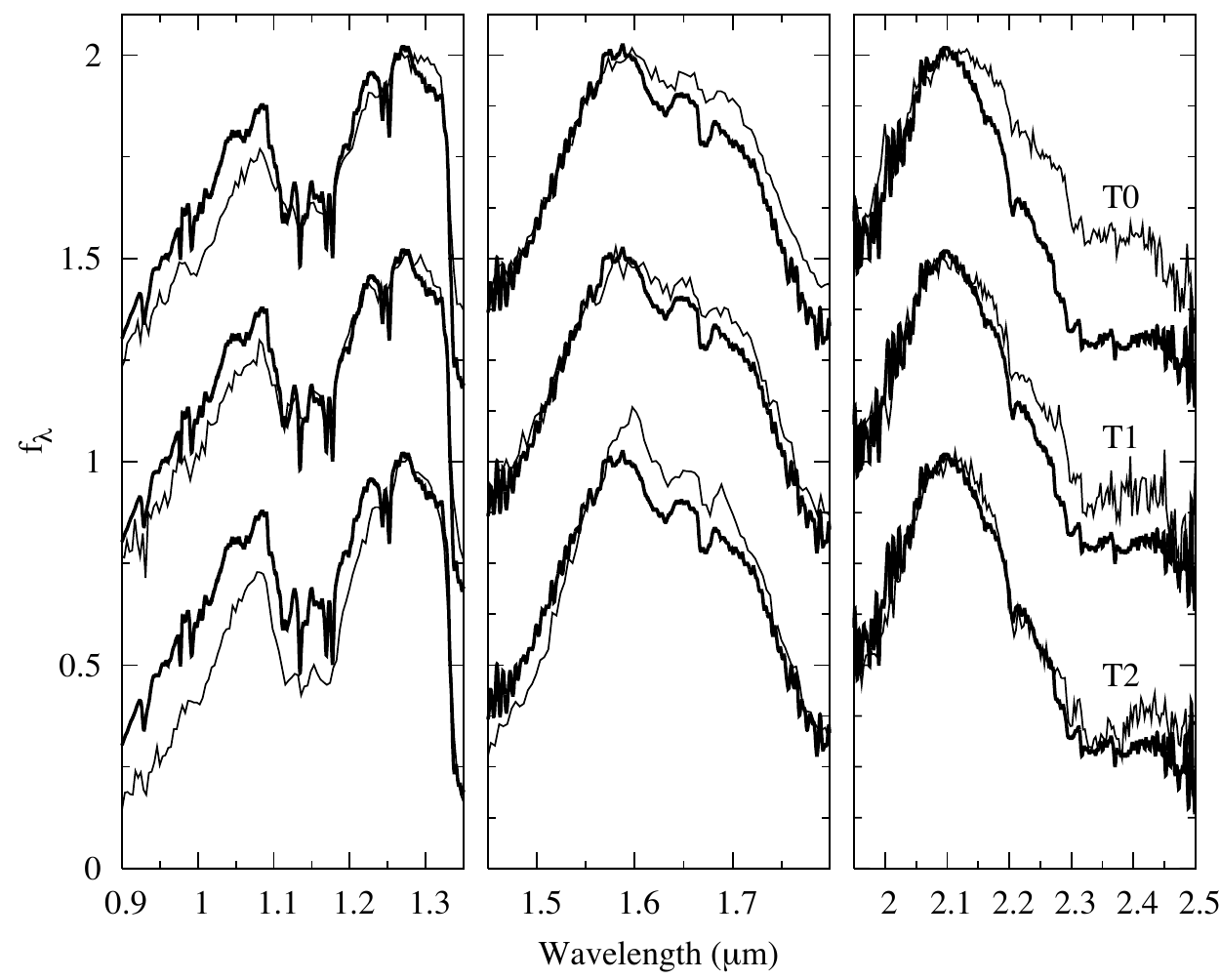}}
    \caption{Smoothed spectrum (30\,\AA\ FWHM) of $\varepsilon$ Indi
      Ba (thick lines) and spectra of the T0 \citep[SDSS
      120747.17+024424.8;][]{Looper:2007}, T1 \citep[alternate
      standard SDSS 015141.69+124429.6;][]{Burgasser:2004}, and T2
      \citep[SDSS J125453.90-012247.4;][]{Burgasser:2004} spectral
      standards. The three panels show the $JHK$ bands omitting the
      lower signal-to-noise H$_2$O absorption regions between the
      bands. Spectra are normalised to unity at 1.27\micron\ in the
      $J$-band, at 1.58\micron\ in the $H$-band, and at 2.10\micron\
      in the $K$-band and offset to show the \epsba\ spectrum against
      each of the spectral standards.}
    \label{Ba_Tstd_SPEX}
\end{figure}

\begin{figure} 
    \resizebox{\hsize}{!}{\includegraphics{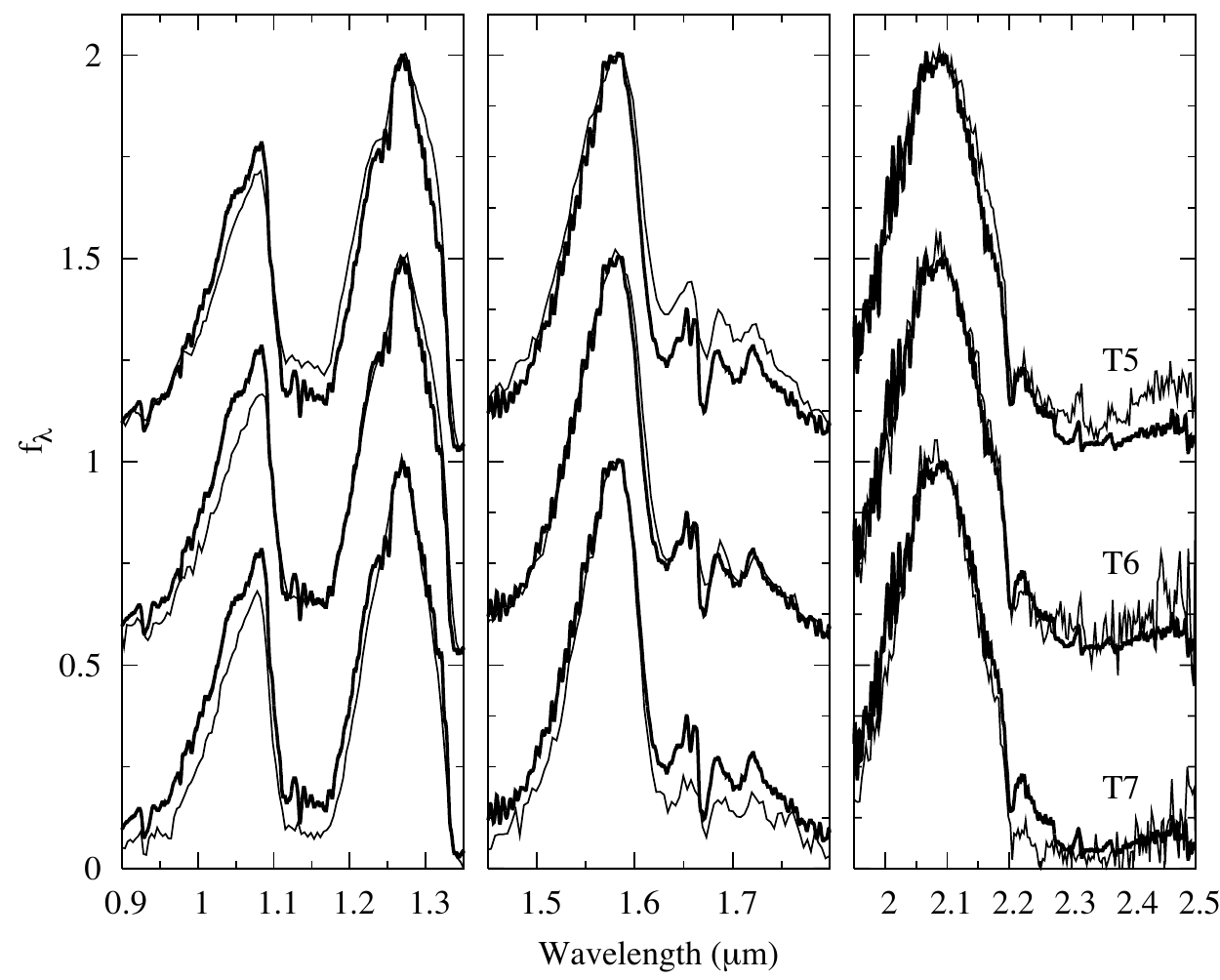}}    
    \caption{Smoothed spectrum (30\,\AA\ FWHM) of \epsbb\ (thick
      lines) and spectra of the T5 \citep[2MASS
      15031961+2525196;][]{Burgasser:2004}, T6 \citep[SDSS
      162414.37+002915.6;][]{Burgasser+B+K:2006}, and T7 \citep[2MASS
      0727182+171001;][]{Burgasser+B+K:2006} spectral
      standards. Plotted regions and normalisation are as in
      Fig.~\ref{Ba_Tstd_SPEX}.}
    \label{Bb_Tstd_SPEX} 
\end{figure}

\subsection{near-IR spectral classification}

\citet{McCaughrean:2004} used the \citet{Geballe:2002} and
\citet{Burgasser:2002} classification indices to provide spectral
classifications for \epsba, Bb based on their $H$-band spectra. They
arrived at spectral types of T1 and T6 for Ba and Bb, respectively, by
employing both of the \citet{Burgasser:2002} $H$-band indices and
the CH$_4$ index of \citet{Geballe:2002} The H$_2$O index of
\citeauthor{Geballe:2002} indicated T0 and T4 and was excluded based
on previous spurious spectral classification of Gl 229 B.


\begin{table*}
  \centering
  \caption{Spectral indices and classifications for \epsba, Bb.  The spectra have been smoothed to two resolutions to emulate the observations of the spectral standard stars defined by \citet{Burgasser:2006}.  The indices derived from our unsmoothed data have been included to quantify the uncertainty when measuring indices from spectra at different resolutions. M1 refers to method 1 of \citeauthor{Burgasser:2006} where the measured spectral indices are compared to that of each of the standard stars, and M2 refers to their method 2 where the measured spectral indices are compared to a range of values for each spectral type.  Direct classification compares the overall spectral shape to the standard star spectra. The spectral type derived from each index is shown in parentheses after each measurement with only M1 being defined for a resolution of R=500.  The average spectral type from each method is shown on the right-hand side. Only direct classification is possible for our unsmoothed spectra.  The range of the H$_{2}$O$-J$ index is ill-defined for the earliest T dwarfs and is therefore omitted for \epsba.}
  \label{spectral_indices}
  \begin{tabular}{ccccccccc}
  \hline
  \hline
  Source & H$_{2}$O$-J$ & CH$_{4}$$-J$ & H$_{2}$O$-H$ & CH$_{4}$$-H$ & CH$_{4}$$-K$ & Direct & M1 & M2\\
  \hline
  \multicolumn{9}{c}{Spectra smoothed to R $=150$} \\	
  \hline	
  $\varepsilon$~Indi~Ba   & 0.650 (T0.0) & 0.656 (T0.0) & 0.592 (T0.0) & 0.873 (T2.0) & 
  0.587 (T2.0) & T1.0 & T1.0 & ...  \\
  $\varepsilon$~Indi~Bb   & 0.172 (T6.0) & 0.305 (T6.5) & 0.308 (T5.5) & 0.295 (T6.0) & 
  0.178 (T5.5) & T6.0 & T6.0 & ...  \\
  \hline		
  \multicolumn{9}{c}{Spectra smoothed to R $=500$} \\	
  \hline
  $\varepsilon$~Indi~Ba   & 0.652 (.../...) & 0.657 (T1.0/T1.5) & 0.589 (T1.0/T0.5) & 
  0.872 (T2.0/T2.0) & 0.588 (T2.0/T2.0) & T1.0 & T1.5 & T1.5  \\
  $\varepsilon$~Indi~Bb   & 0.171 (T6.0/T6.0) & 0.302 (T6.0/T6.0) & 0.303 (T6.0/T6.0) & 
  0.290 (T6.0/T6.0) & 0.180 (T5.5/T5.5) & T6.0 & T6.0 & T6.0  \\
  \hline	
  \multicolumn{9}{c}{Unsmoothed spectra} \\	
  \hline	
  $\varepsilon$~Indi~Ba   & 0.645  & 0.660  & 0.589  & 0.870  & 
  0.589 & T1.0 & ... & ...  \\
  $\varepsilon$~Indi~Bb   & 0.171 & 0.307 & 0.299 & 0.287 & 
  0.178 & T6.0 & ... & ...  \\
  \hline		
\end{tabular}
\end{table*}

Given the greatly improved data here, we employ both methods of
near-IR spectral index classification of \citet{Burgasser:2006},
namely direct comparison of spectral indices with standard indices and
against index ranges defined for each subtype. In addition, we plot
our \epsba, Bb spectra alongside the spectra of the defined standards
to allow direct morphological comparison.

Since the precise value of spectral indices is dependent on the
spectral resolution, we calculate the indices with our spectra
smoothed to resolutions of R=150 and R=500 equivalent to the
resolutions for the standard stars as observed with IRTF/SPEX (at
R$\sim$150) and UKIRT/CGS4 (at R$\sim$500). Table
\ref{spectral_indices} shows the five spectral indices measured from
the different resolution spectra for both \epsba\ and Bb. We also list
the values for the unsmoothed spectra to quantify the effect of
comparing indices for spectra at different resolutions. Spectral types
inferred from each index are shown in parentheses. The range of index
values for each subtype of the T class \citep[method 2
of][]{Burgasser:2006} has only been defined for the R$\sim$500 CGS4
spectra of the spectral standard stars.

Figure~\ref{Ba_Tstd_SPEX} shows the spectrum of $\varepsilon$ Indi Ba
smoothed to R=150 to match the resolution of the spectra of the
standards and normalised within each of the $JHK$ spectral regions,
omitting the low signal-to-noise H$_2$O absorption regions between the
bands. The spectrum is plotted along with the spectra of the T0, T1,
and T2 dwarf standards observed with IRTF/SPEX
\citep{Looper:2007,Burgasser:2004}. The T1 standard provides the best
overall match, however there are some obvious discrepancies. The
$J$-band spectra are normalised to one at 1.27\micron\ and although
the T1 standard fits this peak and the 1.15\micron\ CH$_4$/H$_2$O
absorption band well, the relative strength of the 1.1\micron\ peak is
not well-matched.  The $H$-band is again best matched by the T1
standard with however some deficit in the flux of $\varepsilon$ Indi
Ba in the 1.62--1.74\micron\ region which seems to indicate slightly
differing levels of CH$_4$ absorption. The $K$-band is better fit by
the T2 standard, however all three standard spectra struggle to match
the region beyond 2.3\micron.

Similarly, Figure~\ref{Bb_Tstd_SPEX} shows the smoothed spectrum of
$\varepsilon$ Indi Bb along with the spectra of the T5, T6, and T7
dwarf standards observed with IRTF/SPEX \citep{Burgasser:2004,
  Burgasser+B+K:2006}. The T6 standard provides the best overall
match, however there are again some obvious discrepancies. The
$J$-band spectra are normalised at 1.27\micron\ and although the T6
standard fits this peak and the CH$_4$/H$_2$O absorption band well,
the relative strength of the 1.1\micron\ peak is better matched by the
T5 standard (although the absorption dip then no longer matches).  The
$H$-band is very well-matched by the T6 standard, with the T5 and T7
standards, respectively, over- and under-estimating the flux in the
1.62--1.74\micron\ region. The $K$-band is also best fit by the T6
standard, although again all three standard spectra differ slightly
beyond 2.3\micron.

The discrepancies between the relative strength of the 1.1\micron\ and
1.25\micron\ peaks between $\varepsilon$ Indi Ba, Bb and the spectral
standards shown in the left-most panels of Figs.~\ref{Ba_Tstd_SPEX}
and \ref{Bb_Tstd_SPEX} could possibly be explained by the effects of
different metallicities, surface gravities, or cloud cover. This may
further explain the apparent excess flux in \epsba\ beyond 2.1\micron\
relative to the T1 standard. We discuss the effects of surface gravity
and slightly sub-solar metallicity of model spectra in
Sect.\,\ref{sec:metal_effects}.

Finally, the derived near-IR spectral types for both objects from each
of the three methods are summarised in Table
\ref{spectral_indices}. We adopt final near-IR spectral types of
T1--T1.5 and T6 for \epsba\ and Bb, respectively. The uncertainty on
the spectral type of \epsba\ is not due to poor signal-to-noise in our
data nor the spectra of the standards, but rather is presumably due to
second order spectral variations due to differences in metallicity and
surface gravity.

\subsection{optical spectral classification}

\begin{figure}
    \resizebox{\hsize}{!}{\includegraphics{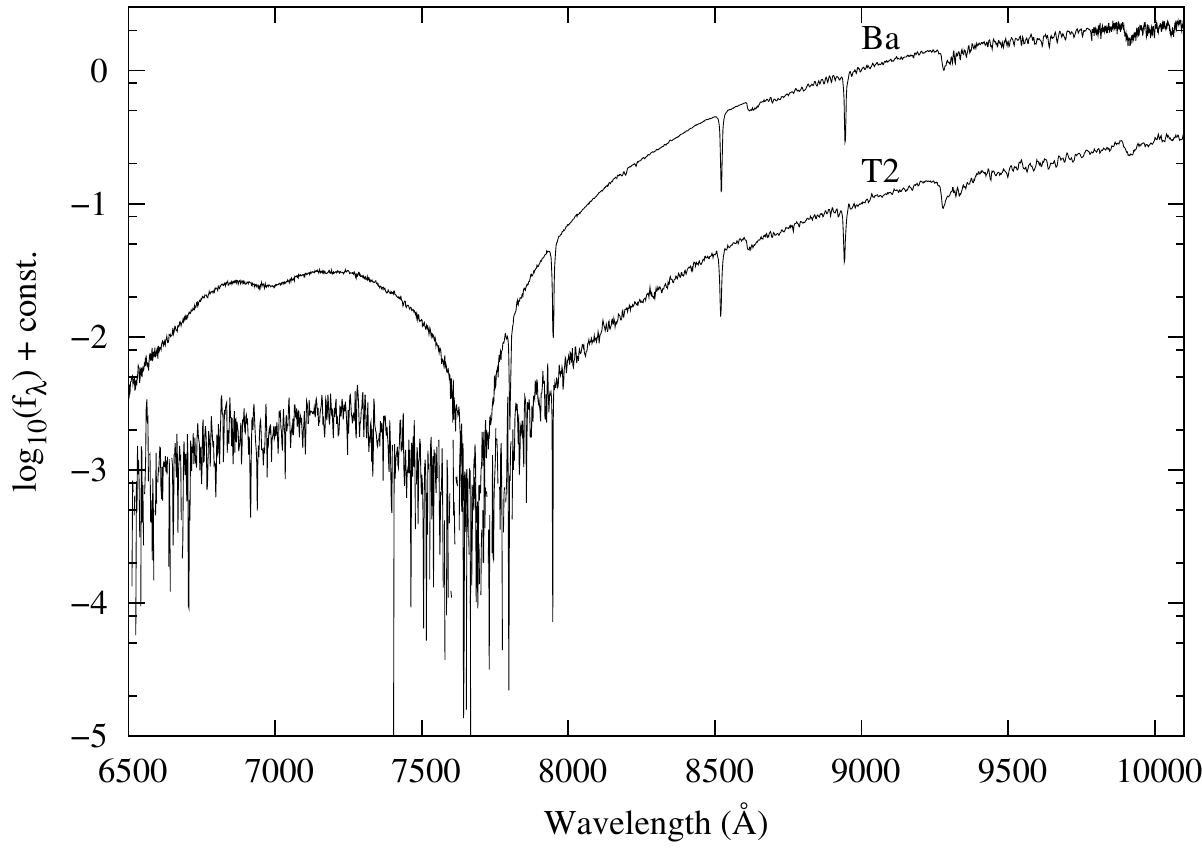}}
    \caption{Optical spectrum of \epsba\ and the T2 (SDSS 1254-0122
      \citep{Burgasser:2003}) optical spectral standard. There is no
      object defined as either the T0 or T1 spectral standard in the
      optical. The \epsba\ spectrum is normalised to unity at
      9000\,\AA\ and the T2 standard spectrum is offset by 1 dex for
      clarity.}
    \label{opt_spectral_stds_Ba}
\end{figure}

\begin{figure}
    \resizebox{\hsize}{!}{\includegraphics{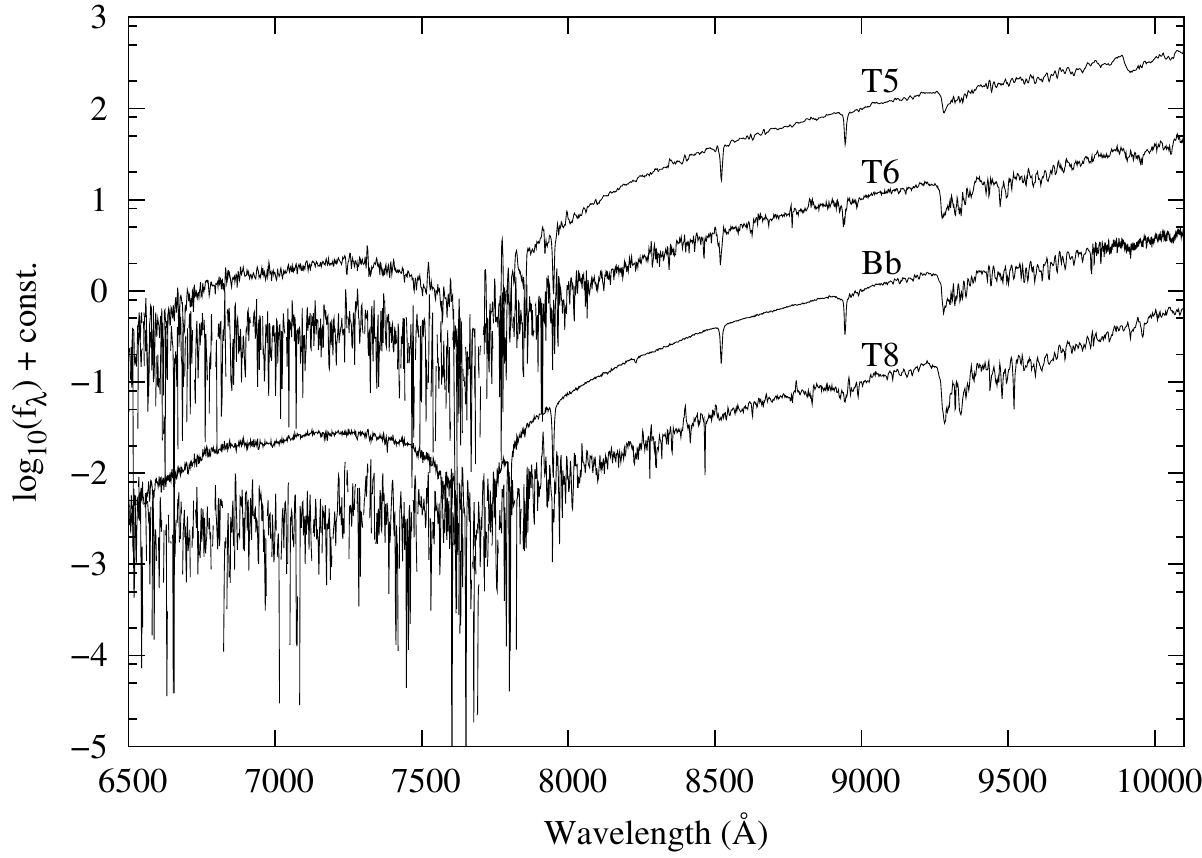}}
    \caption{Optical spectrum of \epsbb\ and the T5 \citep[2MASS
      0559-1404;][]{Burgasser:2003}, T6 \citep[SDSS
      1624+0029;][]{Burgasser:2000}, and T8 \citep[2MASS
      0415-0935;][]{Burgasser:2003} optical spectral standards. There
      is no object defined as the T7 spectral standard in the
      optical. Spectra are normalised at 9000\,\AA\ and offset in
      steps of 1 dex for clarity.}
    \label{opt_spectral_stds_Bb}
\end{figure}

We have also derived spectral types from our optical spectra based on
the optical T dwarf classification scheme of \citet{Burgasser:2003}.
The spectral indices are all defined for wavelengths above
0.90\micron\ and so only use the higher signal-to-noise portion of the
standard star spectra. Table \ref{spectral_indices_opt} shows the
value of the four indices calculated from our full resolution spectra,
along with the derived mean spectral type and the classification from
direct comparison to the spectral standards defined by
\citeauthor{Burgasser:2003} We do not smooth the spectra for this
comparison.  The direct spectral comparison is shown in
Figs. \ref{opt_spectral_stds_Ba} and \ref{opt_spectral_stds_Bb}.

The optical classification broadly agrees with the near-IR
classification for these two T dwarfs with spectral types of T0--T2
for \epsba\ and T6.0--T6.5 for \epsbb.  It is clear that our high
signal-to-noise optical spectra will be useful as standards for future
comparisons of optical T dwarf spectra.

\begin{table}
\centering
\caption{Optical spectral indices and classifications for \epsba, Bb.  The spectra have not been smoothed to derive these indices.  Direct classification compares the overall spectral shape with the standard star spectra. The spectral type derived from each index is shown in parentheses after each measurement. The spectral range of the FeH(B) index is limited to types later than T2 and the range of the colour-e index to types earlier than T2. The approximate classification from direct comparison for Ba is due to the lack of a
 T0 or T1 optical spectral standard.}
\label{spectral_indices_opt}
\begin{tabular}{lll}
  \hline
  \hline
  Index & $\varepsilon$~Indi~Ba & $\varepsilon$~Indi~Bb\\
  \hline
  \ion{Cs}{I}(A) & 2.147 (T2.0$\pm$1.0) & 1.738 (T6.0$\pm$1.0) \\
  CrH(A) / H$_{2}$O & 0.906 (T0.0$\pm$1.0) & 0.452 (T6.0$\pm$1.0) \\
  FeH(B) & 1.237 ($<$ T4?) & 1.109 (T6.5$\pm$0.5) \\
  Colour-e & 3.323 (T0.0$\pm$1.0) & 4.135 ($>$ T2) \\
  \hline
  Mean & T0.5 & T6.0 \\
  Direct & $<$T2.0 & T6.0 \\
  \hline		
\end{tabular}
\end{table}

\section{Constraints from $\varepsilon$ Indi A} 
\label{sec:EpsA_constrain}

The binary \epsba, Bb is a common proper motion companion to the K4.5V
primary, \epsa, which allows us to constrain some of the fundamental
properties of the $\varepsilon$ Indi system. Most obviously, since the
parallax of the primary was measured by HIPPARCOS
\citep{Perryman:1997}, we have an accurately known distance. This was
updated in the reanalysis of \citet{vanleeuwen:2007} to yield
3.6224$\pm$0.0037\,pc. The uncertainty on the distance to \epsba, Bb
is somewhat larger: although we know the projected separation of the
primary and brown dwarf binary to be 0.007\,pc, we do not know the
line-of-sight separation. This added uncertainty should be resolved by
the ongoing absolute astrometric monitoring which along with the mass
ratio of the \epsba, Bb system, allows determination of the parallax
of the brown dwarf binary.

Similarly, determinations of the metallicity of the primary can be
applied to the brown dwarf companions assuming them to be co-eval and
born from the same molecular material. The metallicity of \epsa\ has
been studied by a number of authors who arrive at somewhat differing
results due to the chosen spectral lines and the different models with
which the spectra are fit to derive the abundances. \citet{Abia:1988}
derived a metallicity of [Fe/H]=$-0.23$, while the Geneva group
reported metallicities in the range [Fe/H]=$-0.2$--$+0.06$
\citep{Santos:2001,Santos:2004} with their most recent determination
of [Fe/H]=$-0.2$ reported by \citet{Sousa:2008}.  These results
suggest that \epsa, and by association \epsba, Bb, appear to have
slightly sub-solar metallicity. In Sect.\,\ref{sec:metal_effects} we
show atmospheric models with both [M/H]=$0.0$ and $-0.2$ fit to our
spectroscopic observations. Note however, that [M/H] refers to the
global metallicity and also depends on the abundance of
$\alpha$-elements \citep[cf.][]{Ferraro:1999}, so is not necessarily
equivalent to [Fe/H].

In previous analyses, the age estimate of 0.8--2.0\,Gyr for
$\varepsilon$ Indi A from \citet{Lachaume:1999} was used to make
predictions from evolutionary models. Using this age range
\citet{McCaughrean:2004} derived model masses of 47$\pm$10 and
28$\pm$7\,M$_{\rm{Jup}}$. As we will discuss, we now have direct
determinations of the luminosity of both brown dwarfs and of the
system mass. However, the age of an individual system is a notoriously
difficult parameter to ascertain and so we will return to discuss the
reliability of this age in Sect.\,\ref{sec:age}

\section{Luminosity Determination} 
\label{sec:lum}

\citet{McCaughrean:2004} used their photometry of \epsba, Bb and
estimated bolometric corrections from the spectral type-M$_{\rm{bol}}$
relation of \citet{Golimowski:2004} to derive luminosities of log
L/L$_{\sun}=$ $-4.71$ and $-5.35$ for \epsba\ and Bb, respectively,
with estimated uncertainties of $\pm$20\%. Here we use our flux
calibrated spectra to derive luminosities directly following a similar
process to \citet{Golimowski:2004}.

We sum the 0.63--5.1\micron\ flux with linear interpolation in the
unobserved regions 2.5--2.86\micron\ and 4.19--4.53\micron. This is
extended to the mid-IR using the {\em{Spitzer}}  Infrared
Spectrograph \citep[IRS,][]{Roellig:1998} 5--15\micron\
\citep{Roellig:2004} and 10--19\micron\ spectra \citep{Mainzer:2007} of
the unresolved \epsba, Bb system. To approximate the mid-IR spectrum of
each object and to ensure an accurate representation of the absolute
fluxes, we also used the resolved mid-IR VLT/VISIR photometry of
\citet{Sterzik:2005}. These two determinations of the absolute flux of
the combined system differ significantly. We found that the VISIR
photometry would have to be scaled upward by $\sim$50\% to be fully
consistent with the {\em{Spitzer}} spectrum. We therefore used these
two measurements as the bounds on the total flux in this region and
used the measured flux ratio of $\sim$2.1 at 8.6\micron\ and
10.5\micron\ from \citet{Sterzik:2005} to produce approximate spectra
for each object for our luminosity determination.

Beyond this we used a blackbody tail with T$_{\rm{eff}}$ given by the
model fit to the spectra (see Sect.\,\ref{sec:atm_model_comp}). This
however contributes only 0.5\% and 0.9\% to the total flux for \epsba\
and Bb, respectively, so the precise temperature used is
unimportant. The mid-IR spectrum used to extend our spectroscopic
measurements contributes $\sim$8.5\% and $\sim$14.5\% of the total
flux of \epsba\ and Bb, respectively. We did not attempt to extend our
observed spectra blueward for the luminosity derivation as our
$V$-band photometry shows the flux to have dropped by 2--3 orders of
magnitude relative to the $I$-band flux for both \epsba\ and Bb.

The resulting luminosities are log L/L$_{\sun}=-4.699\pm0.017$ and
$-5.232\pm0.020$ for \epsba\ and Bb, respectively, with the
uncertainties derived from our photometry, 20\% uncertainty on the
mid-IR fluxes, and an assumed 3\% uncertainty on the absolute flux of
the Vega spectrum used for flux calibration \citep{Mountain:1985,
  Hayes:1985}, all of which dominates the distance uncertainty.  The
difference between our determination of the luminosity of \epsbb\ and
that of \citet{McCaughrean:2004} is likely explained by the increased
scatter in the spectral type-BC$_K$ relation of
\citet{Golimowski:2004} at late-T spectral types.

\section{Dynamical Masses}
\label{sec:mass}

Ongoing relative astrometric monitoring of the \epsba, Bb orbit since
May 2004 has allowed a preliminary system mass of
121$\pm$1\,M$_{\rm{Jup}}$ to be determined \citep[][in
prep.]{Cardoso:2008, McCaughrean:2009}. As mentioned previously,
absolute astrometric monitoring is also ongoing \citep[][in
prep.]{Cardoso:2010} which will determine the mass ratio of the system
allowing a model independent determination of the individual masses of
both \epsba\ and Bb. However, we can already provide some constraints
on the individual masses of the brown dwarfs on the basis of our
photometric and spectroscopic observations. Since we know these objects
to be physically bound and so, in all likelihood, co-eval, \epsba\ must
be more massive than its fainter companion. Additionally, we also know
from the low temperatures required to produce T dwarf spectra that both
objects must be substellar, therefore the more massive component must
have a mass below the hydrogen burning minimum mass (HBMM) of
$\sim$0.070\,M$_{\sun}$ (73\,M$_{\rm{Jup}}$) for a cloudy,
approximately solar metallicity source
\citep{Chabrier:2000,Saumon:2008}.

Together then, the masses of these two brown dwarfs are constrained to
be between 73\,+\,47\,M$_{\rm{Jup}}$ and 60\,+\,60\,M$_{\rm{Jup}}$,
with the latter being unlikely due to the different luminosities of the
two brown dwarfs. Therefore, the mass of \epsba\ must be in the range
60--73\,M$_{\rm{Jup}}$ and \epsbb\ in the range 47--60\,M$_{\rm{Jup}}$,
but with a sum of any pairing equal to 121$\pm$1\,M$_{\rm{Jup}}$.

\section{Evolutionary Model Comparisons}
\label{sec:evo_model_comp}

\subsection{photometry}
\label{sec:model_comp_phot}

\begin{figure}
  \resizebox{\hsize}{!}{\includegraphics{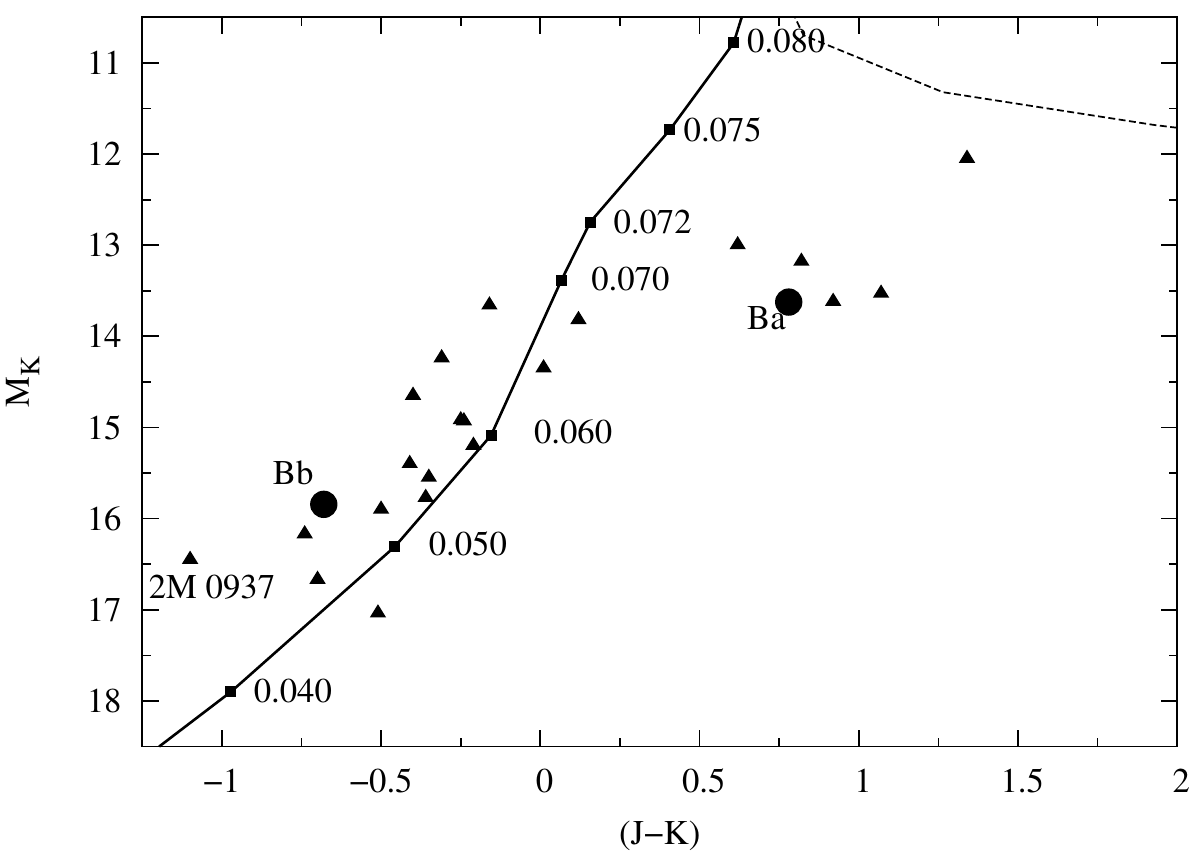}}
  \caption{M$_K$-($J-K$) colour-magnitude diagram with the 5\,Gyr
    COND03 isochrone \citep{Baraffe:2003} with masses (in
    M\,$_{\sun}$) indicated as filled squares. The DUSTY00 5\,Gyr
    isochrone \citep{Chabrier:2000} is also shown (upper, dashed
    line).  The positions of \epsba\ and Bb are marked with filled
    circles, the filled triangles show the T dwarf observations of
    \citet{Knapp:2004} and \citet{Chiu:2006}, and the metal-poor T6
    subdwarf 2MASS 0937+2931 is labelled. All magnitudes are in the
    MKO system. We do not show isochrones for different ages as they
    show little variation in position, except that the same masses
    correspond to different magnitudes.}
    \label{fig:Iso_MKvsJ-K}
\end{figure}

In Table \ref{tab:mags} we presented the apparent magnitudes of both
components of the \epsba, Bb system in the FORS2 $VRIz$, and MKO
$JHKL'M'$ filters. Here we will compare the photometric predictions of
the solar metallicity COND03 evolutionary models \citep[][hereafter
B03]{Baraffe:2003} with the absolute magnitudes derived from our
observations. However, before doing so, it is important to note the
limitations of these models even if they represent the current
state-of-the-art. Although reasonable predictions can be made for the
near-IR colours of mid-late T dwarfs, the colours of early T dwarfs
fall somewhere between the predictions of the COND03 and the DUSTY00
models \citep{Chabrier:2000}. The COND03 evolutionary models neglect
the effect of dust opacity in the radiative transfer, whereas the
DUSTY00 models included dust but once formed it remained in the
atmosphere. The newer BT-Settl atmosphere models account for the
settling of some species from the atmosphere, so once incorporated
into the evolutionary models may provide a more realistic match to
observed colours across the L and T spectral classes.

\begin{figure}
  \resizebox{\hsize}{!}{\includegraphics{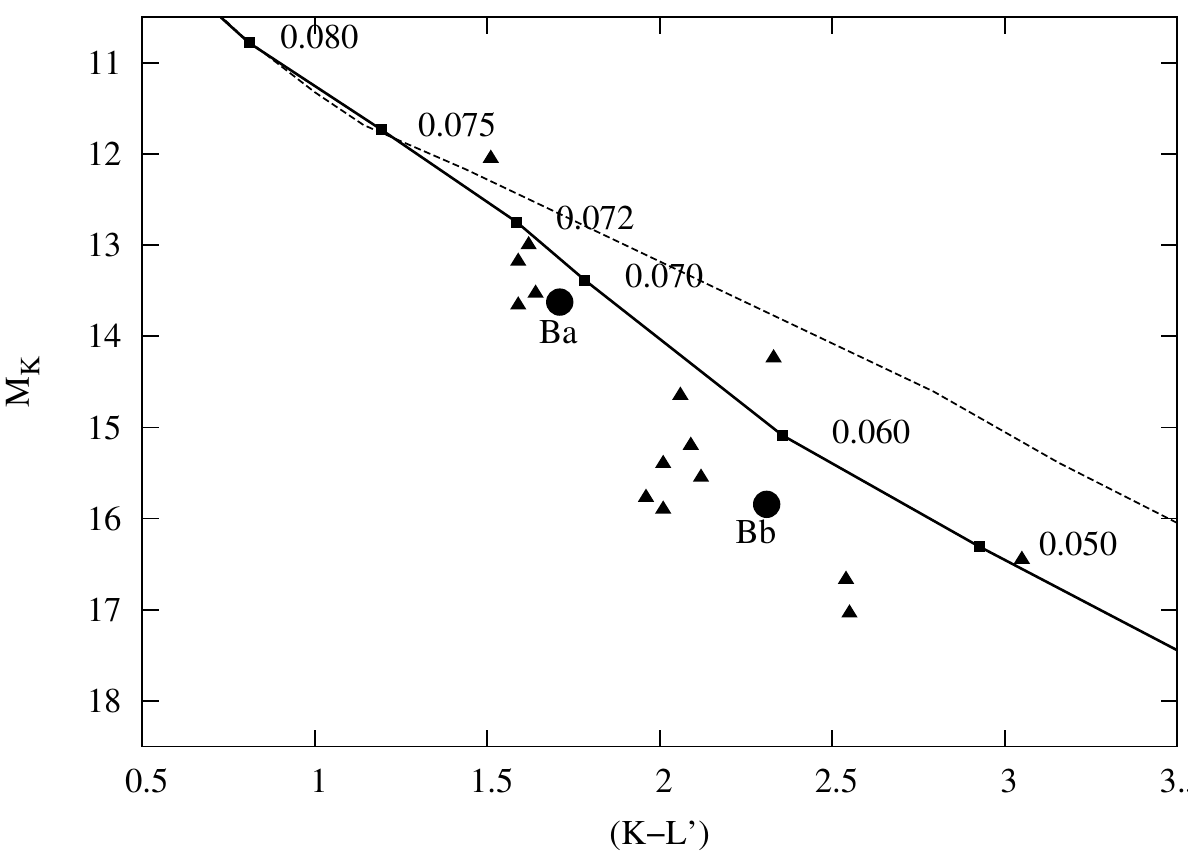}}
  \caption{M$_K$-($K-L'$) colour-magnitude diagram with the 5\,Gyr
    COND03 isochrone with masses (in M\,$_{\sun}$) indicated as filled
    squares. The DUSTY00 5\,Gyr isochrone is also shown (upper, dashed
    line). Symbols are the same as in Fig.~\ref{fig:Iso_MKvsJ-K} with
    the unlabelled data from \citet{Knapp:2004}, \citet{Chiu:2006},
    and \citet{Golimowski:2004}.  The data are in the MKO system,
    whereas the model $L'$ magnitude is in the Johnson-Glass system.}
    \label{fig:Iso_MKvsK-L} 
\end{figure}

\begin{figure}
  \resizebox{\hsize}{!}{\includegraphics{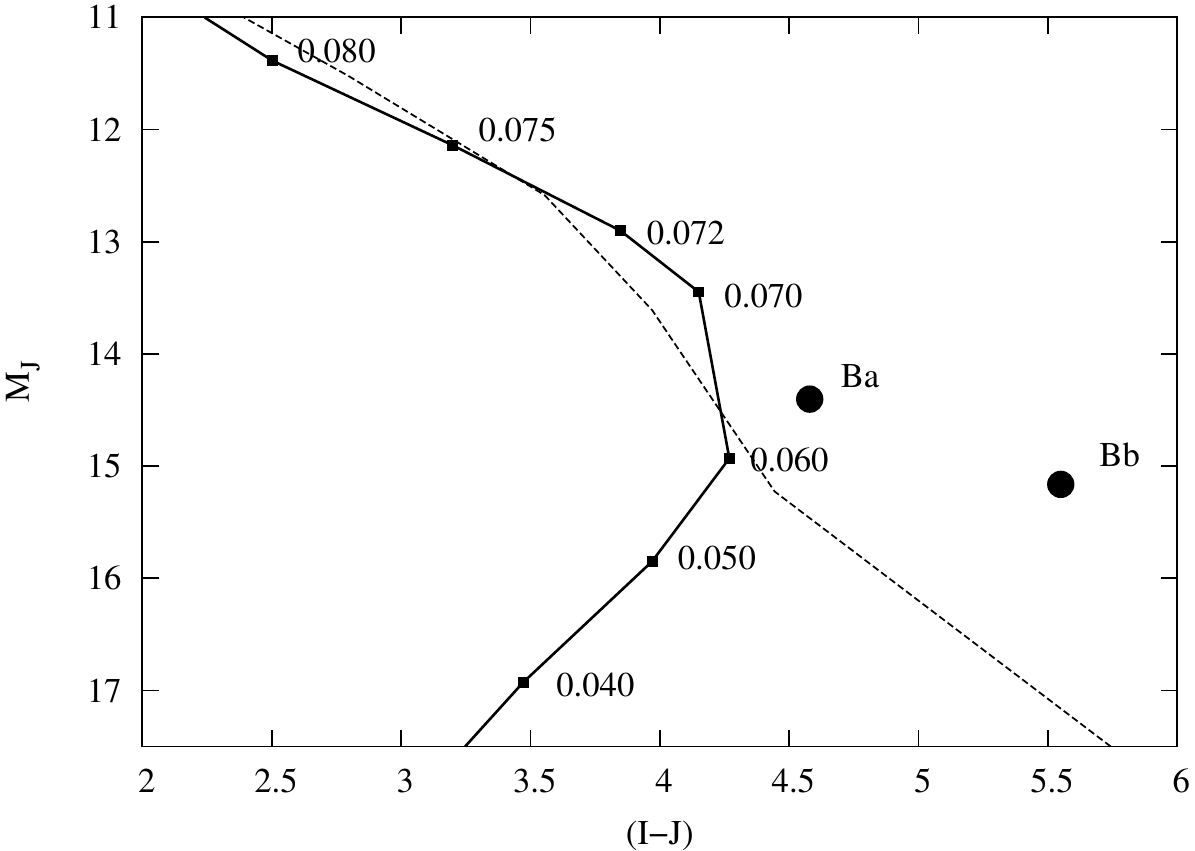}}
  \caption{M$_J$-($I-J$) colour-magnitude diagram with the 5\,Gyr
    COND03 isochrone with masses (in M\,$_{\sun}$) indicated as filled
    squares. The DUSTY00 5\,Gyr isochrone is also shown (dashed
    line). Symbols are the same as in
    Fig.~\ref{fig:Iso_MKvsJ-K}. $J$-magnitudes are in the MKO system
    and $I$-magnitudes use the FORS2 Bessell $I$ filter.}
    \label{fig:Iso_MJvsI-J}
\end{figure}

Figures \ref{fig:Iso_MKvsJ-K}, \ref{fig:Iso_MKvsK-L}, and
\ref{fig:Iso_MJvsI-J} show colour-magnitude diagrams (CMDs) comparing
our \epsba, Bb photometry with the COND03 and DUSTY00 5\,Gyr
isochrones along with other T dwarf observations from the literature.
The model $JHK$ magnitudes have been transformed from the CIT system
to the MKO system using the colour relations of \citet{Stephens:2004},
while the model $L'$ magnitudes are left in the Johnson-Glass system
and the $I$ magnitudes in the Bessell system.

In Fig.\,\ref{fig:Iso_MKvsJ-K}, \epsba\ is seen to lie $\sim$0.5--1.0
mag redward of the model isochrone, while \epsbb\ is $\sim$0.5 mag
blueward. The position of the metal-poor ($-0.4$ $<$ [M/H] $<$ $-0.1$)
T6 subdwarf, 2MASS 0937+2931 \citep{Burgasser:2003}, is marked and
seen to be bluer than \epsbb. \citet{Schilbach:2009} find that 2MASS
0937+2931 falls above an extrapolated [M/H]=$-0.5$
\citet{Baraffe:1998} isochrone in an M$_K$-($J-K$) CMD and is possibly
part of the thick disc or halo population. By contrast, the position
of \epsbb\ suggests only a slightly sub-solar metallicity.

Figure\,\ref{fig:Iso_MKvsK-L} shows that the predicted $(K-L')$
colours are a better match to our and other early T dwarf observations
as the effects of clouds are much reduced in this region, but observed
colours are still consistently bluer than the models implying that
spectral features in these regions are not well reproduced by these
models.

When optical colours are used in a CMD, the mismatch between
observations and models increases. For late T dwarfs such as \epsbb,
as shown in Fig. \ref{fig:Iso_MJvsI-J}, our observed ($I-J$) colours
are significantly redder than model predictions even after accounting
for differences due to the use of different filter systems. These
issues will be addressed when we compare our observations with the
more up-to-date atmospheric models in Sect.\,\ref{sec:atm_model_comp}.

This discrepancy between the isochrones and the observations for the
early T dwarfs is an indication of the complexity of their atmospheres
where the role of clouds is greater than for later types. As brown
dwarfs transition from dusty L to cloud-free mid-late T dwarfs, the
observations are found to lie between the COND03 and DUSTY00 models
\citep{Baraffe:2003}.

\subsection{physical properties}
\label{sec:phys_prop}

\begin{table} 
  \centering

  \caption{Predictions of the parameters of \epsba\ and Bb from the
COND03 evolutionary models. These parameters are derived for the age
range of 3.7--4.3\,Gyr given by the measured total mass (Ba+Bb) and
individual luminosities.}

\label{COND_predict}
\begin{tabular}{ccccc}
  \hline
  \hline
  Source & Mass & Radius & T$_{\rm{eff}}$ & log g \\
  & (M$_{\rm{Jup}}$) & (R$_{\sun}$) & (K) & (cm\,s$^{-2}$) \\
  \hline
  \epsba\  &  67.6--69.1 &  0.080--0.081 & 1352--1385 &  5.43--5.45 \\
  \epsbb\  &  50.0--54.5 &  0.082--0.083 &  976--1011 &  5.27--5.33 \\
  \hline
\end{tabular}
\end{table}

\begin{figure}
    \resizebox{\hsize}{!}{\includegraphics{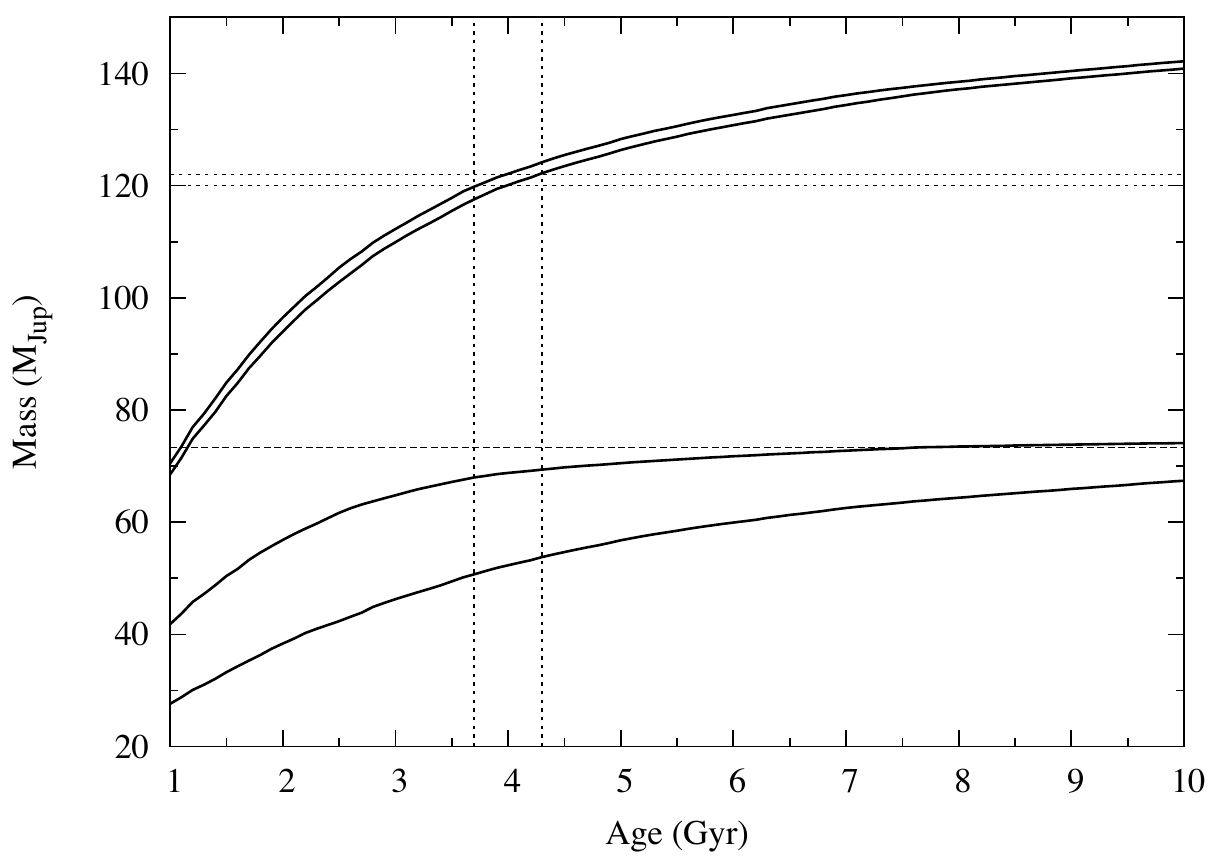}}   
    \caption{Variation of mass with age for lines of constant
      luminosity interpolated from a fine grid of COND03 evolutionary
      models for the observed luminosities of \epsba\ and Bb (lower
      two curves). The upper two curves show the predicted combined
      mass at each age accounting for the uncertainties on the
      individual luminosities, while the constraints on the system
      mass from the dynamical monitoring are indicated by the upper
      two horizontal dotted lines. The lower horizontal dashed line
      indicates the hydrogen burning minimum mass of 0.070\,M$_{\sun}$
      and the 3.7--4.3\,Gyr age predicted by the COND03 models is
      shown by the two vertical dotted lines.}
    \label{fig:Mass-age}  
\end{figure}


Although the previous section has shown that neither the COND03 nor the
DUSTY00 atmospheres necessarily reproduce the observed colours of T dwarfs
faithfully, the comparison of properties such as luminosity, mass, and
effective temperature should be more reliable as they are not so strongly
dependent on the specifics of the atmospheric model used.  That said,
\citet{Chabrier:2000} showed that the luminosity and effective temperature
evolution at a specific mass are not entirely independent of the treatment
of dust in the atmosphere. They find the difference between the DUSTY00
and COND03 model predictions, for a given age, can be up to 10\% in
effective temperature and up to 25\% in luminosity with the COND03 models
predicting systematically higher effective temperatures than the
corresponding DUSTY00 models. With the precision of our mass and
luminosity determinations, these uncertainties are significant.  
However, for the age and mass range of the \epsba, Bb system, the DUSTY00
evolution models predict only slightly less efficient cooling than the
COND03 models, resulting in $\sim$3\% smaller masses, 5\% larger radii and
less than 3\% differences in effective temperature than the COND03
models.

\begin{table*} \centering
\caption{Physical parameters of \epsba, Bb derived using the observed luminosities and the
COND03 models of \citet{Baraffe:2003} for three ages: 1, 5, and 10\,Gyr.}
\label{phys_paras}
\begin{tabular}{|cc|ccc|ccc|ccc|ccc|}
  \hline
  \hline
  Source &  log L/L$_{\sun}$ & \multicolumn{3}{c|}{Mass} & \multicolumn{3}{c|}{Radius} 
  & \multicolumn{3}{c|}{T$_{\rm{eff}}$} & \multicolumn{3}{c|}{log g} \\
  &                   & \multicolumn{3}{c|}{(M$_{\rm{Jup}}$)} 
  & \multicolumn{3}{c|}{(R$_{\sun}$)} & \multicolumn{3}{c|}{(K)} 
  & \multicolumn{3}{c|}{(cm\,s$^{-2}$)} \\
  \hline
  &&  1\,Gyr & 5\,Gyr & 10\,Gyr & 1\,Gyr & 5\,Gyr & 10\,Gyr & 1\,Gyr & 5\,Gyr & 10\,Gyr & 1\,Gyr & 5\,Gyr & 10\,Gyr\\
  \hline
  $\varepsilon$ Indi Ba  & $-$4.699 & 42.0 & 69.8 & 74.1 & 0.093 & 0.080 & 0.079 & 1274 & 1387 
  & 1393 & 5.10 & 5.45 & 5.49 \\
  \epsbb\  & $-$5.232 & 28.0 & 57.2 & 66.2 & 0.097 & 0.081 & 0.077 & 929 & 1019
  & 1051 & 4.88 & 5.36 & 5.47 \\
  \hline
\end{tabular}
\end{table*}

Additionally, \citet{Saumon:2008} have recently compared their
evolutionary model predictions with the Lyon DUSTY00 and COND03
models, concluding that the differences between the corresponding
cloudy and cloud-free models at intermediate ages can be explained by
the lack of electron conduction in the \citet{Saumon:2008} models. The
differences seen when compared to the \citet{Burrows:1997} models is
found to be mainly due to the use of older non-grey atmospheres in the
\citet{Burrows:1997} models. We therefore employ only the Lyon models
in our comparison.

Given the observed luminosities of \epsba\ and Bb, the evolutionary
models can be used to predict the corresponding masses as a function of
the system age, as shown in Fig.~\ref{fig:Mass-age}. It is immediately
apparent that the original age estimate of 0.8--2.0\,Gyr is
inconsistent with the evolutionary models given the newly established
mass of 121$\pm$1\,M$_{\rm{Jup}}$. From the COND03 models, we find that
an age range of 3.7--4.3\,Gyr is necessary to accommodate the
mass and luminosity constraints. Table~\ref{COND_predict} shows that
for this age range, the COND03 models predict effective temperatures of
 1352--1385\,K and 976--1011\,K for \epsba\ and Bb, respectively,
and a mass ratio in the range 0.73--0.78. Similarly,
Table\,\ref{phys_paras} shows the predicted physical parameters of
\epsba, and Bb for the observed luminosities and ages of 1, 5, and
10\,Gyr.

\subsection{age of the $\varepsilon$ Indi system}
\label{sec:age}

This age range of 3.7--4.3\,Gyr for \epsba, Bb is significantly larger
than the age of 0.8--2.0\,Gyr estimated by \citet{Lachaume:1999} for
\epsa\ based on the rotational period given by \citet{Saar:1997}. In
fact, the measured $v$\,sin\,$i$ for \epsa\ was too small to derive
any meaningful rotational period and this period was inferred from the
\ion{Ca}{II} H\&K emission observed by \citet{Henry:1996}, rendering
it less reliable. \citet{Barnes:2007} also calculates an age of
1.03$\pm$0.13\,Gyr from rotation, although this uses the same inferred
22 day period. \citet{Lachaume:1999} also derive an age from the
\ion{Ca}{II} H\&K activity using their R$'_{\mathrm{HK}}$-age
relation, which suggests an age in the range 1--2.7\,Gyr. However, in
choosing a young age based on this activity indicator, they ignored the
much greater estimate of $>$7.4\,Gyr which they derived from the
kinematic properties of the system.

Age constraints are also available if one considers the larger moving
group of which \epsa\ is the eponymous member, and for which
\citet{Cannon:1970} quoted an age of 5\,Gyr derived from the observed
$(B-V)$ colour of an apparent red giant clump. However with only 15
members \citep{Eggen:1958}, it is not clear that there would be enough
objects at the end of their main-sequence lives to fit this
reliably. Furthermore, the moving group members given in
\citet{Eggen:1958} were presumably used in \citet{Cannon:1970},
although no details of the members used are given.  This membership
list was later revised by \citet{Eggen:1971} which excluded five stars
from the original list of members. Thus, the age of this apparently
elderly moving group is not a strong age constraint by itself. However
$\lambda$ Aurigae, another member of the moving group, has an
isochronal age of 5.8$\pm$0.43\,Gyr from \citet{Soubiran:2005} which
adds support to an intermediate age for the entire moving group,
although this technique has been shown to be unreliable when applied
to individual objects. This worsens toward later spectral types where
the isochrones become degenerate, so isochrone fitting would be
unreliable for \epsa, but in the case of $\lambda$ Aurigae the
isochrones are relatively well separated.

Finally, \citet{Rocha-Pinto:2002} classifies \epsa\ as a
''chromospherically young, kinematically old'' star based on the
disagreement between the age from activity indicators and the observed
space velocities. They find a chromospheric age of 0.39\,Gyr, but note
that \epsa\ has no obvious \ion{Li}{I} absorption line and so argue
that it is an older star than activity suggests. The origin of these
stars is as yet unknown.


After further consideration then, an age of $\sim$3.7--4.3\,Gyr for
\epsba, Bb as predicted by the evolutionary models in
Sect.\,\ref{sec:phys_prop} seems plausible at least. The discrepancy
between the various age estimates highlights the problem of deriving ages
for individual objects, as any one indicator cannot be entirely
trusted. To address this issue, we are currently investigating the
feasibility of obtaining high resolution, time-resolved spectroscopy
of \epsa\ to derive an age via asteroseismology.

\section{Atmospheric Model Comparison}
\label{sec:atm_model_comp}

\begin{figure*}
  \resizebox{\hsize}{!}{\includegraphics{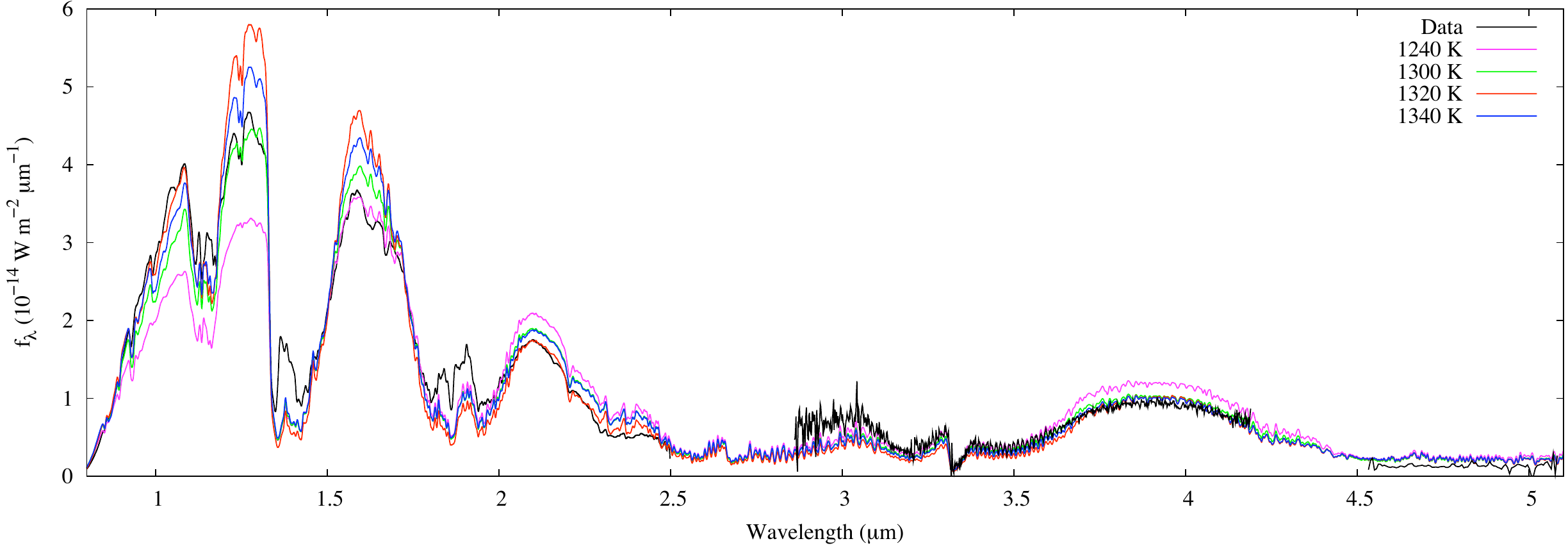}}
  \caption{Near- to thermal-IR spectrum of \epsba\ (black line) with
    sub-solar metallicity ([M/H]=$-0.2$) BT-Settl spectra with
    T$_{\rm{eff}}$=1300, 1320, 1340\,K and log g=5.50 (green, red, and
    blue lines respectively). The T$_{\rm{eff}}$=1240\,K spectrum
    (magenta line) is also shown to indicate the large difference
    between the observations and spectral models for this lower
    effective temperature . All spectra have been smoothed to 60\,\AA\
    FWHM and the observed spectrum median filtered to remove the
    lowest signal-to-noise points.}
    \label{fig:linear_teff_Ba}
\end{figure*}

The comparisons presented in Sect.\,\ref{sec:model_comp_phot} demonstrated
that the COND03 and DUSTY00 models yield a relatively poor match to the
observed photometric properties of \epsba\ and Bb.   Using the newer
BT-Settl atmosphere models \citep[][in prep.]{Allard:2003,Allard:2009},
which account both for the formation and optical effects of dust in the
atmosphere and for settling of condensates under steady-state conditions,
the calculation of synthetic spectra and colours can reproduce these
observations much more accurately. By allowing the description of
partially settled clouds, these model atmospheres are a more appropriate
comparison in particular for early T dwarfs, where the atmospheres
transition from being dust-dominated to cloud-free, and so here we compare
BT-Settl atmosphere models with our detailed spectral observations of
\epsba\ and Bb.

These atmosphere models have yet to be incorporated into a
self-consistent set of interior structure and evolution models, but since
even the differences between the DUSTY00 and the completely cloud-free
COND03 models result in only small discrepancies in cooling rates and
radius evolution, our analysis should not be much compromised by applying
the newer models on top of the older evolution models.


The BT-Settl models are based on version 15 of the \texttt{PHOENIX}
stellar atmosphere code \citep{Hauschildt:1999} with updated opacity
databases, including among others the recent water line list of
\citet{bt2} and extended methane line lists \citep{stdsAtmos}.  The
chemical equilibrium code used in the DUSTY00 and COND03 models
\citep{LimDust,Baraffe:2003} has been modified to include grain settling
effects \citep{Allard:2003}. At each layer in the model atmosphere, the
dust grain number densities are calculated in equilibrium to the gas
phase. The timescales for condensation and growth processes and
gravitational settling are then calculated following \citet{rossow78} and
compared to the turbulent mixing timescale to predict the median size and
the fraction of grains which have settled.  This fraction is then removed
from the composition and a new equilibrium obtained, iterating the process
until the grain density no longer changes. For more detailed results and a
comparison with other cloud models see \citet{cloudcomp08}. Turbulent
mixing in the present models is calculated by interpolation from 2D and 3D
radiation hydrodynamic (RHD) models with the CO5BOLD code, extending the
results of \citet{ludwig-MconvII} to lower temperatures by including a
self-consistent dust module \citep{Freytag:2009}. As such, the BT-Settl
models do not include any adjustable parameters, except for some freedom
in the translation of the hydrodynamic velocity field in the convective
overshoot region into an effective timescale. This conversion has been
chosen such as to optimally reproduce the entire photometric sequence from
early-L to mid-T dwarfs.

For each model spectrum of known effective temperature, we inferred a
radius using our observed luminosities of \epsba\ and Bb, which along
with the known distance, allowed us to convert the emergent flux
density of the atmospheric models to the flux that would be observed
from Earth. Therefore, for each model spectrum we were able to compare
predicted absolute flux levels against those observed without reliance
on evolutionary models. In other words, we were not free to normalise
the model spectra to match our observations.

The model grid sampled effective temperature in steps of 20\,K,
surface gravity in steps of 0.25 dex, for solar and slightly sub-solar
metallicity ([M/H]$=0.0,-0.2$). The metal abundances of
\citet{Grevesse:1993} were employed instead of the more recent
\citet{Asplund:2005} abundances as the validity of the latter
determination of oxygen abundance is the subject of ongoing debate
given its effect on previously well-matched helioseismological theory
and observations \citep[cf.][]{Ayres:2008,Caffau:2008}.

\subsection{effective temperature effects}

Figure~\ref{fig:linear_teff_Ba} shows the BT-Settl near-IR spectral
models with effective temperatures in the range 1300--1340\,K and log
g=5.50, [M/H]=$-$0.2 compared to our spectrum of \epsba. Since
\citet{Kasper:2009} find acceptable fits to low resolution near-IR
spectra of \epsba\ using models with effective temperatures of 1250\,K
and 1300\,K, we include our 1240\,K model to show the large mismatch
such a low effective temperature would have with the BT-Settl models.
Each of the models have an absolute flux scale set by the effective
temperature and the observed luminosity of the corresponding object,
therefore there is no scaling of model spectra to improve the fit to
observations. We find effective temperatures in the range
1300--1340\,K produce the most reasonable fit to the flux level across
the spectrum and to individual features.

\begin{figure}
  \resizebox{\hsize}{!}{\includegraphics{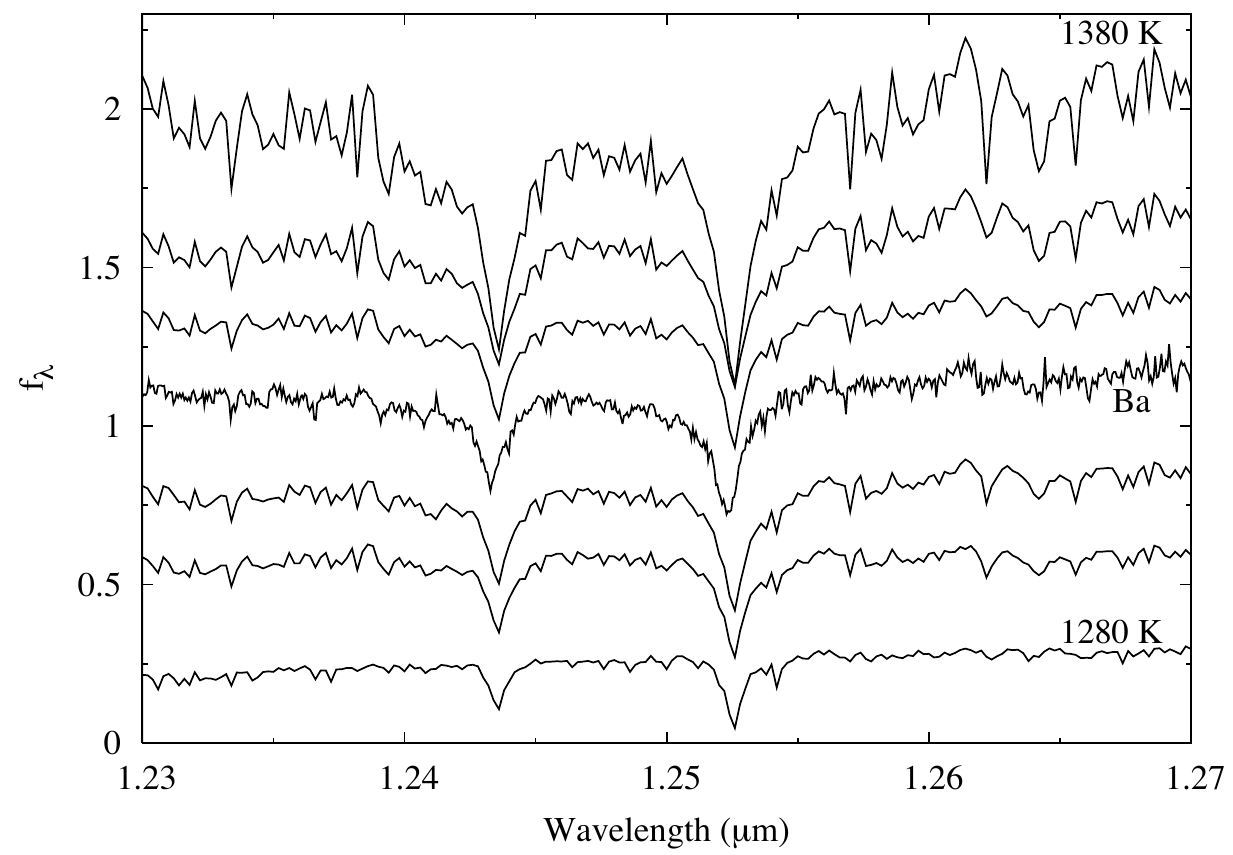}}
  \caption{1.23--1.27\micron\ spectrum of \epsba\ (thicker line, fourth
    from top) and the [M/H]=$-0.2$, log g=5.50 BT-Settl spectra with
    T$_{\rm{eff}}$=1280--1380\,K (increasing bottom to top in steps of
    20\,K). All spectra have been normalised and offset for
    clarity. Model spectra have been smoothed to 2.4\,\AA\ FWHM.}
    \label{fig:ba_teff_seq}
\end{figure}

For higher effective temperature models, the \ion{K}{I} lines at
1.25\micron\ are too strong and the FeH and CrH features around
1\micron\ are too deep, while by 1280\,K the 1\micron\ features are no
longer reproduced and the depth of the \ion{K}{I} doublet is
significantly reduced (see Fig.~\ref{fig:ba_teff_seq}). Additionally,
at lower temperatures, we find that the flux level of the near-IR
peaks become increasingly difficult to reconcile with the
observations. 

\begin{figure}
  \resizebox{\hsize}{!}{\includegraphics{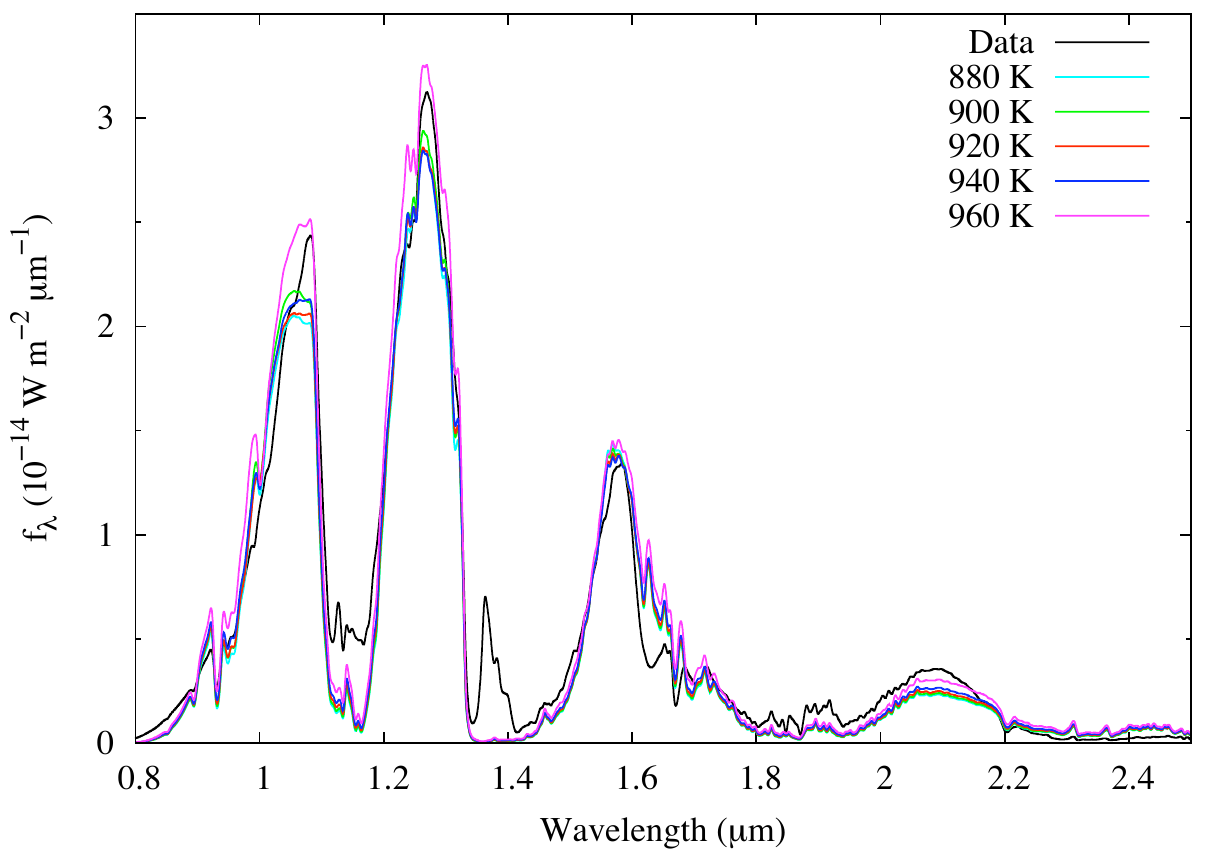}}
  \caption{Near-IR spectrum of \epsbb\ (black line) with sub-solar
    metallicity ([M/H]=$-0.2$) BT-Settl spectra with
    T$_{\rm{eff}}$=880, 900, 920, 940, 960\,K and log g=5.25 (cyan,
    green, red, blue, and magenta lines respectively). All spectra
    have been smoothed to 60\,\AA\ FWHM and the observed spectrum
    median filtered to remove the lowest signal-to-noise points.}
    \label{fig:linear_teff_Bb}
\end{figure}

\begin{figure}
  \resizebox{\hsize}{!}{\includegraphics{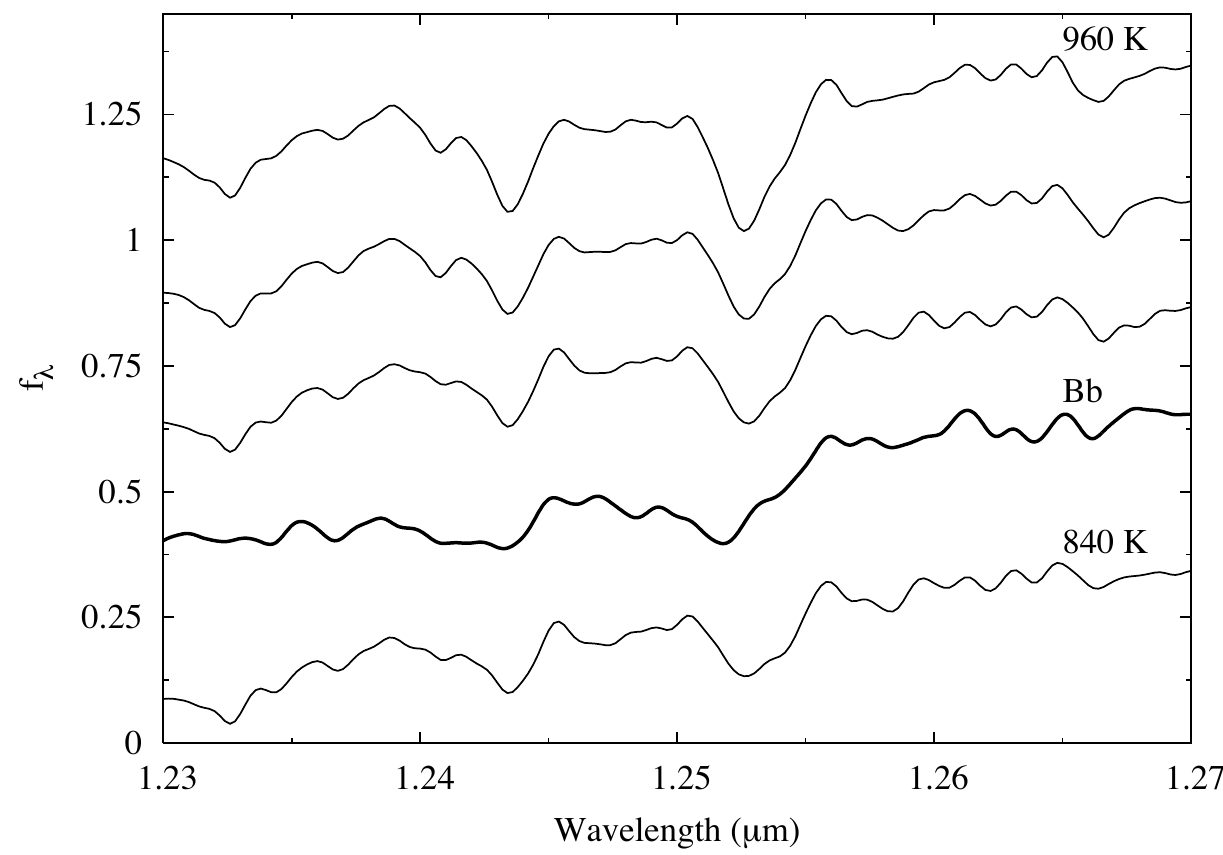}}
  \caption{1.23--1.27\micron\ spectrum of \epsbb\ (thicker line,
    second from bottom) and the [M/H]=$-0.2$, log g=5.25 BT-Settl spectra
    with T$_{\rm{eff}}$=840--960\,K (increasing bottom to top in
    steps of 40\,K). All spectra have been normalised and offset for
    clarity. All spectra have been smoothed to 10\,\AA\ FWHM for
    easier comparison.}
    \label{fig:bb_teff_seq}
\end{figure}

Figure \ref{fig:linear_teff_Bb} shows the BT-Settl near-IR spectral
models with effective temperatures in the range 880--960\,K and log
g=5.25, [M/H]=$-$0.2 compared to the data from \epsbb. While the
optical spectrum is problematic at any effective temperature (see
Sect.\,\ref{sec:alkali}), there is relatively little variation within
this effective temperature range in the thermal-IR. Based on the
near-IR alone then, we find the most reasonable fits to have effective
temperatures in the range 880--940\,K. Figure \ref{fig:linear_teff_Bb}
shows that at temperatures below 900\,K, the flux level of the
1.1\micron\ and the 2.1\micron\ peaks becomes progressively more
depressed, and as seen in Fig.\,\ref{fig:bb_teff_seq}, at higher
temperatures, the 1.25\micron\ \ion{K}{I} lines are considerably
deeper than observed, although the shapes of the tops of the
1.1\micron\ and 1.25\micron\ peaks are not well-matched by any of the
models.


\subsection{metallicity and surface gravity effects}
\label{sec:metal_effects}

In low-mass stars, the effect of decreasing metallicity is to increase
the effective temperature at constant mass \citep{Baraffe:1997}, with
the magnitude of the effect decreasing toward the substellar boundary.
However, it is not clear what happens in the lower-mass regime where
degeneracy effects may alter the situation. Since no evolutionary
models exist for sub-solar metallicity, substellar objects, we cannot
provide a quantitative analysis of the evolutionary effect of slightly
sub-solar metallicity. As discussed in
Sect.\,\ref{sec:EpsA_constrain}, studies of the parent main-sequence
star, \epsa, derive a metallicity in the range
[Fe/H]=$-0.23$--$+0.06$, with the most recent study supporting
[Fe/H]=$-0.2$. The effects of lower metallicity and increasing surface
gravity are somewhat complementary, so we compare our observations
with BT-Settl models of surface gravity from log g=5.00 to 5.50 and
with solar ([M/H]=0.0) and slightly sub-solar metallicity
([M/H]=$-$0.2).

\begin{figure}
  \resizebox{\hsize}{!}{\includegraphics{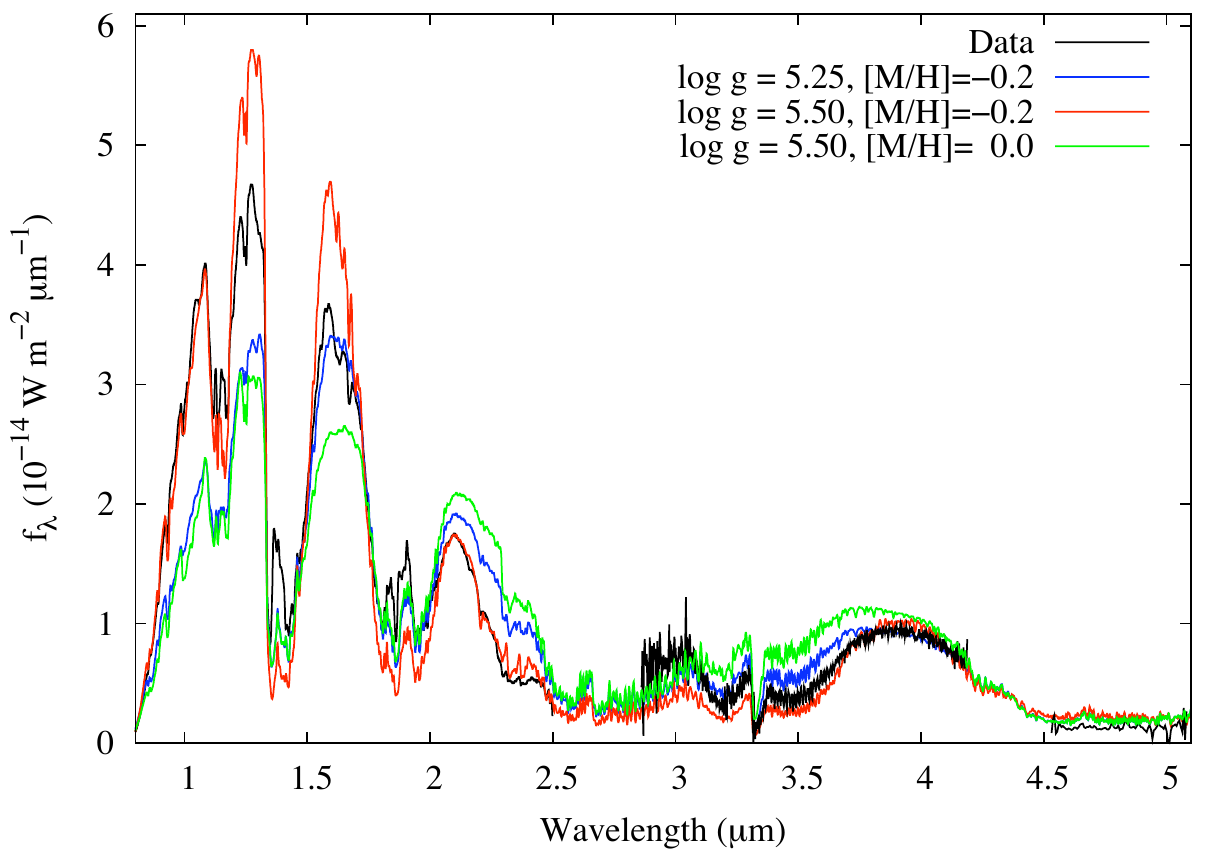}}
  \caption{Near- to thermal-IR spectra of \epsba\ (black line), the
    BT-Settl spectrum (T$_{\rm{eff}}$=1320\,K, log g=5.50,
    [M/H]=$-0.2$, red line), the same with log g=5.25 (blue line), and
    with log g=5.50 but with solar metallicity ([M/H]=0.0, green
    line). All spectra have been smoothed to 60\,\AA\ FWHM and the
    observed spectrum median filtered to remove the lowest
    signal-to-noise points.}
    \label{fig:metal_logg_Ba}
\end{figure}

Figure \ref{fig:metal_logg_Ba} shows the near- to thermal-IR spectrum
of \epsba\ compared to models with an effective temperature of
1320\,K, surface gravities log g=5.25 and 5.50, and
metallicities of [M/H]=0.0 and $-0.2$. The effect of higher
metallicity and lower surface gravity is shown to be most prominent in
the $J$- and $H$-bands. The flux in the 1.1, 1.25, and 1.6\micron\
peaks is suppressed, while the 2.2\micron\ peak flux is
over-estimated. Additionally, the shape of the $K$- and $L$-bands is
inconsistent with the observed spectrum. We find the higher surface
gravity (log g=5.50) and sub-solar metallicity ([M/H]=$-$0.2) to be
the better fit to the spectrum of \epsba.

\begin{figure}
  \resizebox{\hsize}{!}{\includegraphics{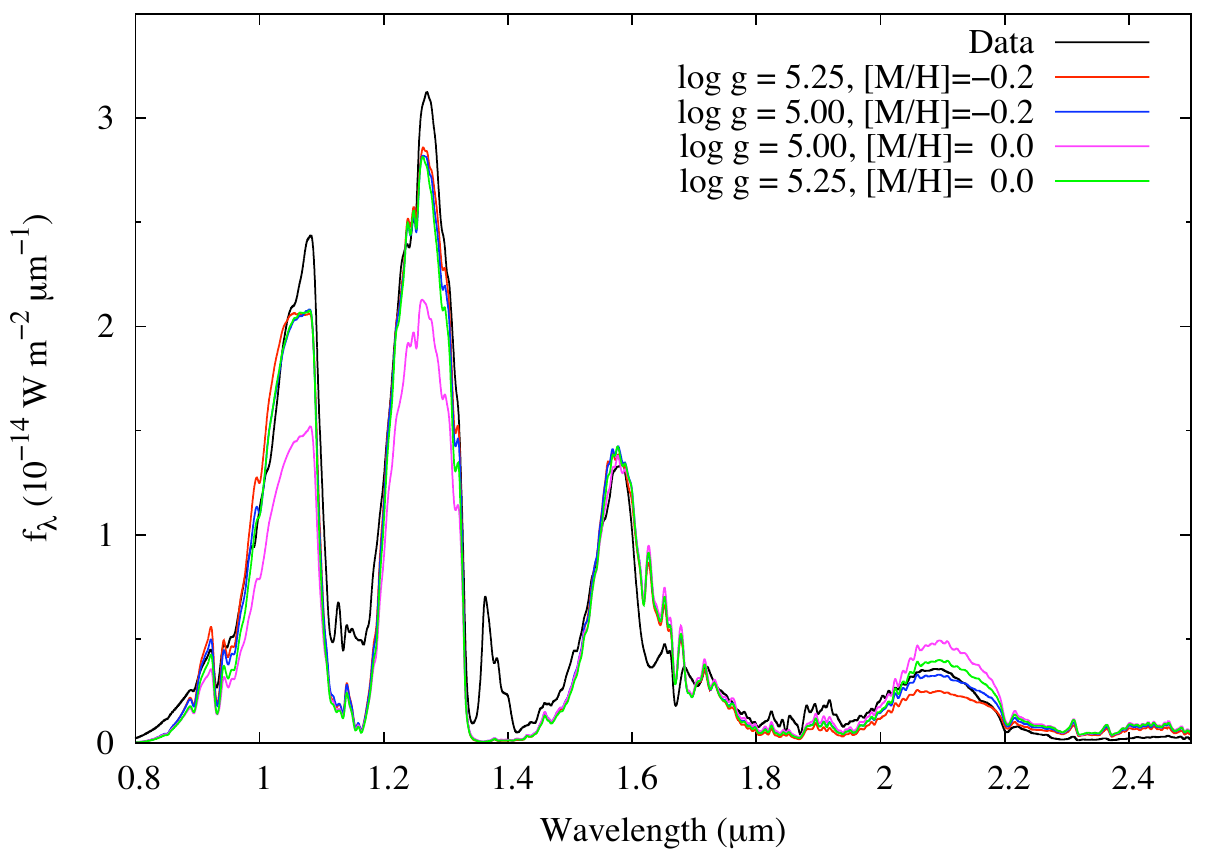}}
  \caption{Near-IR spectra of \epsbb\ (black line), the
    BT-Settl spectrum (T$_{\rm{eff}}$=920K, log g=5.25, [M/H]=$-0.2$,
    red line), the same with log g=5.00 (blue line), with log g=5.00,
    [M/H]=0.0 (magenta line), and log g=5.25, [M/H]=0.0 and (green
    line). All spectra have been smoothed to 60\,\AA\ FWHM and the
    observed spectrum median filtered to remove the lowest
    signal-to-noise points. The thermal-IR spectra have not been shown
    as there is little to distinguish between models.}
    \label{fig:metal_logg_Bb}
\end{figure}

Figure \ref{fig:metal_logg_Bb} shows the near-IR spectrum of \epsbb\
compared to models with an effective temperature of 920\,K, surface
gravities log g=5.00 and 5.25\,cm\,s$^{-2}$, and metallicities of
[M/H]=0.0 and $-0.2$. We neglect the longer wavelength data here as
the models show little variation. As with the comparison of models
with different effective temperatures, we find the $H$-band is
relatively invariant to the different surface gravities and
metallicity. The effects of surface gravity and metallicity are most
apparent in the $J$- and $K$-bands. The general trend for dust-free T
dwarfs such as \epsbb\ is well established
\citep[cf.][]{Leggett:2009}: both the lower metallicity and higher
gravity result in higher atmospheric pressures, favouring the
formation of methane and increasing collision induced absorption (CIA)
opacity around 2\micron.

The solar metallicity model with surface gravity log g=5.00 is the
most inconsistent with our observations as the 1.1 and 1.25\micron\
peaks are significantly under-estimated while the 2.1\micron\ peak is
significantly over-estimated. As explained earlier (see
Sect.\,\ref{sec:atm_model_comp}), we have chosen to use the
\citet{Grevesse:1993} abundances which is likely the cause of the
differences seen between our solar metallicity model spectra and those
used in \citet{Kasper:2009}. The other models make very similar
predictions to one another, except in the $K$-band. Although from
Fig.\,\ref{fig:metal_logg_Bb}, the log g=5.00, [M/H]=$-$0.2 model is
marginally favoured, the comparison to all models within the
effective temperature range of 880--940\,K favours log g=5.25,
[M/H]=$-$0.2. Additionally, none of the models can reproduce the lower
flux seen in the 2.2--2.5\micron\ region.

\begin{figure*}
    \resizebox{\hsize}{!}{\includegraphics{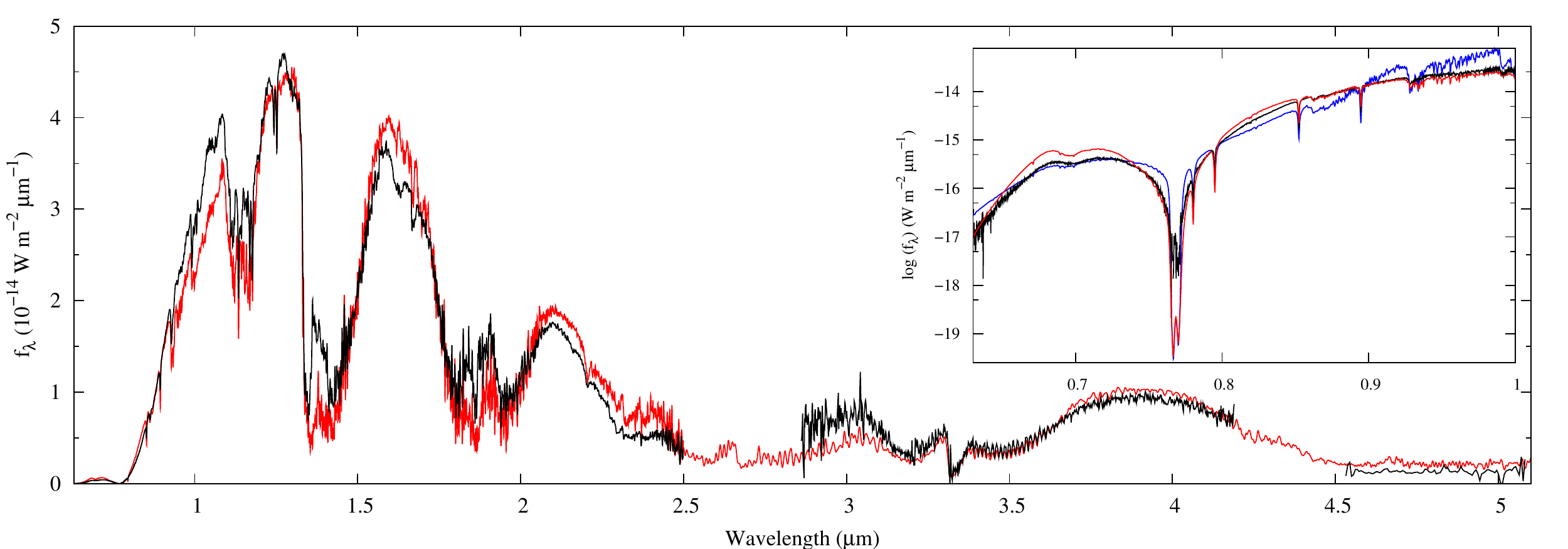}}
    \caption{Optical to thermal-IR spectrum of \epsba\ (black line)
      and the best-fit BT-Settl spectrum (red line,
      T$_{\rm{eff}}$=1300K, log g=5.50, [M/H]=$-0.2$). Here the
      optical and near-IR data have been smoothed to 17\,\AA~FWHM
      and median filtered to remove the lowest signal-to-noise
      points. The model has been smoothed to 60\,\AA~FWHM beyond
      2.5\micron. Inset: the optical portion of the spectra smoothed
      to 6.5\,\AA~FWHM shown with the flux on a logarithmic
      scale. Also shown is a model fit to the optical spectrum with
      depleted alkali abundances (blue line).  For this early T-dwarf,
      the standard alkali abundances provide the better match to the
      observed flux levels and spectral morphology. Additionally, the
      lithium abundance has been reduced by a factor of 1000 to allow
      a comparison with the observations which show no detectable
      lithium at 6707\,\AA.}
\label{fig:linear_full_Ba} 
\end{figure*}

\begin{figure*}
    \resizebox{\hsize}{!}{\includegraphics{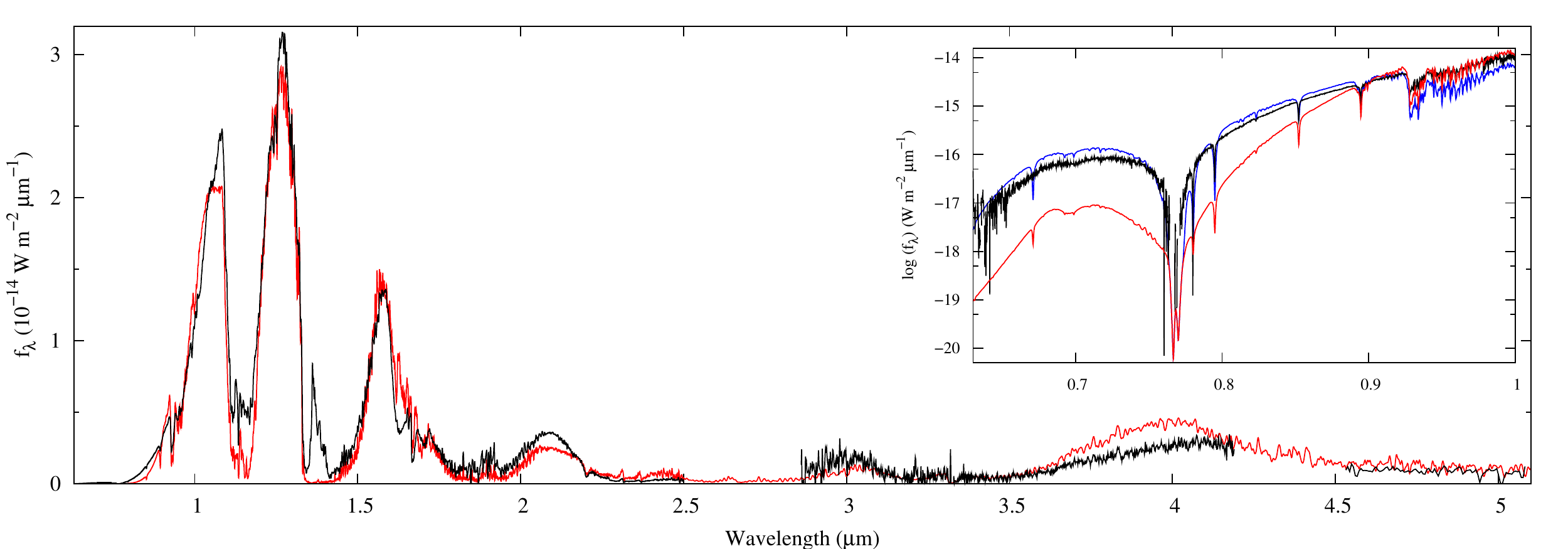}}
    \caption{Optical to thermal-IR spectrum of \epsbb\ (black line)
      and the best-fit BT-Settl spectrum (red line,
      T$_{\rm{eff}}$=920\,K, log g=5.25, [M/H]=$-0.2$). The spectra
      have been smoothed as in
      Fig.\,\ref{fig:linear_full_Ba}. For this late T-dwarf,
      the depleted alkali model provides the better match to the
      observed flux levels and spectral morphology in the optical
      which may be an indication of the formation of feldspars in
      late-T dwarfs which is not evident in earlier types. No lithium
      depletion has been implemented in these models.}
    \label{fig:linear_full_Bb}
\end{figure*}

In summary, we find BT-Settl atmosphere models with effective
temperatures in the range 1300--1340\,K, surface gravity of log
g=5.50, and slightly sub-solar metallicity of [M/H]=$-0.2$ are the
most reasonable fits to the observed optical to thermal-IR spectrum of
\epsba. In Fig.\,\ref{fig:linear_full_Ba} we show the 1300\,K spectral
model compared to the data. It is clear that this model does not
provide a perfect match to the near-IR peaks, however general features
are reproduced.  The inset plot shows the mismatch in the shape of the
\ion{K}{I} profile at 0.77\micron\ which may suppress the continuum to
beyond 1\micron\ \citep{Burrows:2000} and so affect the relative
strengths of the 1.1\micron\ and 1.25\micron\ peaks. Although the flux
in the 2.8--3.2\micron\ region is underestimated, the 3.3--4.2\micron\
region is well-reproduced. However, in the $M$-band the model predicts
higher fluxes than observed, suggesting insufficient understanding of
the sources of opacity.

For \epsbb, we find the BT-Settl models with effective temperatures of
880--940\,K, surface gravity of log g=5.25, and metallicity of
[M/H]=$-0.2$ provide the best match to the observations.
Figure~\ref{fig:linear_full_Bb} shows the 920\,K model does not
provide a perfect match to the near-IR peaks, although this is less
pronounced than for Ba. The inset plot shows the large mismatch in the
shape and flux level of the \ion{K}{I} profile at 0.77\micron\ and the
effect of employing a model with depleted alkali absorption. This
effect is also evident in the optical colour-magnitude diagram of
Fig. \ref{fig:Iso_MJvsI-J} where \epsbb\ falls $\sim$1.5~mag redward
of the COND03 evolutionary models. We return to discuss the cause of
this large effect in Sect.\,\ref{sec:alkali}. The mismatch around
1.62--1.74\micron\ is mainly due to incomplete knowledge of CH$_4$
absorption in this range. In the 2.8--3.5\micron\ region, the shape of
the spectrum due to CH$_4$ absorption is not well fit.  In particular,
the drop in flux predicted at $\sim$4.1\micron\ is not observed, and
the $M$-band flux is again over-estimated.

The KI-H$_2$ quasi-molecular satellite feature at $\sim$0.69\micron\
discussed in \citet{Allard:2007} is seen here in both \epsba\ and Bb
(inset plots of Figs.  \ref{fig:linear_full_Ba} and
\ref{fig:linear_full_Bb}). The shape of the feature is not perfectly
reproduced, mainly because the effects of KI-He absorption have yet to
be included. To a lesser extent, the ill-fitting wings of the very
wide, pressure-broadened absorption lines of \ion{K}{I} at
0.77\micron\ and \ion{Na}{I} at 0.59\micron\ may also play a role.

\subsection{unidentified feature at 1.35--1.40\micron}
\label{1.35_feature}

We note a spectral feature at $\sim$1.35--1.40\micron\ in the spectra
of both \epsba\ and Bb which is poorly reproduced by the BT-Settl
models (see Figs. \ref{fig:linear_full_Ba} and
\ref{fig:linear_full_Bb}) and has not been identified in all spectra of
the T dwarf standard stars. Many observers remove this region from
published spectra due to the high telluric absorption and so there are
few available comparisons. Nevertheless, the region contains valuable
information on the continuum flux level in these deep absorption bands
present in T dwarfs. A similar feature is however seen in the spectrum
of the T1 spectral standard SDSS0151+1244 \citep{Burgasser:2006}, the
T8.5 and T9 dwarfs ULAS1238 and ULAS1335 \citep{Burningham:2008},
and some L dwarfs \citep[e.g. 2MASS J1507$-$1627
(L5),][]{Burgasser:2007}. 

While this feature could be caused be problems with the telluric
correction in this region, we believe that it may be intrinsic to these
objects. We are satisfied that this feature is not an artefact of
our spectral extraction. The rise in flux occurs part way through the
last of the seven spectral orders which were combined to produce our
$J$-band spectrum (see Appendix \ref{sec:phot_calib}), so the shape is
not due to a scaling mismatch between adjoining regions. This region
corresponds to the gaps in our standard star spectrum where high
telluric absorption required replacement by the solar model of Kurucz
(see Sect.\,\ref{sec:nir_spec-JHK}), but this is not the cause of the
feature, as the solar model has only weak spectral features here.

The BT-Settl model for \epsba\ shows some structure in the deep water
bands between the $J$ and $H$ peaks that is qualitatively similar to
the observed feature, though the flux is underestimated by a factor of
$\sim$2. In the \epsbb\ model the disagreement is much worse, with
about an order of magnitude mismatch between the modelled feature
(which is nearly invisible at the scale of
Fig.\,\ref{fig:linear_full_Bb}) and the observation. Still, this
suggests that the feature is due to the structure of the strongest
part of the water absorption bands, which in this part of the spectrum
form fairly high up in the atmosphere, corresponding to optical depths
of $\tau_{\rm{Rosseland}}$=0.1--1.0 and temperatures of 700--1000\,K.
Incidentally, two other features which also have systematically
underestimated flux in the present models form at a similar level,
namely the collisionally-induced absorption (CIA) capping the flux peak
in the $K$ band, and the wings of the potassium doublet centred on
0.77\micron\ (see Sect.\,\ref{sec:alkali}). A possible explanation for
the mismatch in all these cases might be an underestimated local
temperature, and thus source function, at this atmospheric level. A
toy model indicates that an increase in temperature of 200--400\,K
could reconcile the modelled and observed flux levels, but at this
point we have no reasonable idea for the cause of such heating (e.g.
for an additional opacity source causing a corresponding back-warming).
A full explanation of this feature must also account for the apparent
difference in strength between objects of the same spectral type.




\subsection{alkali depletion}
\label{sec:alkali}

The optical spectra of T dwarfs are dominated by alkali resonance
lines which are the result of a balance between the increased
transparency of the atmospheres, due to the progressive sedimentation
of condensates, and the depletion of alkali metals from the gas phase, 
also due to condensation processes. It has been noted
\citep{Lodders:2006, Burrows:2009} that alkali metal depletion cannot
occur above T$\sim$\,1400\,K, below the condensation temperatures of
most refractory species and furthermore, the first condensates are the
feldspars ([Na,K]AlSi$_3$O$_8$) which require aluminium and silicon to
form. However, in a stratified atmosphere, the latter elements can be
expected to have already been depleted by higher-temperature
condensates before feldspars would have a chance to form. In that case,
the alkali elements could only condense into sulfides and halides
(mostly Na$_2$S, KCl) at temperatures around 1000\,K.

In any case, for mid- to late-type T dwarf atmospheres, the depletion
of the sodium and potassium is not adequately accounted for in current
models. At the higher temperatures of early T dwarfs such as \epsba,
the standard BT-Settl model reproduces the shape and flux level of the
\ion{K}{I} doublet at 0.77\micron\  (and the red edge of the \ion{Na}{D}
line at 0.589\micron) reasonably well as shown in
Fig.~\ref{fig:linear_full_Ba} (inset), but over-estimates the
absorption in the later type \epsbb\ by an order of magnitude
(Fig.~\ref{fig:linear_full_Bb}, inset, red line).

The line wings of the \ion{K}{I} resonance doublet at 0.77\micron\  
extend several 1000 \AA\ and thus potentially suppress the flux out to
the 1.1 and even 1.25\micron\ peaks. As pointed out already by
\citet{Burrows:2000}, these line profiles show strong deviations from
classical Lorentzian wings and thus require a more sophisticated
broadening theory. Detailed spectral models of the alkali lines have
been successfully used to model the optical spectrum of \epsba\ by
\citet{Allard:2007}, based on improved interaction potentials with
H$_{2}$ and He for the line profiles, and an earlier version of the
settling framework to calculate the depth-dependent abundances of
neutral alkali atoms. They were also able to identify the
quasi-molecular satellite of \ion{K}{I} seen in our resolved spectra
(see Figs.\,\ref{fig:linear_full_Ba} and \ref{fig:linear_full_Bb}).
While the overall agreement of the modelled optical spectrum for
\epsba\ is fair, the Bb model, though showing the same general
features, underestimates the observed flux by up to an order of
magnitude. Since the lines in both brown dwarfs should form at quite
similar temperature levels, it is extremely unlikely that the line
profiles are off by such a large amount for the cooler component only.
One possible explanation for the mismatch is that the \ion{K}{I} in the
gas phase is much less abundant than expected from the settling model.

If we therefore consider additional condensation of the alkali metals
into feldspars, which would result in alkali depletion from the gas
phase at somewhat higher temperatures than in current models, we can
bring the \epsbb\ spectrum into reasonable agreement with observations
as shown in Fig.~\ref{fig:linear_full_Bb} (inset, blue line).  However
in this case the fit to the \epsba\ spectrum becomes worse. The
formation of feldspars implies that  Al and Si would still have to be
present in sufficient quantities at the feldspar condensation level
and, indeed, recent calculations using results from updated RHD
simulations \citep[][in prep.]{Freytag:2009, Homeier:2009} imply that
the upmixing of these species might be more efficient than previously
assumed. Thus we suggest that feldspar formation can efficiently
deplete these species at the temperatures of mid-late T dwarfs.



\subsection{lithium}
\label{sec:lithium}

\begin{figure} 
\resizebox{\hsize}{!}{\includegraphics{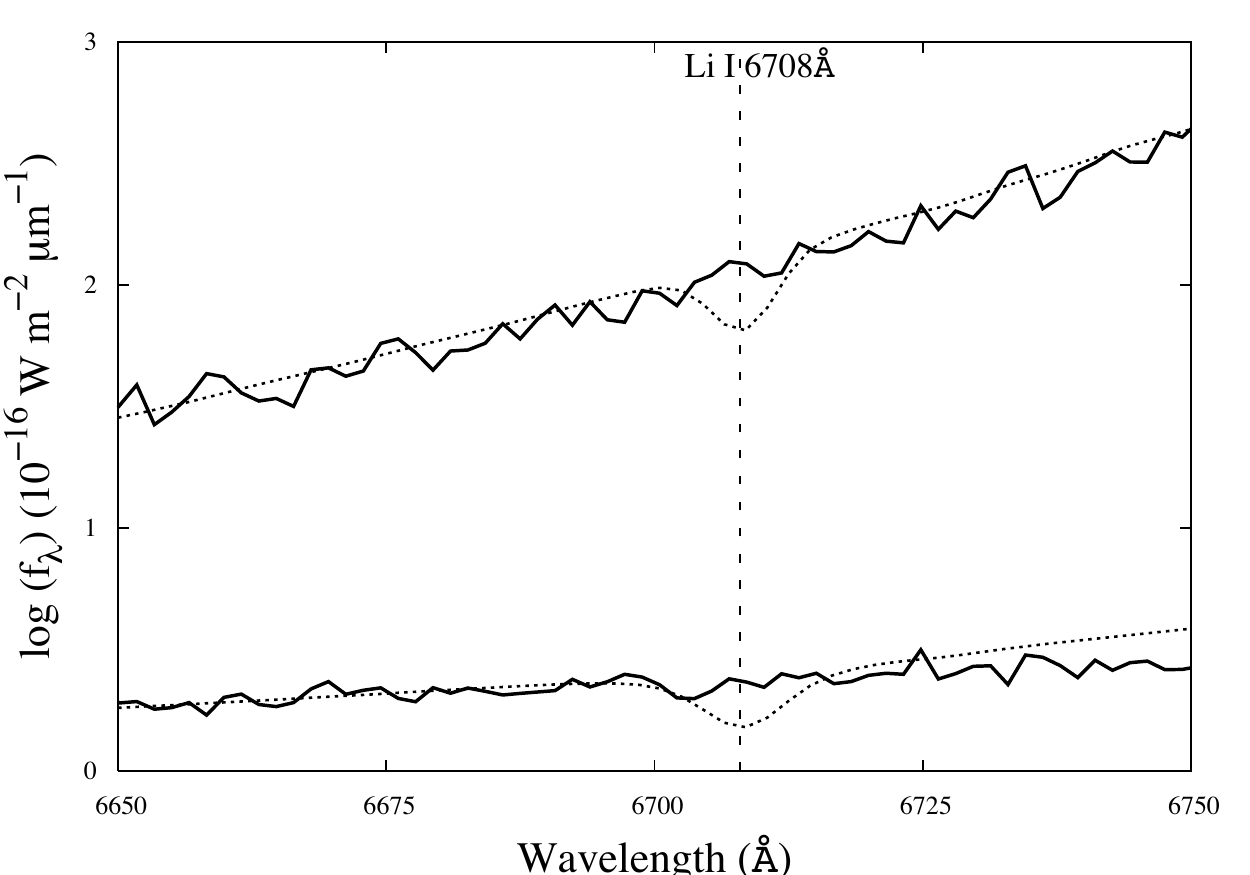}}
\caption{The 6650--6750\,\AA\ spectra of \epsba\ and Bb (upper and lower
  thick black lines, respectively) and the BT-Settl models with T$_{\rm{eff}}$=1300\,K,
  log g=5.50, and [M/H]=$-$0.2 (upper dotted line) and T$_{\rm{eff}}$=920\,K,
  log g=5.25, and [M/H]=$-$0.2 (lower dotted line). Both models have
  been smoothed to 6.5\,\AA~FWHM to match the observations. The
  lithium abundance has been reduced by a factor 1000 in the
  T$_{\rm{eff}}$=1300\,K model in order to  provide a meaningful upper
  limit to the observed levels of \ion{Li}{I} absorption in the spectrum of
  \epsba. }
    \label{fig:lithium}
\end{figure}

The presence and state of lithium in the atmospheres of brown dwarfs
is governed by mass, present effective temperature, and age. The
models of \citet{DAntona:1994}, \citet{Chabrier:1996}, and
\citet{Burke:2004} predict that a brown dwarf must have a mass greater
than 0.060--0.065\,M$_{\sun}$ to be capable of reaching the core
temperature of $\sim$3$\times$10$^6$ K \citep{Bildsten:1997} required
for lithium depletion ($^{7}$Li + $p$ $\rightarrow$ $^{4}$He +
$^{4}$He). Therefore, since objects less massive than
$\sim$0.4\,M$_{\sun}$ are fully convective, all lithium should have
been processed for any brown dwarf above 0.065\,M$_{\sun}$.
\citet{Chabrier:1996} predict that an object at this mass boundary
will deplete its primordial lithium by a factor of 100 in
$\sim$1\,Gyr, with faster depletion in higher-mass objects.

Furthermore, the substellar chemistry models of \citet{Lodders:2006}
show that \ion{Li}{I} is the dominant form of lithium down to
$\sim$1520\,K at 1 bar pressure (with the temperature limit increasing
with increasing pressure), beyond which lithium is bound into
molecules (LiCl, LiOH, etc.). \citet{Kirkpatrick:2000} and
\citet{Kirkpatrick:2008} presented spectra of a number of L dwarfs
with and without \ion{Li}{I} absorption. They showed that the strength
of the \ion{Li}{I} 6708\,\AA~line peaks at L6 and is observed in some
L7 and L8 dwarfs with the strength of the line and number of
detections decreasing toward later types.


Here we have higher resolution and higher signal-to-noise data and,
from the observed spectral type-equivalent width relation of
\citet{Kirkpatrick:2000}, we would have expected to detect \ion{Li}{I}
at 6708\,\AA\ in \epsba\ if it were present, but not necessarily in
\epsbb. Indeed, \citet{Burrows:2002} state that there is no apparent
reason why \ion{Li}{I} would not be seen in objects as late as T6 with
high enough quality data.

On the other hand, the presence of \ion{Li}{I} absorption in our
current atmospheric models may be due to unrealistic modelling of the
removal of solid species from the atmosphere after formation (Marley,
pers. comm.). As can be seen in the Fig.\,\ref{fig:lithium}, we find
no evidence for absorption by monatomic lithium at 6708\,\AA\ in the
spectrum of \epsba, while the spectrum of \epsbb\ may have inadequate
signal-to-noise at this wavelength to judge its presence or absence at
the low abundance levels expected for T6 dwarfs. To approximately
quantify the level of lithium depletion, our models for \epsba\ must
have the proto-solar lithium abundance reduced by a factor of at least
1000 (see Fig.~\ref{fig:lithium}) in order to reproduce the data.

The level of lithium depletion in \epsba\ is compatible with a mass in
excess of 0.065\,M$_{\sun}$ (68\mjup). The less luminous \epsbb\ on
the other hand, must be less massive than this limit, and given its
lower temperature, has most probably lost its atomic lithium to
molecules.




\subsection{chemical equilibrium departures}
\label{sec:chem}

\begin{figure}
\resizebox{\hsize}{!}{\includegraphics{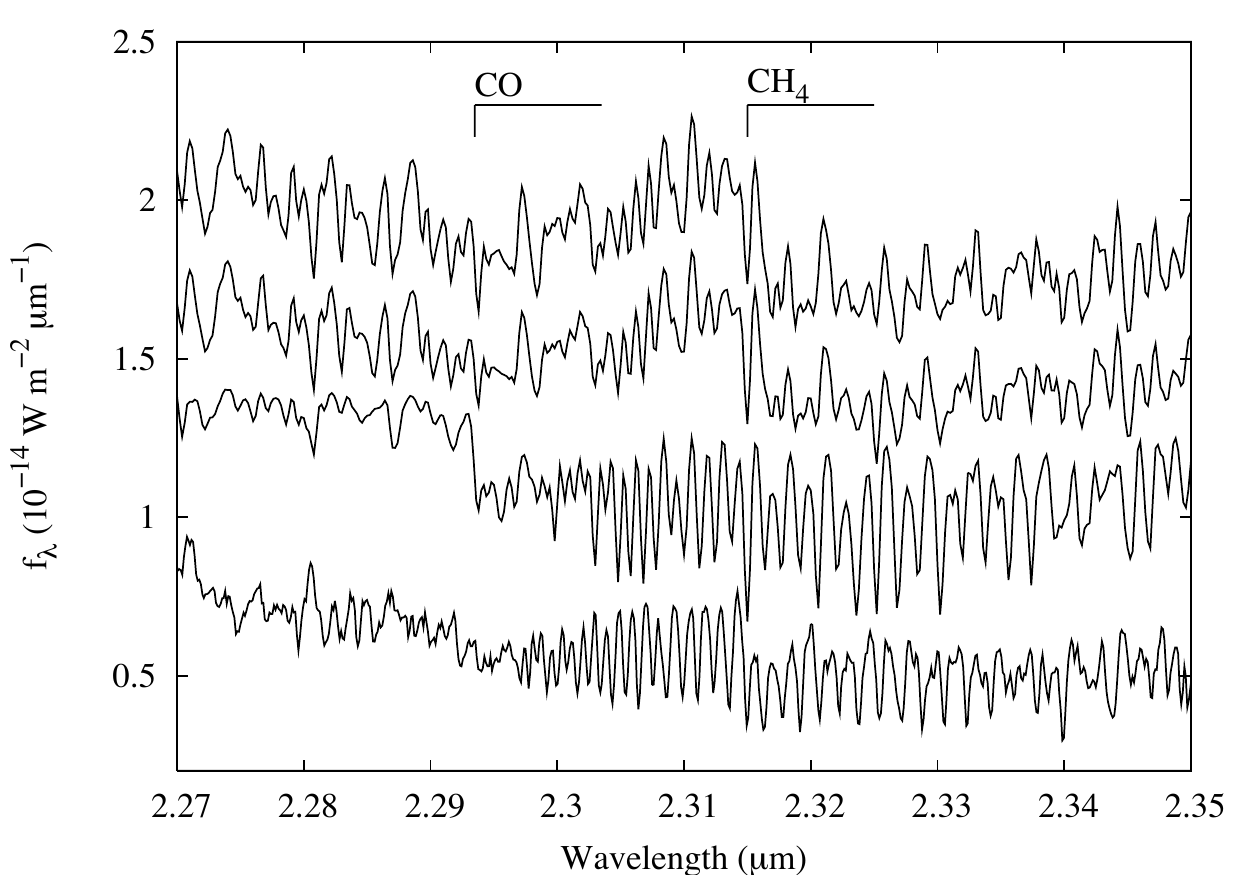}}

\caption{The 2.27--2.35\micron\ spectrum of \epsba\ (bottom line)
showing the CO 2--0 overtone along with models employing different
chemical equilibrium (CE) timescales. All models have
T$_{\rm{eff}}$=1320\,K, log g=5.50, and [M/H]=$-$0.2. A chemical
equilibrium model is shown at the top, the \citet{Prinn:1977} model second from
bottom, and a model following the \citet{Yung:1988} chemistry second from top. All
models have been smoothed to 4.9\,\AA~FWHM to match the observations.
 The \epsba\ spectrum shows the absolute flux scale, while the
\citet{Prinn:1977}, \citet{Yung:1988} and CE models have been offset
for clarity by 0.1, 0.8, and 1.0, respectively.}

\label{fig:chem_kband} 
\end{figure}

At the effective temperatures of \epsba\ and Bb, carbon would be
expected to be predominantly locked in carbon monoxide (CO) in the
hotter, lower layers of the atmosphere, and in methane (CH$_4$) in the
cooler, upper regions under chemical equilibrium (CE) conditions.
However, since the detection of the CO fundamental band in Gl\,229B by
\citet{Noll:1997}, observational evidence has accumulated that CO
persists in the upper atmospheres of T dwarfs in excess of its CE
abundance.

\citet{Griffith:1999} and \citet{SaumonIAU211} have shown that this excess
can be explained by the upmixing of CO from the warm deeper layers, since
if one assumes sufficiently efficient turbulent mixing in the upper
atmosphere, the kinetic rates for the conversion reactions from CO to
CH$_4$ are too slow to adjust the mixing ratios to CE abundances. The
BT-Settl models employ a similar calculation of these CE departure
effects, but using diffusion coefficients derived from the CO5BOLD
radiation hydrodynamics simulations \citep{Freytag:2009} rather than
adjusting them as a free parameter \citep[][in prep.]{Homeier:2009}. 
The diffusion coefficients thus calculated are therefore height-dependent
and not directly comparable to a single choice of an eddy diffusion
coefficient, with typical values in our models ranging between 10$^5$ and
10$^9$ cm$^2$/s. Since the CE departure is most sensitive to the mixing at
the temperature level relevant for the transition from CO to CH$_4$, i.e.
between 1000 and 1500\,K, the diffusion coefficient in this part of the
overshoot region would best characterise our model, corresponding to
10$^7$--10$^8$ cm$^2$/s in the \epsba\ model.  In addition, we consider
several reaction pathways and timescales besides the one from
\citet{Prinn:1977}, on which the \citet{SaumonIAU211} models are based,
notably the revised time scale for the scheme of \citet{Prinn:1977}
suggested by \citet{Griffith:1999}, and the reaction scheme of
\citet{Yung:1988} (see also \citet{Griffith:1999}). Among these, the
\citet{Yung:1988} model generally predicts the fastest conversion rates
from CO to CH$_4$, and the \citet{Prinn:1977} model with the modified
rates of \citet{Griffith:1999} the slowest rates, implying the strongest
CE departure effects.

Since the velocity field derived from the CO5BOLD simulations is used
for the description of both the cloud dynamics and the CE departures,
these models employ a high degree of self-consistency, with any change in
the mixing properties immediately affecting both the dust content and the
gas-phase chemistry of the atmosphere. On the other hand, other potential
uncertainties in our cloud model would also feed back into the thermal
profile of the atmosphere and might thus affect the domains of different
carbon chemistry discussed above. However, any major changes in cloud
opacity would also inevitably change the spectral energy distribution and
thus produce inconsistencies with the observed IR photometry. As the
overall fit to the spectral energy distribution is good, we thus feel
confident that our chemistry model is not affected by major uncertainties
in the thermal structure due to backwarming from the cloud deck.

\begin{figure} \resizebox{\hsize}{!}{\includegraphics{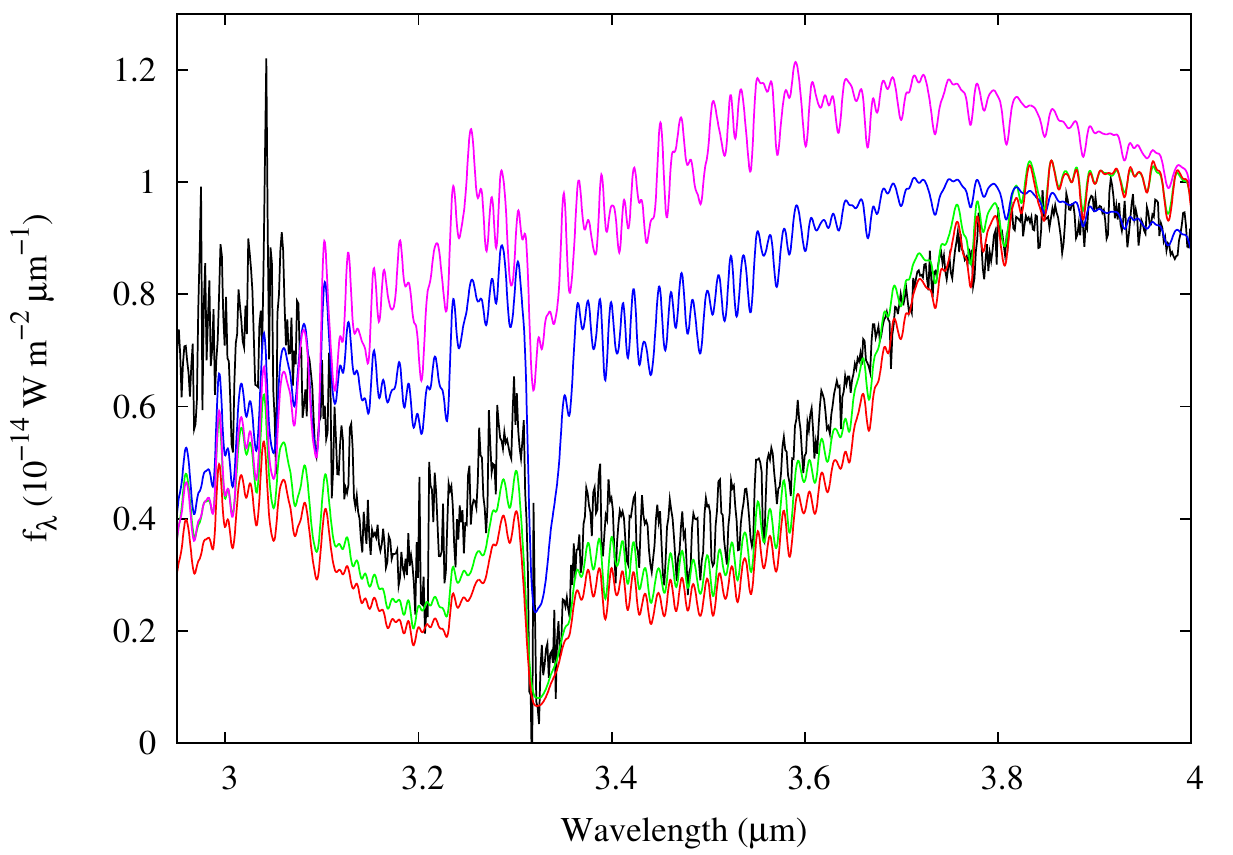}}
  \caption{The $L$-band spectrum of \epsba\ (black line) and different
    chemistry models with T$_{\rm{eff}}$=1320\,K, log g=5.50,
    [M/H]=$-$0.2. A chemical equilibrium model is shown in green, the
    \citet{Prinn:1977} model in blue, the \citet{Prinn:1977} model
    with the revised reaction rates of \citet{Griffith:1999} in
    magenta, and a model following the \citet{Yung:1988} chemistry in red. All models have
    been smoothed to 60\,\AA~FWHM to match observations.}
    \label{fig:chem_lband}
\end{figure}

As an L--T transition object, marking also the transition from
CO-dominated to CH$_4$-dominated chemistry, \epsba\ is particularly
sensitive to the reaction details discussed above. The high quality of the
present $K$- and $L$-band spectra thus allow quantitative estimates of the
CO and CH$_4$ mixing ratios, enabling us to directly test these models of
the non-equilibrium chemistry and constrain the relevant timescales.
Fig.\,\ref{fig:chem_kband} shows the observed 2.27--2.35\micron\ spectrum,
where the CO 2--0 overtone band (starting at 2.2935\micron) can be
identified even on top of the octad band of CH$_4$. For comparison, a
chemical equilibrium model and non-CE model spectra, based on the reaction
models of \citet{Prinn:1977} and \citet{Yung:1988} respectively, are
shown. The \citet{Prinn:1977} model better reproduces the rotational
series of the CO 2--0 R branch extending redwards from 2.2935\micron, 
but the CE and \citet{Yung:1988} models match the overall morphology
better, including the 2.315\micron\  CH$_4$ bandhead.  However, as
seen in Fig.~\ref{fig:chem_lband}, the \citet{Prinn:1977} model does not
match the shape of the $L$-band spectrum, whereas the chemical equilibrium
model and the \citet{Yung:1988} model provide reasonable matches to the
shape and absolute flux levels - the CH$_4$ absorption predicted by both
models is identical within the internal uncertainties of the model
atmospheres. Also shown is a model employing the \citet{Prinn:1977} model
with the revised reaction rates of \citet{Griffith:1999} which shows an
even poorer match to the spectral morphology, since in this model even
less CH$_4$ has formed. However, the depth of the CH$_4$ absorption at
$\sim$3.3\micron\ falls between that seen in the spectra predicted
by the \citet{Yung:1988} and \citet{Prinn:1977} models, just as the
2.3\micron\ spectrum indicates a CO abundance intermediate between
those models. This would suggest that the correct reaction timescale might
have a value slightly below that of \citet{Yung:1988}, or alternatively,
the vertical mixing could be more efficient than assumed in the BT-Settl
models.  \citet{Freytag:2009} suggest such additional mixing can be
produced by convectively driven gravity waves, though the mixing
efficiency of such waves in terms of an equivalent diffusion coefficient
is still under investigation.

Our observations provide only limited coverage and resolution of the
strongest CO absorption feature, the 1--0 fundamental band at 4.55\micron\
(see Fig.~\ref{fig:Bab_1.9-2.5}). However, the resolution of our $M$-band
spectra does not allow us to favour any model over the other. Although the
models differ in the predicted absolute flux levels, none of these match
the observed flux levels.

We have shown that the observed $K$-band spectrum of \epsba\ shows
evidence of non-CE processes and the $L$-band spectrum supports a
CO--CH$_4$ reaction rate intermediate between that predicted by the
\citet{Yung:1988} and \citet{Prinn:1977} models.  However, we postpone a
more detailed analysis of the non-chemical equilibrium signatures to
a future paper.


\section{Mass limits on further system members}
\label{sec:Bc_mass_limits}

Using the previously derived flux limits from the deep imaging of the
system (see Sect.\,\ref{deep_comp_search}) and an approximate system
age of 5\,Gyr, we can derive mass limits of further members using
models. The COND03 models rule out any system members of mass greater
than 15\,M$_{\rm{Jup}}$ ($H$=19.1$^{\rm{m}}$) in the field around
\epsba, Bb (7--39.7\,AU), and also rule out objects more massive than
34\,M$_{\rm{Jup}}$ ($H$=15.8$^{\rm{m}}$) from 7\,AU down to 0.8\,AU
from either Ba or Bb. If the system were as young as 1\,Gyr, the
corresponding mass limits would be 6.0 and 13\,M$_{\rm{Jup}}$,
respectively.

At separations less than 1--2 FWHM ($\sim$0.4--0.8\,AU), our ability
to distinguish companions of Ba or Bb is limited by the high object
flux and pixelation of the PSF. However, if either \epsba\ or Bb had
substantial unresolved companions, then their combined observed
spectra would be notably different compared to the spectral standards.
The combined spectrum of \epsba\ and Bb is drastically different to
that of Ba alone, so any additional companion must be significantly
fainter, and hence less massive, than Bb.



\section{Comparison of  Model Predictions}
\label{sec:model_diffs}

\subsection{previous determinations}

Observations of \epsba, Bb have previously been compared to different
atmospheric and evolutionary models. \citet{Smith:2003} used
R$\sim$50\,000 spectra in the ranges 1.553--1.559\micron\ and
2.308--2.317\micron\ to derive an effective temperature for \epsba\
from comparisons of the observed spectra with spectral models to fit
bands of CO and H$_2$O. They derived effective temperatures of 1400\,K
and 1600\,K from the two regions and so adopted 1500\,K as the
effective temperature of \epsba. As they note, this is significantly
in excess of that derived in other analyses, and the temperatures
derived from the two regions are inconsistent. Moreover, the effective
temperature of 1600\,K derived for the 2.308--2.317\micron\ region may
be affected by the CO/CH$_{4}$ chemistry.

\citet{Roellig:2004} and \citet{Mainzer:2007} observed the combined
spectrum of the \epsba, Bb system using IRS on {\em{Spitzer}}. Using
the luminosities of \citet{McCaughrean:2004} and an age of
$\sim$1\,Gyr, they derived evolutionary model parameters of 
T$_{\rm{eff}}$=1210\,K, log g=5.10 (for a cloudy evolutionary model)
and T$_{\rm{eff}}$=840\,K, log g=4.89 for \epsba\ and Bb, respectively.
Using these parameters along with radii of 0.094 and 0.100\,R$_{\sun}$
for \epsba\ and Bb, respectively, \citet{Roellig:2004} and
\citet{Mainzer:2007} generated a composite model spectrum which agreed
well with the observed combined spectrum. However, \citet{Sterzik:2005}
presented mid-IR photometry of the individual sources from VLT/VISIR
which they suggest is not compatible with the absolute fluxes of the
individual models from \citet{Roellig:2004}. Indeed, as pointed out in
Sect.\,\ref{sec:lum}, the {\em{Spitzer}} and VISIR fluxes differ
significantly. Using an assumed age of 1\,Gyr, \citet{Sterzik:2005}
found an effective temperature of 1100\,K for \epsba. However, our
comparison of the optical to thermal-IR spectrum excludes such low
effective temperatures.


Finally, \citet{Kasper:2009} used low resolution near-IR spectroscopy
and the models of \citet{Burrows:2006} to yield effective temperatures
of 1250--1300\,K and 875--925\,K and surface gravities of 5.2--5.3 and
4.9--5.1, for \epsba\ and Bb, respectively. These temperatures are
broadly comparable to ours, although in more detail we find the
BT-Settl spectral models cannot match the observed \epsba\ spectra at
effective temperatures as low as 1250\,K. Their grid of surface gravity
was finer than used here and they derive values significantly lower
than ours. For \epsba, we have compared our observations to models with
log g=5.25 and 5.50, and specifically prefer the higher value.
Similarly, for \epsbb, we have tested models with log g=5.00 and 5.25,
and again prefer the higher value. The various determinations of
the effective temperatures of \epsba\ and Bb are summarised in
Table\,\ref{model_prediction_compare}.

The situation is complicated by complementary effects on the spectral
morphology due to surface gravity, metallicity, and elemental
abundances. We do not believe that the surface gravity can be
constrained to better than $\sim$0.25 dex. More precise determinations
of the surface gravity can lead to substantially inaccurate
predictions of the mass as in \citet{Kasper:2009}, which may do the
evolutionary models an injustice. Indeed, \citet{Burrows:2006} suggest
that the lack of a detailed understanding of brown dwarf meteorology
may lead to ambiguity in derived effective temperatures of
$\sim$50--100\,K and surface gravities and $\sim$0.3.

\citet{Kasper:2009} used their fitted effective temperatures and
surface gravities to derive ages and masses from the evolutionary
models of \citet{Burrows:1997}. For \epsba, they found an age of
1.0--2.3\,Gyr and mass of 46--62\mjup, and for \epsbb\ an age of
1.0--2.0\,Gyr and a mass of 29--39\mjup. Although the objects appear
to be co-eval and consistent with the age estimate of
\citet{Lachaume:1999}, this age range is clearly lower than we have
derived in the present paper using the observed luminosities and
dynamically derived system mass. Additionally, their predicted total
maximum system mass (101\mjup) is considerably lower than the measured
system mass (121\mjup). This suggests that either there are large
systematic errors in the evolutionary model predictions, or the
effective temperatures and surface gravities derived from observed
spectra are inaccurate. This comparison highlights the problems
associated with deriving the physical properties of other field
objects for which we do not have a comprehensive set of observations
or external constraints from a companion star.



\begin{table} 
  \centering

  \caption{Comparison of our model predictions with previous
studies.  The last two entries show the determinations of the effective
temperatures of the two T dwarfs from this work, both from fitting
atmospheric models to the observed spectra, and by comparing the
derived luminosities and dynamical system mass with evolutionary
models.}

\label{model_prediction_compare}
\begin{tabular}{ccccc}
  \hline
  \hline
  Study & \epsba\ T$_{\rm{eff}}$ & \epsbb\ T$_{\rm{eff}}$ \\
  \hline
  \citet{Smith:2003} & 1400,1600 & ... \\  
  \citet{Roellig:2004} & 1250 & 800 \\ 
  \citet{Sterzik:2005} & 1100 & ...  \\
  \citet{Mainzer:2007} & 1250 & 800 \\  
  \citet{Kasper:2009} & 1250--1300 & 875--925  \\
  This work (atm. models) & 1300--1340 & 880--940  \\
  This work (evo. models) & 1352--1385 & 976--1011  \\
  \hline
\end{tabular}
\end{table}

\subsection{our model predictions}

Our comparison of the observed luminosities and measured total mass of
\epsba, Bb has allowed us to derive a predicted age range of
 3.7--4.3\,Gyr from the COND03 evolutionary models. While this age
range is higher than the previously used age estimate of 0.8--2.0\,Gyr
from \citet{Lachaume:1999}, it is younger than other age indicators
for \epsa\ in the literature discussed in Sect.\,\ref{sec:age}. From
this age range and the individual luminosities we have extracted the
predicted individual masses, effective temperatures, radii, and
surface gravities (see Table\,\ref{COND_predict}). The effective
temperatures predicted by these models are 1352--1385\,K for \epsba,
and 976--1011\,K for Bb with surface gravities of 5.43--5.45 and
5.27--5.33.

We have also independently derived the effective temperature and
surface gravity of both objects through a direct comparison of the
observed optical to thermal-IR spectra with the BT-Settl atmospheric
models. For \epsba, we found models with effective temperature in the
range 1300--1340\,K and surface gravity of log g=5.50 are the most
appropriate, while for \epsbb\ we find effective temperatures in
the range 880--940\,K with surface gravity of log g=5.25, both with
slightly sub-solar metallicity of [M/H]=$-0.2$.

We cannot presently reconcile the effective temperatures derived from
atmospheric modelling with those derived from evolutionary models using
the measured system mass at the observed luminosities. However, it must
be noted that the evolutionary models do not presently incorporate the
new BT-Settl atmospheres. The difference between the predicted
effective temperatures for \epsbb\ show the atmospheric models to be
inconsistent with the evolutionary models. The upper limit on the
effective temperature from the comparison with atmospheric models is
940\,K, while the evolutionary models predict effective temperatures of
at least 975\,K. For \epsba, the difference is less severe. There is
only a 10\,K difference between the limits on the effective temperature
from the evolutionary and atmospheric models. This is midway between
the effective temperature steps in our grid, although it is clear that
by 1360\,K the atmospheric models are no longer consistent with the
spectroscopic observations. These differences may be resolved by the
inclusion of the BT-Settl atmospheric models in the next generation of
evolutionary models, as the effect on the effective temperature would
presumably be intermediate between that of the COND03 and DUSTY00
models.  However, we note that \citet{Dupuy:2009} also find that
the effective temperatures predicted using atmospheric and evolutionary
models are in disagreement for the components of the L4+L4 binary
HD\,130948BC.

The derived surface gravities are consistent between the evolutionary
and atmospheric models. However, due to the grid step chosen (0.25 dex)
and the complementarity with metallicity, the surface gravities derived
from the comparison of our observed spectra with the atmospheric models
do not provide a strong test of the atmospheric and evolutionary model
predictions.  Additionally, any mass estimated from the radius
determined for each spectral model, the fitted surface gravity and our
derived luminosities, will give a large range of possible masses.

As described earlier, once the individual masses have been
determined from the ongoing absolute astrometric monitoring
\citet{Cardoso:2010}, we will be able to test the evolutionary models
using the individual masses, luminosities, and the same age for both T
dwarfs. Additionally, once more reliable age determinations become
available we will be able to directly test the evolutionary models and
determine if the luminosities are overestimated for intermediate age
brown dwarfs as suggested by \citet{Dupuy:2009} for the young system
HD\,130948BC.

For the observed luminosity of \epsba\ and effective temperature range
of 1300--1340\,K derived from the comparison to atmospheric models, the
COND03 evolutionary model predicts a mass for \epsba\ in the range
46--64\mjup, and for \epsbb\ in the range 16--37\mjup\ from its
observed luminosity and the effective temperature range of
880--940\,K. Given the preliminary dynamical system mass of
121$\pm$1\mjup\ \citep{Cardoso:2008}, it therefore appears that with
current theoretical models and spectroscopically derived effective
temperatures, one cannot obtain reliable mass predictions for T dwarfs
such as these even when precise luminosity constraints are available.

\section{Conclusions}
\label{sec:conc}

We have presented the results of a comprehensive photometric and
spectroscopic study of the individual components of the nearest known
binary brown dwarf system, \epsba, Bb. The relative proximity of these
T1 and T6 dwarfs to the Earth resulted in very high quality data,
while archival results for the well-studied parent star, \epsa,
provide invaluable additional information. We find the spectra of
these brown dwarfs are best matched by the BT-Settl spectral
models with T$_{\rm{eff}}$=1300--1340\,K and log g=5.50 for \epsba\
and 880--940\,K and 5.25 for \epsbb, both with a metallicity of
[M/H]=$-0.2$.

COND03 evolutionary model predictions for the masses are significantly
inconsistent with the measured system mass if the young age range of
0.8--2.0\,Gyr suggested by \citet{Lachaume:1999} is used. We find that
a system age of 3.7--4.3\,Gyr is necessary for the COND03 evolutionary
models to be consistent with the measured system mass at the observed
luminosities, and a review of the literature finds evidence supporting
an age of $\sim$5\,Gyr for \eps\ A\@. In the age range 3.7--4.3\,Gyr,
the COND03 models predict effective temperatures in the range
1352--1385\,K and 976--1011\,K, for Ba and Bb, respectively.

It is clear that there are several areas in which the atmospheric
models currently do not reproduce observations and a more detailed
analysis of these issues will be the subject of future work. They
include the strength and shape of the wide absorption by \ion{K}{I}
and \ion{Na}{I} in the optical, the possible formation of feldspars in
mid-late T dwarfs, and the reaction rates of CO and CH$_{4}$. In
addition, the spectral shape in the $L$-band caused by CH$_4$
absorption is poorly reproduced, as is also the case for CH$_4$
absorption at $\sim$1.6\micron. The $M$-band spectra, although low
resolution, also show that the atmospheric models significantly
over-estimate the flux in this region. While the flux levels of the
near-IR peaks can be reasonably reproduced, the level of absorption
between the peaks tends to be problematic. In particular, we find a
feature at 1.35--1.40\micron\ in both our object spectra which is not
predicted in the atmospheric models.


Neither source has detectable atomic \ion{Li}{I} absorption at
6708\,\AA. The absence of lithium in the more massive component is
consistent with the revised, higher age estimates coupled with its
probable dynamical mass, while the lack of absorption in the cooler
source is expected from its low effective temperature, where lithium
is incorporated into molecules.

Although there is significant room for improvement in the atmospheric
models, the current match to \epsba\ and Bb is nevertheless
impressive. When new data on methane opacities become available, we
will be able to better reproduce the observed spectra and more
reliably compare these spectral models to spectra of objects with less
well-constrained physical parameters. Additionally, when these updated
atmospheric models are incorporated into the evolutionary models, a
fully self-consistent comparison will be possible.

Finally, when the individual dynamical masses become available and if
we can obtain a reliable estimate of the age of this system, based on
asteroseismological observations of the parent star \epsa, then
\epsba\ and Bb will become invaluable benchmark objects with a full
set of physical parameters which newer models will have to reproduce,
making them more reliable for analysing the properties of isolated
ultra-cool field dwarfs.

The predictions of the evolutionary models using luminosity and mass
constraints are somewhat different to the derived effective
temperature and surface gravity from fitting atmospheric models to
observed spectra. These differences may be resolved when the newer
atmosphere models are incorporated into the evolutionary models.
However, it seems that derivations of the mass of cool brown dwarfs
are uncertain even where estimates of the effective temperature,
surface gravity, and luminosity exist. We therefore caution against
the over-analysis of predicted brown dwarf masses at this time.

\begin{acknowledgements} 

NSO/Kitt Peak FTS data used here were produced by NSF/NOAO. RRK
acknowledges the support of an STFC studentship.  Part of this work was
funded by the European Commission Marie Curie Research Training Network
CONSTELLATION (MRTN-CT-2006-035890).  RRK would like to thank Adam
Burgasser, Mark Marley, Davy Kirkpatrick, and Sandy Leggett for useful
discussion.  This research has benefited from the SpeX Prism
Spectral Libraries, maintained by Adam Burgasser at
http://www.browndwarfs.org/spexprism. We thank Isabelle Baraffe for
supplying the grid of evolutionary models and also the referee for
helpful suggestions which improved the paper.  

\end{acknowledgements}

\bibliographystyle{aa}
\bibliography{12981}

\appendix

\section{Image Fitting with Analytic Functions}
\label{sec:image_fitting}


The 2-dimensional profile, or the point spread function (PSF), of
ground-based optical/IR images is a superposition of several effects
\citep{Racine:1996}, most importantly the spreading effect of the
Earth's atmosphere caused by the mixing of air of different
temperatures, leading to different refractive indices. In addition,
there is the contribution of the optics and detectors used and any
unintentional telescope motions, for example, due to imperfect
tracking, which can cause elliptical profiles.

Various authors have attempted to produce an analytical function which
represents the shape of the PSF \citep[eg.][]{King:1971, Bendinelli:1987}.
In a study of the two-dimensional profile of stellar images on
photographic plates, \citet{Moffat:1969} found that a Gaussian profile,
usually assumed to be a good match to the seeing, under-estimated the flux
from the star at large radial distances. He proposed an analytical profile
(now termed the Moffat profile) to be a better match to photographic
stellar images. Similarly, \citet{King:1971} found that a Gaussian profile
was not sufficient to match observations, proposing instead a profile
composed of a Gaussian core, falling to an exponential which tails to a
inverse-square aureole. \citet{Franz:1973} presented a further analytical
representation of the PSF which is often referred to as a modified
Lorentzian.  This was later successfully applied to CCD images of stellar
profiles by \citet{Diego:1985}.


We used the Levenberg-Marquardt technique to extract the best-fit
parameters for each of three different analytical PSF models:
Gaussian, elliptical Moffat, and elliptical modified Lorentzian
profiles, with the aim of extracting the ratio of the fluxes of the
two objects, thus allowing individual magnitudes to be found. The
pixelation of the profile was accounted for in the fitting routine by
pixelating the model values to the same resolution as the data.  The
Moffat profile used is an elliptical version of the profile proposed
by \citet{Moffat:1969}:
\begin{eqnarray*}
  M(r) & = & \frac{A}{\left[{1+B_{12}}\right]^{\beta}} \\
  B_{ij} & = & \left(\frac{r}{\alpha_{ij}}\right)^2  =  \left(\frac{\cos^2\theta}{\alpha_{i}^2} + \frac{\sin^2\theta}{\alpha_{j}^2}\right) (x-x_0)^2 \\
  & & +~\left(\frac{\sin^2\theta}{\alpha_{i}^2} + \frac{\cos^2\theta}{\alpha_{j}^2}\right) (y-y_0)^2  \\
  & & +~\left(\frac{1}{\alpha_{i}^2} - \frac{1}{\alpha_{j}^2}\right)2\cos\theta \sin\theta (x-x_0)(y-y_0) \\
	\label{eqn:modlor}
\end{eqnarray*}

where $A$ is the peak amplitude, $(x_0,y_0)$ the central co-ordinates,
and $\alpha_1$, $\alpha_2$, and $\theta$ define the semi-major and
semi-minor axes, and the position angle of the ellipse. The elliptical
modified Lorentzian used followed that of \citet{Diego:1985}:
\begin{eqnarray*}
  	L(x,y) & = & \frac{A}{\left[1 + B_{12}^{0.5 \gamma\left(1+B_{34}\right)} \right]}, \\
	\label{eqn:moffat}
\end{eqnarray*}

where $\alpha_1$ and $\alpha_2$ are scale factors which relate to the
full-width at half-maximum (Diego's RX and RY), and $\alpha_3$ and
$\alpha_4$ are similar scale factors which allow the exponent of
$B_{ij}$ to vary with position (Diego's PRX and PRY).


In Table \ref{tab:ratio_errors} we showed the flux ratio and
uncertainty derived in each band for both objects along with the
central wavelengths and widths of the observed filters. The
uncertainty on the fitted amplitudes for each image was found from the
covariance matrix calculated by the Levenberg-Marquardt fit, which is
a reasonable estimate of the standard error when, as in this case, the
$\chi^{2}$ function is quadratic.  However, the uncertainties on the
reported flux ratios are the standard errors of the fits for all
images for each filter, which were comparable to the uncertainties
from the covariance matrix. The uncertainties on the flux ratios are
not the dominant uncertainties in the final photometry of the two
objects and so the anti-correlation is negligible.

\begin{figure}
  \centering{}
  \resizebox{\hsize}{!}{\fbox{\includegraphics{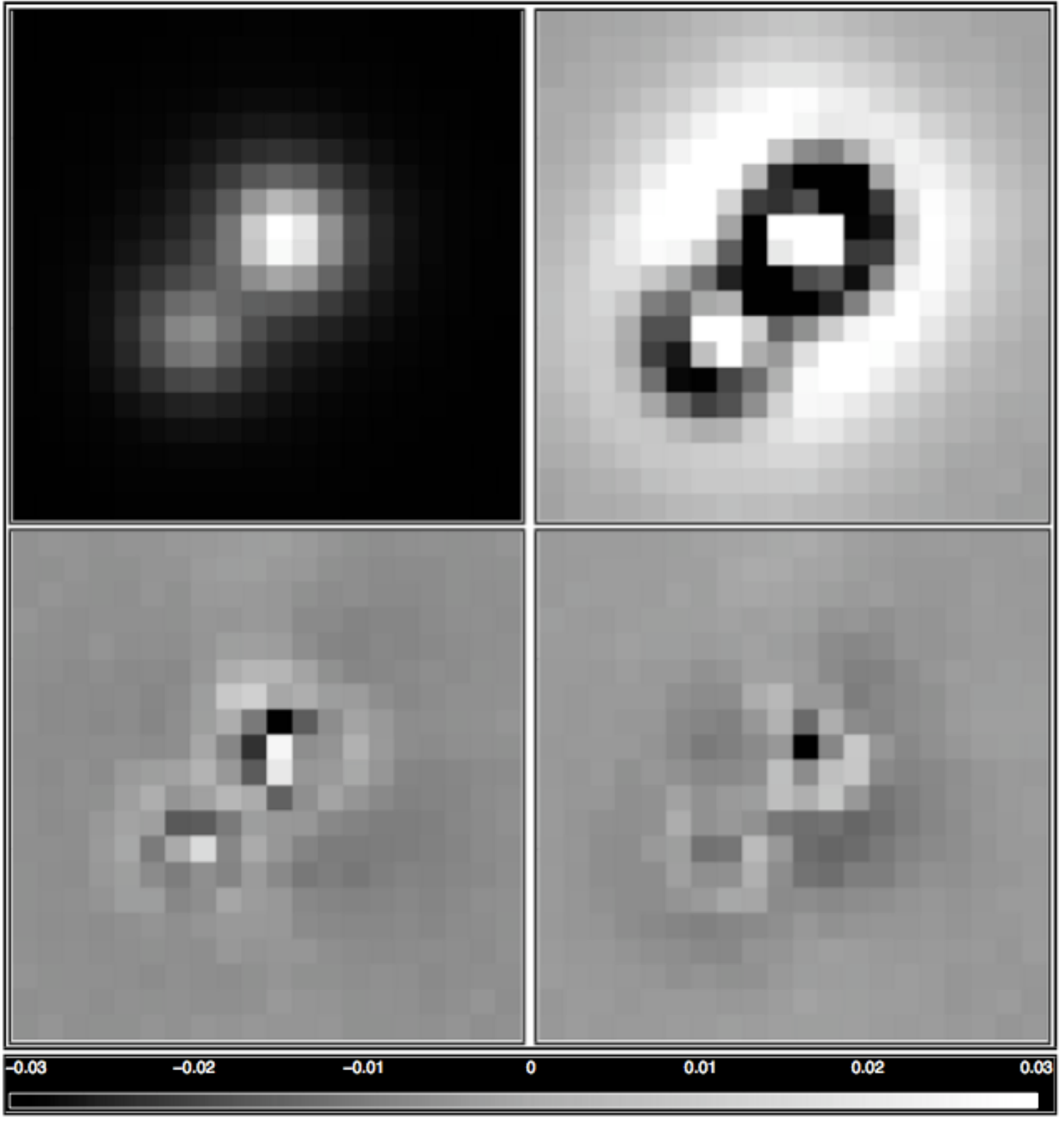}}}
  \caption{An ISAAC $J$-band image of the \epsba, Bb system (top left)
    and the residuals of the Gaussian (top right), modified Lorentzian
    (bottom left), and Moffat profile fits (bottom right). Each image
    is a 20\,$\times$\,20 pixel (2.96\arcsec\,$\times$\,2.96\arcsec)
    sub-section of the full image. North is up, East left.  The scale
    applies to the residual images and shows the flux as a fraction of
    the observed peak pixel flux. It is apparent that the Gaussian
    profile is the worst fit, while the Moffat profile is somewhat
    better than the Lorentzian.}
    \label{fig:j1_20_res}
\end{figure}

The residuals of the best-fit profiles for each of the three functions
are shown in Fig.~\ref{fig:j1_20_res} for an example observation and
demonstrate the suitability of each profile to match the data. The
Gaussian is seen, as expected, to provide the worst fit,
under-estimating the peak and wings while over-estimating in between
as it attempts to match the entire profile. If this profile were used
to fit the relative fluxes of isolated stars in an image, one would
expect it to return reasonable results, but the inability of the
Gaussian to determine the flux in the wings of an object which blends
with the profile of another makes it ineffective at extracting
relative photometry of blended objects. The residuals of the modified
Lorentzian profile however, show much reduced levels, with the Moffat
profile residuals even lower.


As confirmation of the validity of our PSF-fitting routine, it was
employed to fit our optical images which we had previously fit with
the DAOPHOT/IRAF algorithm of \citet{Stetson:1987}. DAOPHOT uses an
analytical profile to model the core of the PSF and builds up an
empirical determination of the wings from the chosen model
stars. Although not completely independent of our method, in that it
too uses an analytical profile, it is the matching of the model PSF to
the wings that allows DAOPHOT to accurately extract the flux of
blended objects. We fit 5 images in the $I$- and $z$-band and 12
images in the $R$-band. Our fitting routine finds flux ratios of
4.71$\pm$0.08, 4.99$\pm$0.05, and 3.78$\pm$0.08 for the $R$-, $I$-,
and $z$-bands respectively, while with DAOPHOT we find flux ratios of
4.81$\pm$0.07, 5.04$\pm$0.05, and 3.85$\pm$0.03. The results of both
PSF-fitting routines agree within the uncertainties in all cases, confirming
the validity of our wholly analytical routine. Although the fitted
flux ratio from DAOPHOT is larger than that from our fitting routine
in these three examples, this is not true for all fitted images.



\section{Spectral Fitting Routine}
\label{sec:spec_fit}

As with the imaging, the spectra of the two brown dwarfs were
partially blended in the spatial direction and so a bespoke fitting
algorithm was implemented to extract the individual spectra. The
process was similar to the image fitting routine.  We iteratively fit
the parameters of a double Gaussian profile to the spatial direction
at each column along the spectra. Initially all parameters are free,
but after the first fit we calculate the mean of the separations of
the two peaks at each wavelength step, weighted by the total flux at
each wavelength, from the $\sim$1000 profiles in each image and refit
the profiles with the separation fixed. We then do the same to derive
a global fit to the spatial FWHM which we take to be constant with
wavelength over each spectral window. The absolute position of the
peaks at each wavelength was then constrained by a trace of the
spectrum across the detector so that an accurate centre would be found
even for wavelengths with little signal-to-noise. With these
constraints, the remaining parameters were refit and the amplitudes
were used along with the wavelength calibration to construct the
individual wavelength-calibrated spectra.

\begin{figure}
  \resizebox{\hsize}{!}{\includegraphics[angle=90]{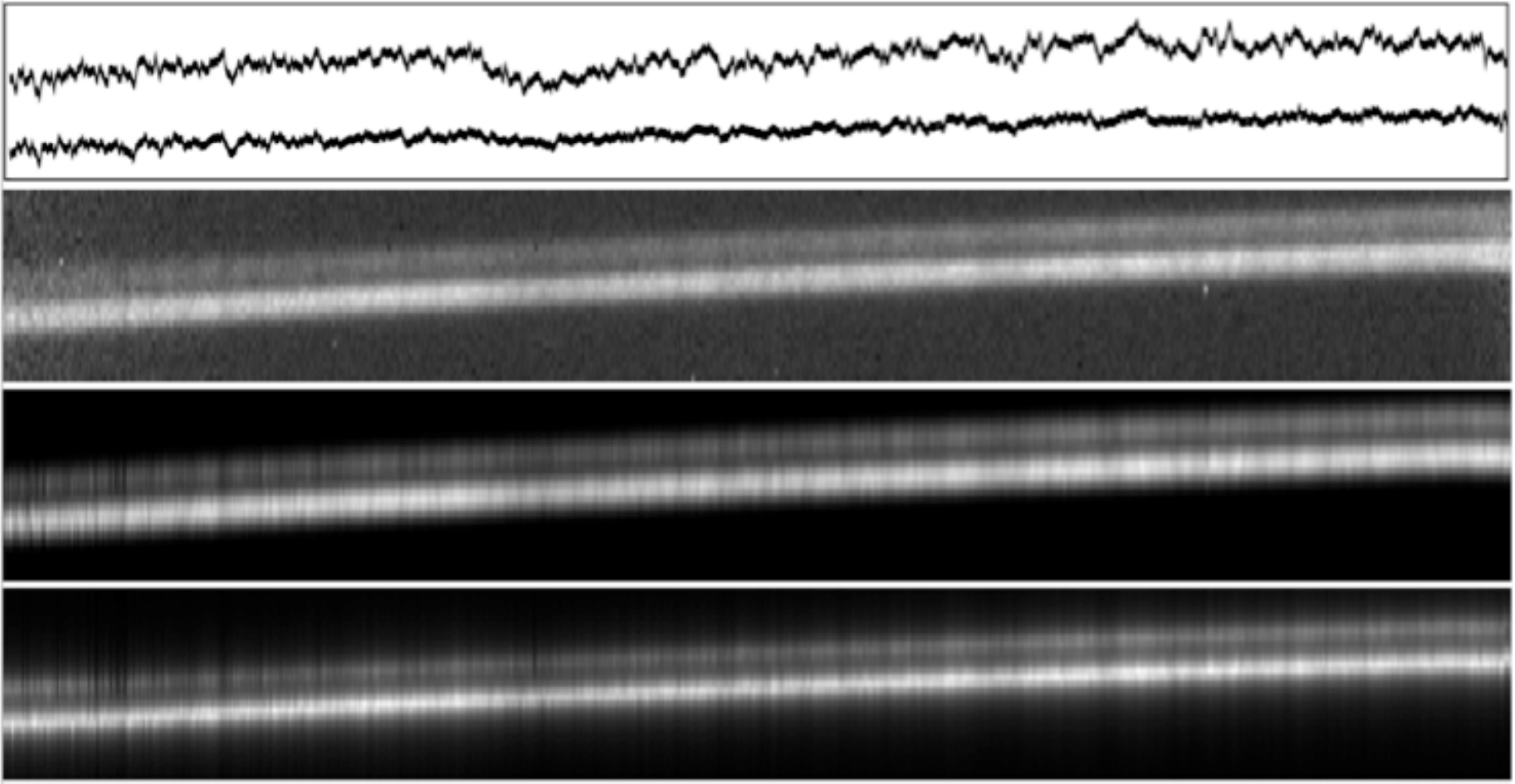}}
  \caption{The top panel shows the spectrum of \epsba\ (upper line)
    and Bb in the region 0.975--1.022\micron\ where absorption by FeH
    and CrH is present. The second panel from the top shows an image
    of the observed spectrum (\epsba\ is the lower spectrum) sloping
    across the detector. The third panel from the top shows the
    corresponding fitted double Gaussian profile and the bottom panel
    shows the fitted double Moffat profile. The vertical striping seen
    in the fits is an effect of utilising the signal of all the pixels
    in the spatial direction to increase the signal-to-noise and so
    detect fine spectral features. }
    \label{fig:spec_fits}
\end{figure}

We found that a double Gaussian profile provided the best fit to our
spectroscopic data, unlike in our broadband imaging where a Moffat
profile was preferred. Figure~\ref{fig:spec_fits} shows an image of
the spectrum of \epsba, Bb in the region 0.975--1.022\micron, along
with fitted spectra, one using Gaussian profiles, the other Moffat
profiles. The apparent vertical striping in the fitted spectra are
real spectral features revealed by using the entire profile to
increase the signal-to-noise.

\begin{figure}
  \resizebox{\hsize}{!}{\includegraphics[angle=90]{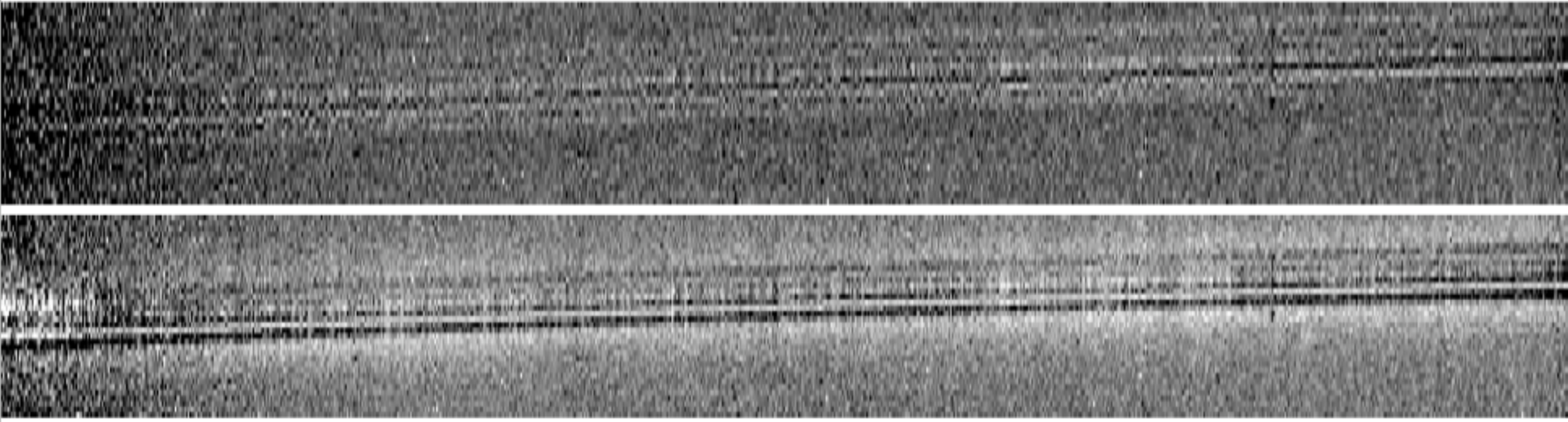}}
  \caption{Residuals of the Gaussian (top) and Moffat profile fits to
    the spectral region shown in Fig\,\ref{fig:spec_fits}. The
    ill-fitting wings of the Moffat profile suggest that it is not a
    good approximation for the spatial profile in our spectroscopic
    images, while the Gaussian profile residuals are almost consistent
    with the background noise.}
    \label{fig:spec_residuals}
\end{figure}

The resulting residuals for each profile
(Fig.~\ref{fig:spec_residuals}) show clearly that the Gaussian profile
is the better match. The Moffat profile under-estimates the peak of
the profile while over-estimating the wings, consequently the
contribution of the flux from one object to the other is not well
determined. While an analytical Gaussian profile should not in
principle be an exact match to the observed stellar spectral profile,
the residuals are almost consistent with the background noise. It may
have been expected that the same profile would be preferred for the
spectroscopy and broadband imaging, but if we consider imaging to be
the integral of many Gaussian profiles with the FWHM varying as a
function of wavelength, then the result would not be truly Gaussian.
Nevertheless, the residual images and spectra clearly favour different
profiles. The resulting full resolution spectra are presented in
Figs.~\ref{fig:Bab_0.6-1.3}--\ref{fig:Bab_1.9-2.5}.

\section{Photometric Calibration}
\label{sec:phot_calib}

\begin{figure} 
\centering{}
\includegraphics[keepaspectratio=true,width=\columnwidth]{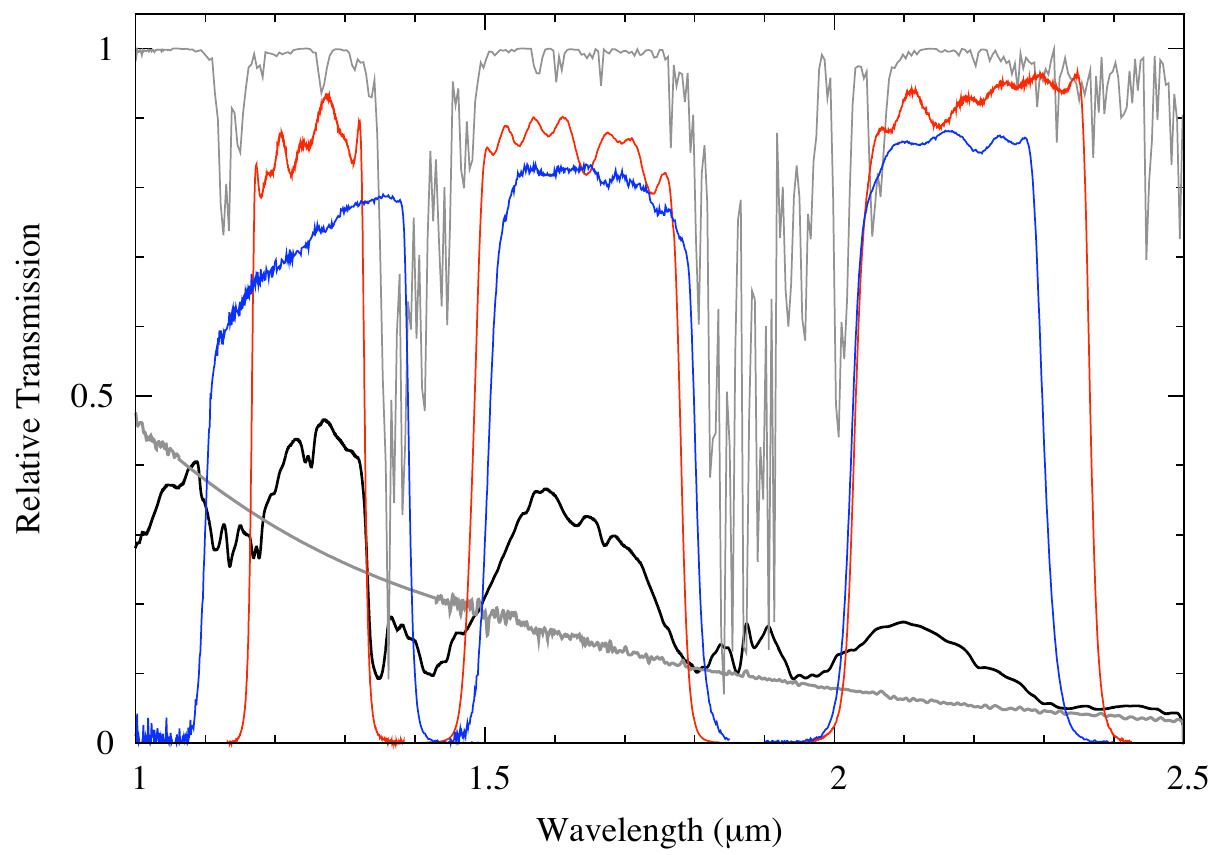}
\caption{The $JHK$ band-passes of the MKO filter-set (red lines) and
  the ISAAC $JHK_{S}$ filter-set (blue lines). The top-most line
  (thin, grey line) traces a typical atmosphere at Paranal showing the
  deep absorption bands between the filters. Also shown is the
  spectrum of a G8V star (thick grey line) from the Pickles spectral
  library \citep{Pickles:1998} along with a smoothed spectrum of
  $\varepsilon$ Indi Ba (thick, black line). This highlights the
  difference between the fraction of each object's flux in the regions
  of high atmospheric absorption. Importantly, we see that the MKO
  filters mostly avoid these regions, while the ISAAC filters,
  especially the $J$-band, extend into them.  Photometry using filters
  which extend into these regions are susceptible to variations in the
  atmosphere and mismatches between target and standard star spectra.}
\label{fig:filter+atm} 
\end{figure}

To allow a meaningful comparison of our sources with other observed
brown dwarfs, it is necessary to place the magnitudes on a common
photometric system due to the differences between filter systems. In
the near- to thermal-IR, the Mauna Kea Consortium filter-set
\citep{Tokunaga:2002} has been chosen for this purpose by several
groups \citep{Stephens:2004,Golimowski:2004,Hewett:2006} since these
filters do not extend into the water absorption bands of the
atmosphere and so avoid the large differences in relative spectral
response between sites and from varying atmospheric conditions. The
differences between the MKO $JHK$ filters and the ISAAC $JHK_{\rm{S}}$
filters used for our observations are shown in
Fig.~\ref{fig:filter+atm}. Many observations have also been reported
in the 2MASS system \citep{Carpenter:2001} due to the large number of
ultra-cool dwarfs found in the 2MASS database, although this system
still encroaches on the regions of telluric absorption. We therefore
present our near-IR photometry of \epsba, Bb in both MKO (Table
\ref{tab:mags}) and 2MASS systems (Table \ref{tab:mags_2mass}) to
allow easy comparison with other data. In the optical, we observed
\epsba, Bb in the FORS2 Bessell $V$ and $I$, the FORS2 $R$ special,
and the FORS2 Gunn $z$ filters and report our photometry in the FORS2
system.

As none of our standard stars were of similar colour to our targets,
when transforming our photometry into standard systems, we could not
use standard colour equations, nor could we ignore colour terms due to
the filter differences. This is clear from Fig.~\ref{fig:filter+atm}
where we see that a T dwarf has relatively little flux in the regions
of high absorption in the Earth's atmosphere (due to the dominance of
H$_2$O in both cases), but the standard star (in this case a
solar-type star) will lose a large fraction of its flux in these
regions. As a result, slight differences in the near-IR profiles of
different systems can result in magnitude differences as large as
$0.1^{\rm{m}}$ \citep[cf.][]{Stephens:2004}.

In the near-IR, photometric calibration used the solar type star,
S234-E \citep{Persson:1998}, as our standard star in the $JHK_{\rm{S}}$
bands with the magnitudes known in the LCO (Las Campanas Observatory)
system. HD205772 (A3) and HR8042 (G3IV) were used in the $L$-band and
$M_{\mathrm{NB}}$-bands, respectively. The $L$-band standard star
magnitude was known in the old UKIRT $L$-band filter and was assumed to
be unchanged on transformation to the MKO $L'$-band as for other A-type
stars \citep[see Fig.\,2 of][and the UKIRT photometric calibration
web-pages\footnote{http://www.jach.hawaii.edu/UKIRT/astronomy/calib/phot\_cal-/ukirt\_stds.html}]{Leggett:2003},
with the spread in the $L/L'$ magnitude differences of A stars being
used as an estimate of the uncertainty on the transformation. For the
$M_{\mathrm{NB}}$-band, the standard star magnitude was known in the
ESO system of \citet{vanderbliek:1996}.

With the flux calibrated spectra of \epsba\ and Bb, we extracted
synthetic photometry in the MKO $JHKL'M'$ filters and the 2MASS
$JHK_S$ filters by convolving the flux-calibrated spectra with the
 appropriate filter profiles and atmosphere for each site. In
deriving our synthetic magnitudes, we used the ISAAC filters convolved
with the model atmosphere given in the ISAAC user
manual\footnote{http://www.eso.org/sci/facilities/paranal/instruments/isaac/doc/}
for typical conditions over Paranal.  For the 2MASS magnitudes, we
used the relative spectral responses of \citet{Cohen:2003} and applied
the zero-point offsets of $+0.001$, $-0.019$, $+0.017$ for the $J$,
$H$, and $K_S$-bands respectively. The MKO filters used were convolved
with the 1.2\,mm PWV (precipitable water vapour) ATRAN model atmosphere
of \citet{Lord:1992}, as used by \citet{Stephens:2004}.  We did not
include the quantum efficiency of the detector nor the transmission
profiles of any other optical elements.  These were assumed to be
practically flat across each near-IR filter as discussed in
\citet{Stephens:2004}. However, although the reflectivity of aluminium
(used to coat the VLT mirrors) is relatively constant across the
near-IR\footnote{http://www.gemini.edu/files/docman/press\_releases/pr2004-5/images/comparison\_AgAl\_02.GIF},
we note that it is strongly wavelength-dependent in the optical regime.
We therefore did not attempt to derive our optical photometry in the
same manner.

That said, the precise optical filters used significantly affect the
derived photometry as the flux of a T dwarf rises by more than three
orders of magnitude over the range 0.6--1.0\micron. Since our observed
spectra do not cover the full $V$- and $R$-bands, we have not
attempted to transform our optical photometry via synthetic
photometry. Instead, we assumed that the magnitudes of our standard
stars, as given by \citet{Landolt:1992} ($VRI$ in the system of
\citet{Bessell:1990}) and SDSS ($z$), were the same as that in the
FORS2 system as would be the case for an A0 star. The spread in
derived zero-points for the different standard stars were smaller than
the other uncertainties which leads us to believe that the precise
spectral type of the standard stars did not significantly affect the
derived photometry. However, our optical photometry of \epsba, Bb is
not what would be measured through the standard Bessell filters and so
is not readily compared with other T dwarf observations.




\end{document}